\documentclass[preprint2]{aa} 
\usepackage{graphicx}
\usepackage{txfonts}
\usepackage{natbib}
\bibpunct{(}{)}{;}{a}{}{,} 
\newcommand{\1}{{~\sc i}}
\newcommand{\2}{{~\sc ii}}
\newcommand{\3}{{~\sc iii}}
\newcommand{\4}{{~\sc iv}}
\newcommand{\5}{{~\sc v}}
\newcommand{\6}{{~\sc vi}}

\newcommand{\wmsr}{{\,W\,m$^{-2}$\,sr$^{-1}$}}
\newcommand{\kms}{{\,km\,s$^{-1}$}}
\newcommand{\cc}{{\,cm$^{-3}$}}
\newcommand{\mic}{{\,$\mu$m}}

\begin{document}

 \title{Chemical enrichment and physical conditions in I\,Zw\,18}

   
   	\authorrunning{Lebouteiller et al.}

   \author{Vianney Lebouteiller\inst{1}, Sara Heap\inst{2}, Ivan Hubeny\inst{3}, Daniel Kunth\inst{4}
          }

   \institute{$^1$ Laboratoire AIM, CEA/DSM-CNRS-Universit\'e Paris Diderot DAPNIA/Service d'Astrophysique B\^at. 709, CEA-Saclay F-91191 Gif-sur-Yvette C\'edex, France
              \email{vianney.lebouteiller@cea.fr} \\
              $^2$ Laboratory of Astronomy and Solar Physics, NASA Goddard Space Flight Center, Greenbelt, MD 20771 \\
              $^3$ Department of Astronomy and Steward Observatory, University of Arizona, Tucson, AZ 85721 \\
              $^4$ Institut d'Astrophysique, Paris, 98 bis Boulevard Arago, F-75014 Paris}

   \date{Received ...; accepted ...}

 
  \abstract
   {Low-metallicity star-forming dwarf galaxies are prime targets to understand the chemical enrichment of the interstellar medium (ISM). The H\1\ region contains the bulk of the mass in BCDs, and it provides important constraints on the dispersal and mixing of heavy elements released by successive star-formation episodes. The metallicity of the H\1\ region is also a critical parameter to investigate the future star-formation history, as metals provide most of the gas cooling that will facilitate and sustain star formation. }
   {Our primary objective is to study the enrichment of the H\1\ region and the interplay between star-formation history and metallicity evolution. Our secondary objective is to constrain the spatial- and time-scales over which the H\1\ and H\2\ regions are enriched, and the mass range of stars responsible for the heavy element production. Finally, we aim to examine the gas heating and cooling mechanisms in the H\1\ region. }
   {We observed the most metal-poor star-forming galaxy in the Local Universe, I\,Zw\,18, with the Cosmic Origin Spectrograph onboard \textit{Hubble}. The abundances in the neutral gas are derived from far-ultraviolet absorption-lines (H\1, C\2, C\2*, N\1, O\1, ...) and are compared to the abundances in the H\2\ region. Models are constructed to calculate the ionization structure and the thermal processes. We investigate the gas cooling in the H\1\ region through physical diagnostics drawn from the fine-structure level of C$^+$. }
   {We find that H\1\ region abundances are lower by a factor of $\sim2$ as compared to the H\2\ region. There is no differential depletion on dust between the H\1\ and H\2\ region. Using sulfur as a metallicity tracer, we calculate a metallicity of $1/46$\,Z$_\odot$ (vs.\ $1/31$\,Z$_\odot$ in the H\2\ region). From the study of the C/O, [O/Fe], and N/O abundance ratios, we propose that C, N, O, and Fe are mainly produced in massive stars. We argue that the H\1\ envelope may contain pockets of pristine gas with a metallicity essentially null. Finally, we derive the physical conditions in the H\1\ region by investigating the C\2* absorption line. The cooling rate derived from C\2* is consistent with collisions with H$^0$ atoms in the diffuse neutral gas. We calculate the star-formation rate from the C\2* cooling rate assuming that photoelectric effect on dust is the dominant gas heating mechanism. Our determination is in good agreement with the values in the literature if we assume a low dust-to-gas ratio ($\sim2000$ times lower than the Milky Way value).   } 


   \maketitle

\section{Introduction}

Numerical simulations by \cite{Cen99}, predicting the evolution of the metal content of the Universe, show that metallicity is a stronger function of density than age. It is therefore plausible that gas-rich dwarf galaxies in the Local Universe may remain in an early stage of chemical enrichment, possibly preserving pockets of pristine gas \citep{Kunth86,Kunth94}. Blue Compact Dwarf (BCD) galaxies represent a major testbed as they include the most-metal poor star-forming galaxies known \citep{Kunth00}. 
Among these, I\,Zw\,18 (Mrk\,116) continues to attract considerable interest and to feed intense debates in extragalactic research. For more than three decades, I\,Zw\,18, discovered by \cite{Zwicky66} and first studied by \cite{Sargent70} held the record as the most metal-deficient galaxy known, with an oxygen abundance $12+\log {\rm (O/H)} = 7.17\pm0.01$\footnote{This translates into a metallicity of $\approx1/31$\,Z$_\odot$ with the solar oxygen abundance taken from \cite{Asplund09}.} in its NW component and $7.22\pm0.02$ in its SE component \citep{Izotov99b}. Only recently has it been displaced by two more metal-deficient BCDs, SBS\,0335--052W with $12+\log {\rm (O/H)} = 7.12\pm0.03$ \citep{Izotov05} and DDO\,68 with $12+\log {\rm (O/H)} = 7.14\pm0.03$  \citep{Izotov07}.

It was postulated that the difficulty of finding more metal-poor objects could be explained by thorough local mixing of the metals produced and released over a single starburst episode \citep{Kunth86}. This hypothesis, implying the existence of quasi-pristine gas beforehand, has been since challenged by observations of uniform abundances in disconnected star-forming regions within a single dwarf galaxy (e.g., \citealt{Skillman93,Kobulnicky97,Noeske00}). Although evidence of local metal enrichment has already been observed (e.g., \citealt{Walsh93}), it is usually associated with the presence of Wolf-Rayet (WR) stars \citep{Walsh89,Thuan96,Lopez07}. This hypothesis was later confirmed by \cite{Brinchmann08}, with the study of a large sample of WR galaxies from the Sloan Digital Sky Survey.

Chemical abundances and metallicity are key parameters for the evolution of galaxies. Abundances in BCDs are usually probed in the ionized gas of the H\2\ regions, as derived from optical emission-lines. While H\2\ regions correspond to the present star-formation episode, they can also be related to previous episodes via the triggered star-formation mechanism, in which star formation take place following compression of molecular clouds by massive stellar winds and supernovae (e.g., \citealt{Vanhala98,Bhattal98,Fukuda00,Pustilnik01b,Lee07,Elmegreen11}). H\2\ regions represent however a small fraction of the galaxy in mass as compared to the surrounding H\1\ region \citep{Thuan81}. 

The H\1\ region provides another way of measuring the metallicity, which accounts for the bulk of the mass of a galaxy. The H\1\ region sets a reference abundance pattern, caused by metal enrichment over long time scales and large spatial scales, that needs to be understood and constrained. With this in mind, the comparison of abundances between the H\1\ region and the H\2\ regions allows us to examine the dispersal and mixing of heavy elements from recent starburst episodes, and to constrain stellar evolution models. 
Furthermore, since future star formation thrives on the H\1\ gas reservoir, the H\1\ region makes it possible to study the interplay between star-formation history and metallicity. While past and present star-formation episodes enrich the ISM in heavy elements, the latter are, in turn, able to facilitate and sustain star-formation through several gas cooling mechanisms (e.g., [C\2] 157\mic, CO, and dust radiative cooling). The main questions that drive the studies of the H\1\ region in BCDs are the following (1) What mechanisms control the H\1\ region metallicity? (2) What is the influence of metallicity on the star-formation history?

BCDs are ideal targets because they display large amounts of H\1\ gas \citep{Thuan81} and because the massive stars provide a strong far-ultraviolet (FUV) continuum on which absorption-lines from neutral species located along the line of sight are superimposed. 
\cite{Kunth94} used the GHRS instrument on board the \textit{Hubble} Space Telescope to observe interstellar absorption lines  toward the many massive stars in I\,Zw\,18. The authors detected the O\1\ line at $1302.2$\,\AA\ and, together with the neutral hydrogen content measured from the 21\,cm line, they estimated the metallicity of the neutral envelope to be at least 10 times lower than that of the ionized gas in the H\2\ regions. Later, \cite{Thuan97} obtained a \textit{GHRS} spectrum for SBS\,0335--052E and found the metallicity of the neutral gas to be similar to that of the ionized gas. 
These results remained however inconclusive because of possible saturation effects of the O\1\ absorption line  \citep{Pettini95}. 

A further step forward was then achieved with the launch of the Far Ultraviolet Spectroscopic Explorer (FUSE; \citealt{Moos00}), which allowed the observation of H\1\ absorption-lines together with many metallic species such as N\1, O\1, Si\2, P\2, Ar\1, and Fe\2. The spectral resolution of FUSE ($\lambda/\Delta\lambda\sim20000$) and its sensitivity made it possible to detect numerous absorption-lines, some of them apparently not saturated. The neutral gas chemical composition was derived in several BCDs, spanning a wide range in ionized gas metallicity, from $1/30$ to $1/3$\,Z$_\odot$, with I\,Zw\,18 (\citealt{Aloisi03}, hereafter A03; \citealt{Lecavelier04}, hereafter L04), SBS\,0335--052 \citep{Thuan05a}, I\,Zw\,36 \citep{Lebouteiller04}, Pox\,36 \citep{Lebouteiller09}, Mark\,59 \citep{Thuan02}, NGC\,625 \citep{Cannon05}, and NGC\,1705 \citep{Heckman01,Lee04}. These investigations showed that the neutral gas of BCDs is not pristine; it has already been enriched with metals, with a minimal metallicity of $\sim1/50$\,Z$_\odot$ (see the FUSE sample analysis in \citealt{Lebouteiller09}). The second most important result is that the metallicity of the neutral gas is generally lower than that of the ionized gas $-$ except for the two lowest-metallicity BCDs, I\,Zw\,18 and SBS\,0335--052E $-$ implying that the H\1\ region has been probably less processed than the H\2\ region. 

However the interpretation of these results has been limited by the lack of constraints on the properties of the gas detected in absorption in the FUV (physical location along the line of sight, density structure, temperature). FUV observations select out dust-free lines of sight toward the massive stars, and the intervening gas might have peculiar properties as compared to the global ISM. 
 Uncertainties also remain in deriving column densities from absorption-lines in complex systems. Most FUSE studies of BCDs assumed a single absorption component arising from a single homogeneous line of sight with the exception of Mark\,59 (\citealt{Thuan02}; see also \citealt{Lebouteiller06} for the giant H\2\ region NGC\,604 in M\,33). A more realistic approach would be to consider multiple lines of sight toward massive stars, each line of sight intersecting clouds with possibly different physical conditions (such as turbulent and radial velocity) and chemical properties. Whenever individual absorption components are unresolved, it is almost impossible to tell whether they are saturated, in which case the column density inferred from the global absorption line can be severely underestimated (see e.g., \citealt{Lebouteiller06}). Following the suggestion of \cite{Kunth86}, \cite{Bowen05} got around the problem by deriving the abundances in the neutral gas of the dwarf spiral galaxy SBS\,1543+593 along a quasar line of sight. The authors found that the abundances agree with those in the ionized gas of the brightest H\2\ region in the arms, contrasting with the FUSE results in BCDs. 

With the advent of the \textit{Hubble}'s Cosmic Origins Spectrograph (COS; \citealt{Green12}) we have the opportunity to look anew at the problem of the chemical discontinuity between the ionized and neutral gas in BCDs, with in particular the observation of the S\2\ multiplet, which \cite{Pettini95} have proposed as a tracer of the neutral gas enrichment.
The cornerstone galaxy I\,Zw\,18 is therefore once again put to the test, not only because of its extremely low metallicity content. As it turns out this object has led to discordant analysis of the same dataset obtained with FUSE by two independent groups of scientists. A03 used an individual line fitting method and found abundances globally lower in the neutral gas as compared to the ionized gas, while L04 find no significant difference between the two phases. L04 find O\1/H\1$=-4.7^{+0.8}_{-0.6}$, which is consistent with the O/H ratio observed in the H\2\ regions ($-4.83\pm0.03$; \citealt{Izotov99b}), while A03 report a significantly different value with O\1/H\1$=-5.4\pm0.3$, a discrepancy with the former result that awaits clarification. This is indeed worrisome since this object has also a considerable weight in  the study of depletion and that of a possible threshold for chemical enrichment. Most importantly, the discrepancy between the observational results complicates the interpretation of chemical evolution models. \cite{Recchi04} found that abundances in the H\1\ medium (defined by a temperature lower than $7000$\,K) should be similar to those in the H\2\ gas. A definitive observational constraint is needed to challenge the models.

While providing accurate measurements of chemical abundances in the neutral gas, the COS observations also shed a new light on the physical conditions and thermal processes in the H\1\ region. The absorption line C\2* $\lambda1335.7$, arising from the fine-structure level of C$^+$, is intimately related to the far-IR (FIR) cooling line [C\2] 157\mic\ that was recently detected in I\,Zw\,18 with the \textit{Herschel} space telescope (Lebouteiller et al.\ in prep.). We investigate in the present study the heating and cooling mechanisms in the H\1\ region in order to understand the present and future star-formation history of the galaxy. 

Our study is organized as follows. We present the observations in Sect.\,\ref{sec:observations}. Column densities are determined in Sect.\,\ref{sec:cds}. After the description of the nebular models in Sect.\,\ref{sec:nebular}, we examine the physical conditions in the H\1\ region in Sect.\,\ref{sec:ciiratio}. We then derive and interpret elementary abundances in Sect.\,\ref{sec:abs_discuss}. Finally, we discuss chemical evolution scenarios in Sect.\,\ref{sec:discussion}.

\section{Observations and data processing}\label{sec:observations}

\subsection{Datasets}\label{sec:datasets}

\begin{figure}
\centering
\includegraphics[angle=0,width=9cm,clip=true]{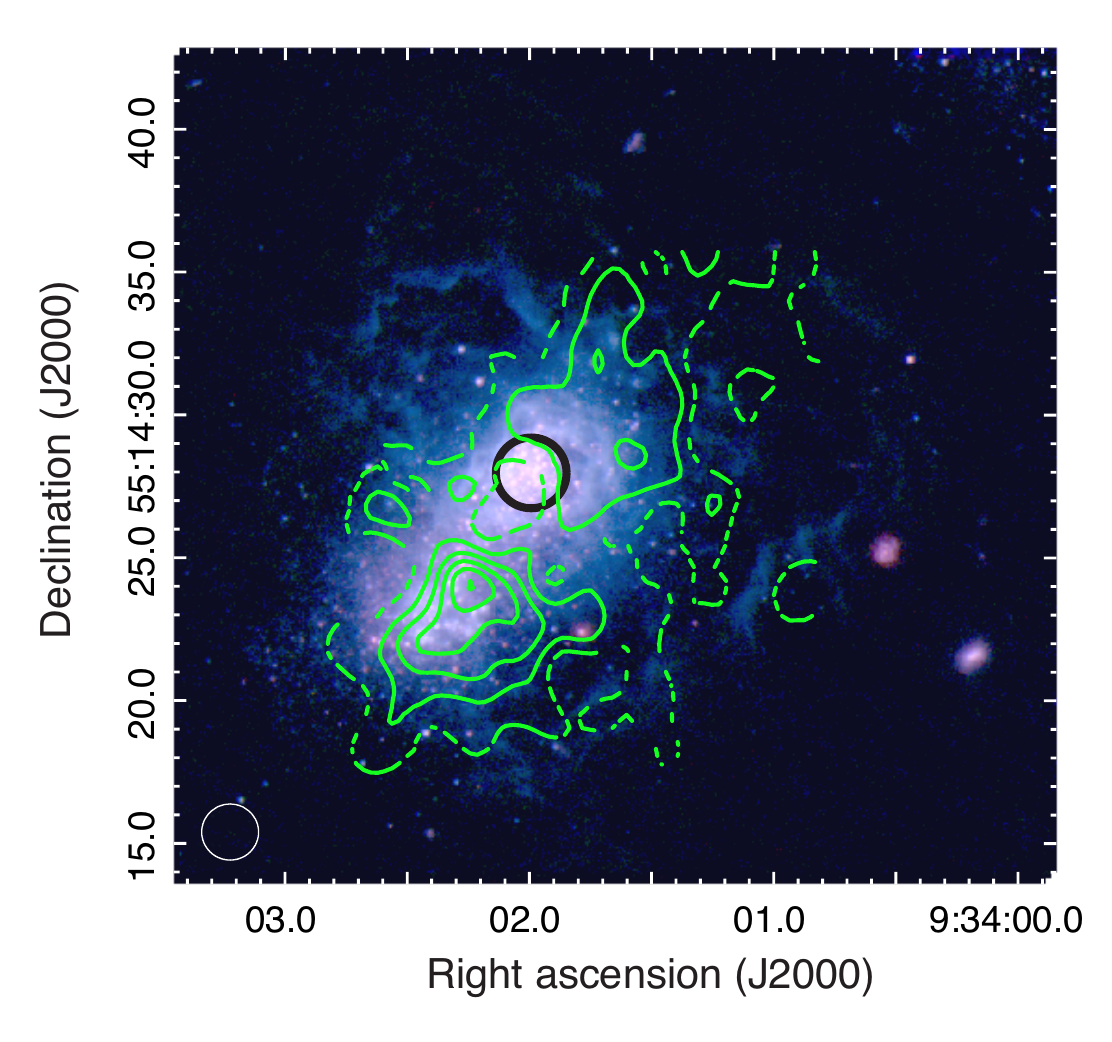}
\caption{\textit{Hubble}/ACS image of I\,Zw\,18 (main body), with RBG colors defined as $I$, $V$, and $B$ filters. 
Images were downloaded from the \textit{Hubble} Legacy Archive (\textit{http://hla.stsci.edu/}) and later combined. The green contours show the H\1\ column density distribution from \cite{Lelli12} with $2$\arcsec\ resolution (beam size in the bottom left). Contours are drawn for $3$ (dashed), $6$, $9$, $12$, and $15\times10^{21}$\,cm$^{-2}$. The COS observation, represented by the black circle ($2.5\arcsec$ diameter), is centered on the NW star cluster. 
}
\label{fig:acs}
\end{figure}

\begin{table}
\caption{COS datasets.\label{tab:obs_log}}
\centering
\begin{tabular}{lllll}
\hline\hline
\# & $\alpha$ & $\delta$ & Grating/$t^{\rm a}$  \\
 & (J2000) &  (J2000) &    \\
 \hline
 11523 & 09:34:02.00 & +55:14:28.0 & G130M/3.4, G160M/7.3,    \\
          &                   &                    &  G185M/8.1    \\ 
 12028 & 09:34:02.00 & +55:14:28.0 &  G130M/8.2, G160M/11.4  \\ 
 11579 & 09:34:01.97 & +55:14:28.1 &  G130M/15.5     \\ 
 \hline
 \end{tabular}\\
\tablefoot{The exact coordinates for observations 11523 and 12028 are unknown because a peak-up was performed beforehand to target the UV brightest region. Coordinates are J2000. }
\tablefoottext{a}{Exposure time in units of $\times10^3$\,s.}
\end{table}

To first approximation, I\,Zw\,18 is composed of two young, massive star clusters, NW and SE, which ionize the surrounding medium. We used COS on the \textit{Hubble} Space Telescope (HST) to observe the brightest region in the FUV, NW, in April 2010 (program 11523) and later in December 2010 (program 12028). Both observations were made with the circular aperture, with a diameter $D\sim 2.5\arcsec$, corresponding to $\sim230$\,pc at the distance of I\,Zw\,18 ($\approx19$\,Mpc; \citealt{Aloisi07}). Observations were centered on NW (see Fig.\,\ref{fig:acs} and the observation log in Table\,\ref{tab:obs_log}). 
Target acquisition consisted of a slew to the specified coordinates of NW followed by a peak-up in ultraviolet light. Although both observing programs made use of the FUV gratings G130M and G160M, the two sets of exposures cover somewhat different wavelength range. Together, they span $\sim 1135-1760$ \AA. We also use the archival dataset from program 11579 which observed with the G130M grating ($\sim1132-1433$\AA). No peak-up was performed for the latter observation and the telescope was slewed to the requested coordinates of NW. The G185M spectrum was not used because of the low data quality\footnote{The G185M spectrum contains in particular the Si\2\ $\lambda1808.0$ line, with an oscillator strength about $50$ times lower than the weakest Si\2\ line in our study, $\lambda1546.7$. However, the low S/N of the spectrum,  due to the almost fully illuminated aperture (Sect.\,\ref{sec:broadening}) and to cross-contamination between the spectral stripes of the grating on the detector, does not allow setting a useful constraint on the Si\2\ column density.}.

\begin{figure}
\centering
\includegraphics[angle=0,scale=0.7,clip=false]{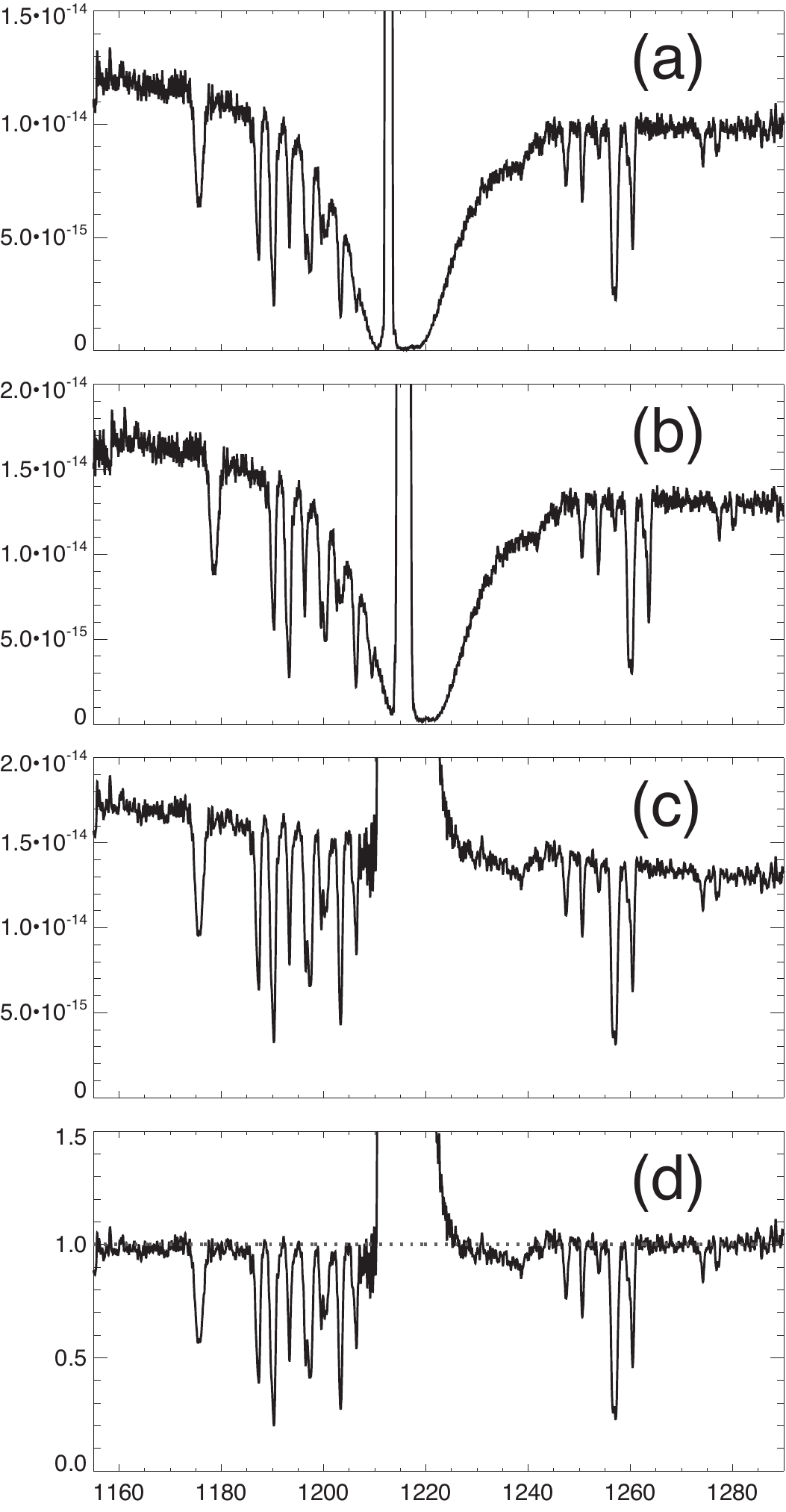}
\caption{Processing of the observed spectrum \textit{(a)} consists of correcting for reddening and for Ly$\alpha$ absorption in the Milky Way and in the high-velocity cloud \textit{(b)}; fitting and removing Ly$\alpha$ from I\,Zw\,18 (c); and normalizing the spectrum \textit{(d)}. The residual emission at Ly$\alpha$ wavelength is due to geocoronal H\1. }
\label{fig:models}
\end{figure}

The pipeline reduction procedure is continually being improved, and we obtained our spectra well after the observations were made (July 2011) from the \textit{Hubble} Data Archive, which makes an ``on-the-fly" reduction using the best calibrations available. We combined exposures from a given program to form a single, co-added spectrum. We did not merge the spectra from the three programs because the pointing was slightly different. This results in differences in the spectra, with for instance the C\4\ $\lambda1548$, $\lambda1550$ doublet, which is formed in stellar winds and in the H\2\ region.

Among the 3 available datasets, the 11579 observation has the longest exposure time below $\sim1433$\AA, while the 12028 observation has the longest exposure time above (Table\,\ref{tab:obs_log}). The 12028 observations should be superior to the 11523 observations by virtue of their longer exposure times. Indeed, the higher signal-to-noise ratio (S/N) in the 12028 spectra enabled us to discern broad, shallow stellar absorption lines that we had not detected before in the other datasets. Also, the 12028 observations were accomplished in a single 5-orbit visit, so we are assured of a constant telescope pointing throughout the visit, whereas in program 11523, the G130M exposures and G160M exposures were obtained in different visits. 

In some exposures, geocoronal emission (in particular from the O\1\ $1302,1305,1306$\AA\ multiplet) is contaminating the interstellar line profiles. We ignored the corresponding exposures for the wavelength range around the geocoronal emission.

As a preparation for the fitting of the interstellar lines, we prepared each COS spectrum in several steps as illustrated in Fig.\,\ref{fig:models}. In the four steps, we:
\begin{itemize}
\item Corrected the observed spectrum for foreground Galactic extinction by dust, using the color excess $E(B-V)=0.032$ \citep{Schlegel98}. This correction is used for the normalization by the stellar model (see below); 
\item Corrected for Ly$\alpha$ absorption in the galaxy and for Ly$\alpha$ absorption by the high-velocity cloud in the line of sight to I\,Zw\,18, following \cite{Kunth94};
\item Measured the column density of Ly$\alpha$ in I\,Zw\,18 and removed Ly$\alpha$ (Sect.\,\ref{sec:hicd});
\item Normalized by the stellar spectrum (continuum and absorption in photospheres; Sect.\,\ref{sec:stellar}). Figure\,\ref{fig:spectra} shows the spectra along with the normalized stellar model. Final spectra are shown in the Appendix.
\end{itemize}

\begin{figure*}
\centering
\includegraphics[angle=0,scale=0.45,clip=true]{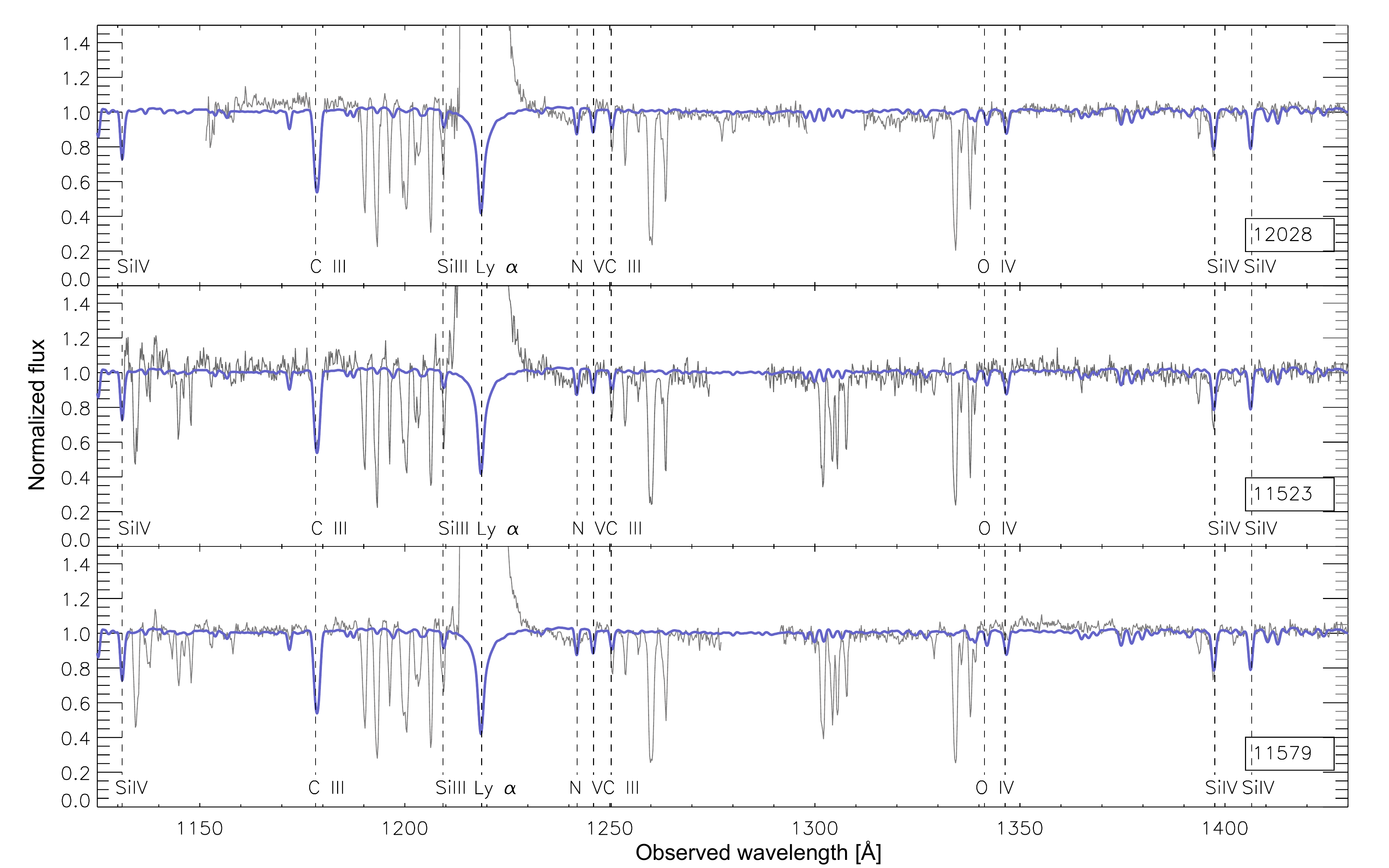}
\includegraphics[angle=0,scale=0.45,clip,trim=0 200 0 0]{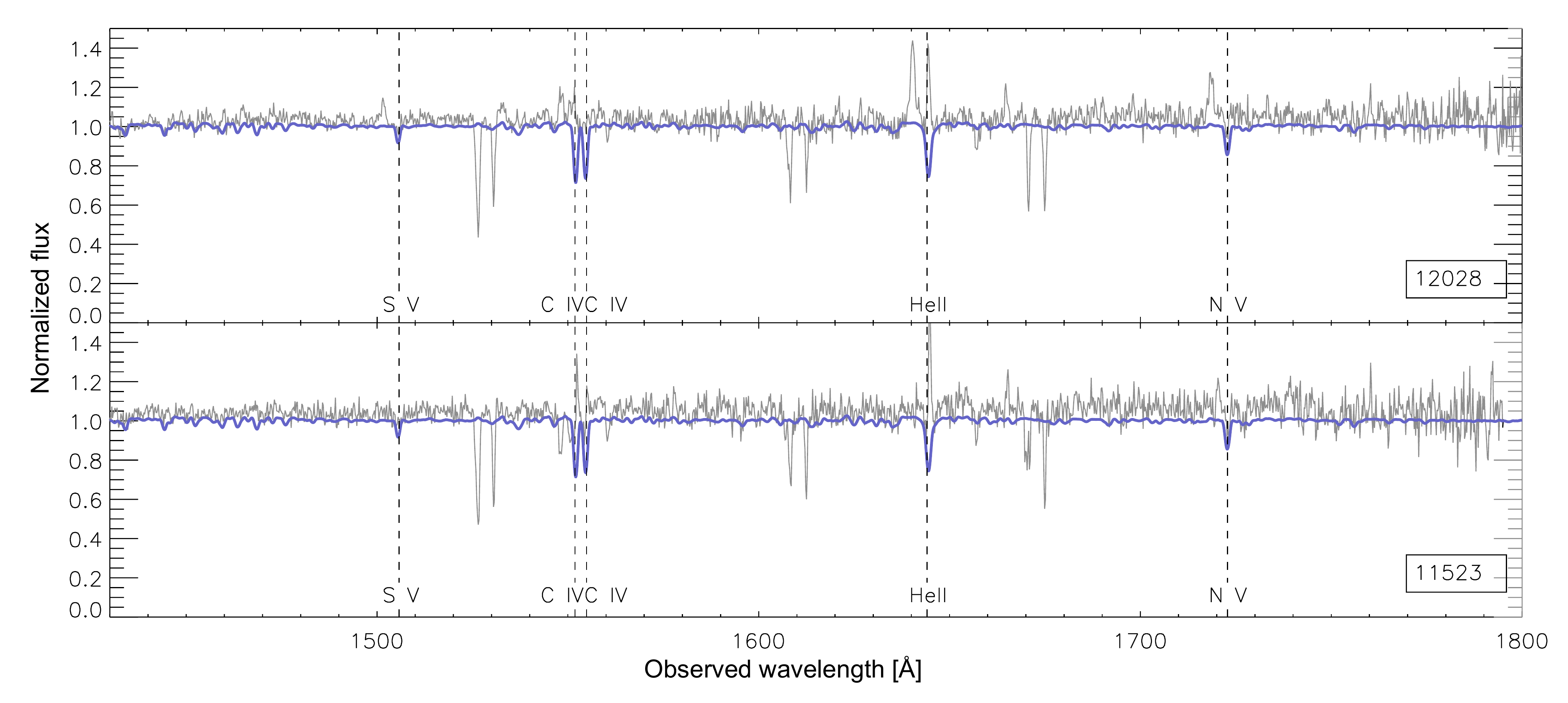}
\caption{COS spectra of I\,Zw\,18. The Ly$\alpha$ absorption (from I\,Zw\,18, the Milky Way, and the high-velocity cloud) was removed and the flux was normalized to the local continuum. 
The numerous absorption features often come in
pairs: the weaker $-$ redshifted $-$ absorption feature is from I\,Zw\,18; the stronger 
feature arises in the Galaxy. A non-LTE, photospheric model spectrum
(no winds) of the stellar population (10\,Myr age with a continuous star formation hypothesis) and $\log (Z/Z_\odot)=-1.7$
is shown in blue. While many features such as C\3\ $\lambda1175$ and the Si\4\ $\lambda1400$ doublet 
appear to be well matched by the stellar model, other high-ionization features such
as He\2\ $\lambda1640$ and the C\4\ $\lambda1549$ doublet are not well matched. The origin of these
lines (e.g., WR stars, H\2\ region...) is unclear. }
\label{fig:spectra}
\end{figure*}

\subsection{Contamination by stellar absorption}\label{sec:stellar}

The FUV spectrum of I\,Zw\,18  is clearly composite, with contributions from stars, H\2\ region, and H\1\ region. 
Absorption in stellar photospheres needs to be removed since they may contaminate interstellar absorption-lines.
We calculated model spectra of young stellar populations using grids of high-resolution (0.2 \AA) non-LTE photospheric spectra (900-2000 \AA) by \cite{Lanz03,Lanz07} and Geneva evolutionary tracks \citep{Lejeune01}.  

High-resolution imaging studies \citep{Hunter95,Heap13} indicate that  I\,Zw\,18 is undergoing continuous star formation rather than an instantaneous burst. Also, it is clear from deep HST/ACS observations that I\,Zw\,18 has a widely extended, older stellar population \citep{Aloisi07,Contreras11}. However, it is equally clear that young, massive stars are the dominant contributors to the UV flux of the NW component. Observations of individual stars in I\,Zw\,18 with WFC2 \citep{Hunter95} and STIS \citep{Brown02,Heap13} on \textit{Hubble} indicate that the NW massive star cluster has a range of ages. The brightest stars in the UV are also the youngest, and their inferred masses range up to $150$\,$M_{\odot}$. 
We therefore assumed continuous star formation and a Salpeter initial mass function with lower and upper mass limits, $1-150$\,$M_{\odot}$, to calculate the spectrum of the stellar population for ages (i.e., the duration of star formation up until now) ranging from $0$ to $100$\,Myr. We then convolved each calculated spectrum with a rotational broadening function ($v \sin i=150$\kms) and with the instrumental function (Sect.\,\ref{sec:broadening}). From a comparison of these models to photospheric lines in the COS spectrum, we derive a continuous-star-formation age of $10-15$\,Myr and a stellar metallicity $\log (Z/Z_{\odot}) \sim -1.7$. 

Using the age and metallicity derived above, we calculated the overall spectral energy distribution (SED) of stars in the NW component using the grids of low-resolution, LTE spectra by \cite{Castelli03}. This grid and the \texttt{TLUSTY/synspec} grid only cover effective temperatures up to $50000$\,K, somewhat lower than the highest temperatures of massive stars on the zero-age main sequence ($65000$\,K), and definitely lower than highly evolved stars such as WR stars. It is therefore likely that our model SED underestimates the ionizing flux of the stellar population.

Figure\,\ref{fig:spectra} shows the stellar models overplotted on each dataset. We find that the contamination from stellar lines is negligible for most of the interstellar lines from I\,Zw\,18 and also from the local absorption systems (Sect.\,\ref{sec:abs_sys}). One exception is the contamination of S\2\ $\lambda1250.6$ in the Milky Way by the stellar C\3\ line. The stellar model was used to normalize the observed spectrum so as to isolate the purely interstellar absorption.

\section{Column density determination}\label{sec:cds}

\subsection{Method}\label{sec:abs_sys}

Once the spectrum has been corrected for stellar absorption (Sect.\,\ref{sec:stellar}), several interstellar absorption line systems can be distinguished along the line of sight to the massive stars in I\,Zw\,18 (Table\,\ref{tab:abs_sys}). 
The main interstellar lines are listed in Table\,\ref{tab:lines}. The atomic data, in particular the oscillator strength values, were taken from \cite{Morton03}, except for P\2\ and S\2\ for which we used \cite{Federman07} and \cite{Podobedova09} respectively. The I\,Zw\,18 component is visible at a velocity of $\approx755$\kms. Two local components are required at $\approx-10$\kms\ and $\approx-90$\kms. In addition, a component at $\approx-60$\kms\ is detected that corresponds to ionized species in the Milky Way. Finally, the high-velocity cloud ``A\6" in complex A is detected at $\approx-171$\kms\ \citep{Hulsbosch88,Wakker97}.

\begin{table}
\caption{Absorption systems.\label{tab:abs_sys}}
\centering
\begin{tabular}{lllll}
\hline\hline
System & species & $v^{\rm a}$ & $b^{\rm b}$ \\
         &   & (\kms) & (\kms) & \\
 \hline
 I\,Zw\,18 & neutral$^{\rm c}$ & $\approx755$ & $\approx17.8$  \\ 
 Milky Way &  neutral$^{\rm c}$ & $\approx-10$ & $\approx22$ \\ 
       & neutral$^{\rm c}$  & $\approx-90$ & $\approx7$  \\ 
 Milky Way & ionized (S\3, Si\3, P\3) & $\approx-60$ & $\approx85 $  \\
 A\6\ cloud$^{\rm d}$ & neutral$^{\rm c}$ & $\approx-171$ & $\approx21$ \\ 
 \hline
 \end{tabular}\\
 \tablefoottext{a}{Radial velocity.}
 \tablefoottext{b}{Turbulent velocity.}
 \tablefoottext{c}{The species in the neutral component are those listed in Table\,\ref{tab:cds}.}
 \tablefoottext{d}{High-velocity cloud.}
\end{table}

We noticed a relative shift in the velocity of the neutral component within I\,Zw\,18 between the datasets: $756\pm3$\kms\ for 12028, $750\pm2$\kms\ for 11523, and $768\pm3$\kms\ for 11579 (Table\,\ref{tab:velocities}). The same relative shifts are observed for the Milky Way component, suggesting an instrumental origin. We attribute this difference to a different distribution of the UV flux in and outside the aperture. As a comparison to our velocity determinations, the H$\alpha$ observation of the NW component, probing the ionized gas in the H\2\ region, is $742\pm7$\kms \citep{Petrosian97} while the H\1\ 21\,cm line  has a velocity of $\approx740$\kms\ \citep{Lelli12}. Table\,\ref{tab:velocities} also shows the radial velocities inferred from each species individually.

The column densities were calculated with the line fitting routine \texttt{Owens} \citep{Lemoine02}. \texttt{Owens} was originally developed  to solve the complexity of many interleaved absorption systems in the FUSE spectra. Although line blending is not a significant issue in the COS spectrum of I\,Zw\,18, we used \texttt{Owens} because of its ability to perform a simultaneous fit of the absorption line profiles from any number of absorption systems. Each gaseous system is defined by (1) a radial velocity $v_{\rm rad}$, (2) a temperature $T$, (3) a turbulent velocity $b$, and (4) a set of column densities from selected species. In practice, the species within a given absorption system are assumed to share the same set of parameters ($v_{\rm rad}$, $T$, $b$).

We assume in the following that the absorption arises from a single broad component as opposed to a collection of narrow components. Within this hypothesis, the column density determined from saturated lines highly depends on the $b$ parameter. Systematic uncertainties on $b$ include errors on the instrumental resolution (Sect.\,\ref{sec:broadening}) or the fact that species in the neutral gas might not share exactly the same $b$ value. If we choose to neglect these systematic uncertainties, the simultaneous fit of lines with a wide range of optical depths provides constraints on both $b$ and the column density. We verified that the additional presence of saturated lines in the simultaneous fit does not bias the column density determination if the column density is otherwise already well constrained by weak lines.

Other effects might also be at work that introduce systematic errors on the column density. For example, the velocity distribution could include a number of individual narrow absorption components that we cannot resolve. The integrated, unresolved, line profile may appear optically thin even though narrow components are saturated (the so-called ``hidden saturation"). 
Although the absorption spectra of metal-poor objects, with their low metal column densities, is in principle less affected than more metal-rich sources, the hidden saturation effect is non-linear and can result in underestimating the total column density by several factors (e.g., \citealt{Lebouteiller05thesis,Lebouteiller06}). 
Observationally, this effect can be minimized by including lines with a low oscillator strength $f$ (e.g., C\2*, S\2, P\2\ in our spectra). Another method is to verify that, for a given species, lines with a wide range of $f$ values can be fitted by a single column density and that the apparent optical depth scales with $f$ (e.g, N\1, S\2, Fe\2\ in our spectra). Nevertheless, the column density of some other species was determined only from strong lines. We note in particular C\2\ $\lambda1334.5$ (optical depth in the line center $\tau_0\approx76$), O\1\ $\lambda1302.2$ ($\tau_0\approx34$), and all the Si\2\ lines, with the weakest being $\lambda1526.7$ ($\tau_0\approx9$). For these species, we rely on the hypothesis that the absorption does not arise from a collection of narrow components, and we keep in mind that the quoted errors could be underestimated.

The absorption lines from the various systems are blended with each other throughout the COS range. The following lines in I\,Zw\,18 are contaminated by local absorption systems, Fe\2\ $\lambda1142.4$, Si\2\ $\lambda1190.4$, $\lambda1193.3$, Mn\2\ $\lambda1199.4$, and S\2\ $\lambda1250.6$. The \texttt{Owens} algorithm allows us to use these lines to determine column densities under certain conditions, either other lines from the same species are isolated and/or the column density of the local component is well constrained by other lines. In the FUV spectrum of I\,Zw\,18, Fe\2, Si\2, Mn\2, and S\2\ have other isolated lines and the local absorption systems are well constrained by a variety of lines throughout the COS range. 

In order to derive column densities, \texttt{Owens} performs a $\chi^2$ minimization over a number of spectral windows containing one or several lines. While \texttt{Owens} enables the parametrization of the local continuum around the lines, the normalization we performed (Sect.\,\ref{sec:datasets}) results in a flat continuum that does not need to be further parametrized. For each specie, we included every line in the COS wavelength domain. This includes the lines that are not detected, since the latter help providing a reliable upper limit to the column density. The fits were performed on the 3 observation sets separately because of possible systematic differences in the spectra.

\begin{table}
\caption{Main interstellar lines used in this study.\label{tab:lines}}
\centering
\begin{tabular}{llllll}
\hline\hline
Line & $\lambda$ & $f$  &  Line & $\lambda$ & $f$ \\
       & (\AA)   &     &        &  (\AA)   &    \\
 \hline
    C \1*    &       1657.907 &       4.71e-02  &   S \1     &       1425.030 &       1.25e-01   \\
    C \1*    &       1657.379 &       3.56e-02  &   S \1*    &       1323.516 &       3.06e-02 \\
    C \1     &       1656.928 &       1.49e-01  &     S \1     &       1316.543 &       3.45e-02 \\
    C \1*    &       1656.267 &       5.88e-02  &   S \1*    &       1310.194 &       2.89e-02 \\
    C \1*    &       1560.682 &       9.59e-03  &    S \1*    &       1277.216 &       4.93e-02 \\
    C \1     &       1560.309 &       1.28e-02  &     S \1     &       1270.780 &       5.51e-02 \\
    C \1*    &       1329.123 &       1.60e-02  &   S \1*    &       1253.325 &       2.90e-02 \\
    C \1*    &       1329.100 &       2.60e-02  &    S \1     &       1247.160 &       3.24e-02 \\
    C \1*    &       1329.085 &       2.13e-02  &     S \2    &       1259.518 &       1.66e-02 \\
    C \1     &       1328.833 &       7.58e-02  &      S \2    &       1253.805 &       1.21e-02 \\
    C \1     &       1280.135 &       2.63e-02  &     S \2    &       1250.578 &       5.43e-03 \\
    C \1*    &       1279.891 &       1.43e-02  &   Cl\1     &       1347.240 &       1.53e-01 \\
    C \1*    &       1277.513 &       2.10e-02  &    Cl\1     &       1335.726 &       3.13e-02 \\
    C \1*    &       1277.283 &       7.05e-02  &    Cl\1     &       1188.774 &       7.01e-02 \\
    C \1     &       1277.245 &       8.53e-02  &      Mn\2    &       1201.118 &       1.21e-01 \\
    C \1*    &       1261.122 &       1.47e-02  &    Mn\2    &       1199.391 &       1.69e-01 \\
    C \1*    &       1260.926 &       1.35e-02  &     Mn\2    &       1197.184 &       2.17e-01 \\
    C \1     &       1260.735 &       5.07e-02  &      Mn\2    &       1162.015 &       1.02e-02 \\
    C \1     &       1193.030 &       4.09e-02  &     Fe\2    &       1611.200 &       1.38e-03 \\
    C \1*    &       1193.008 &       2.91e-02  &    Fe\2*   &       1618.468 &       2.20e-02 \\
    C \1     &       1157.910 &       2.12e-02  &      Fe\2    &       1608.451 &       5.77e-02 \\
    C \1*    &       1157.770 &       1.83e-02  &    Fe\2*   &       1267.422 &       1.69e-02 \\
     C \2*   &       1335.708 &       1.15e-01  &     Fe\2*   &       1266.677 &       8.99e-03 \\
     C \2*   &       1335.663 &       1.28e-02  &     Fe\2    &       1260.533 &       2.40e-02 \\
     C \2    &       1334.532 &       1.28e-01  &        Fe\2*   &       1148.277 &       8.28e-02 \\
    N \1     &       1200.710 &       4.32e-02  &      Fe\2    &       1144.938 &       8.30e-02 \\
    N \1     &       1200.223 &       8.69e-02  &       Fe\2    &       1143.226 &       1.92e-02 \\
    N \1     &       1199.550 &       1.32e-01  &       Fe\2    &       1142.366 &       4.01e-03 \\
    N \1     &       1134.980 &       4.16e-02 &        Fe\2    &       1133.665 &       4.72e-03 \\
    N \1     &       1134.415 &       2.87e-02 &   Ni\2    &       1773.949 &       6.22e-03 \\
    N \1     &       1134.165 &       1.46e-02 &   Ni\2*   &       1754.813 &       1.59e-02 \\
    O \1*     &       1304.858 &       4.78e-02  & Ni\2    &       1751.916 &       2.77e-02 \\ 
    O \1     &       1355.598 &       1.16e-06  &     Ni\2    &       1709.600 &       6.88e-02 \\
    O \1     &       1302.168 &       4.80e-02  &     Ni\2*   &       1788.491 &       2.52e-02 \\
     Al\2    &       1670.787 &       1.74e+00  &    Ni\2*   &       1748.289 &       1.40e-01 \\
     Si\2    &       1526.707 &       1.33e-01  &      Ni\2    &       1741.553 &       4.27e-02 \\
     Si\2    &       1260.422 &       1.18e+00  &     Ni\2    &       1703.412 &       6.00e-03 \\
     Si\2    &       1193.290 &       5.82e-01  &     Ni\2    &       1502.148 &       1.33e-02 \\
     Si\2    &       1190.416 &       2.92e-01  &     Ni\2    &       1467.756 &       9.90e-03 \\
    P \1     &       1787.648 &       6.04e-02  &     Ni\2*   &       1500.434 &       4.77e-02 \\
    P \1     &       1782.829 &       1.13e-01  &     Ni\2    &       1467.259 &       6.30e-03 \\
    P \1     &       1774.949 &       1.69e-01  &     Ni\2    &       1454.842 &       3.23e-02 \\
    P \1     &       1679.697 &       5.36e-02  &     Ni\2    &       1415.720 &       3.19e-03 \\
    P \1     &       1674.595 &       3.25e-02  &     Ni\2    &       1412.866 &       3.55e-03 \\
    P \1     &       1381.476 &       3.16e-01  &     Ni\2*   &       1411.065 &       6.28e-02 \\
    P \1     &       1379.428 &       2.19e-01  &     Ni\2    &       1381.685 &       5.08e-03 \\
    P \1     &       1377.073 &       1.10e-01  &     Ni\2*   &       1381.286 &       1.15e-01 \\
     P \2    &       1152.818 &       2.72e-01  &     Ni\2    &       1370.132 &       7.69e-02 \\
    S \1     &       1473.994 &       7.30e-02  &     Ni\2    &       1345.878 &       7.69e-03 \\
    S \1*    &       1433.309 &       5.85e-02  &     Ni\2*   &       1335.201 &       1.63e-01 \\
    S \1*    &       1433.278 &       1.69e-01  &     Ni\2    &       1317.217 &       1.46e-01 \\
  S \1     &       1425.188 &       3.65e-02  &       Ni\2    &       1308.866 &       5.41e-03 \\
\hline
\end{tabular}
\tablefoot{$\lambda$ is the rest wavelength, $f$ is the transition oscillator strength. The $*$ denotes the transitions from the fine-structure level. }
\end{table}

\begin{table}
\caption{Radial velocities ($2\sigma$ error bars). \label{tab:velocities}}
\centering
\begin{tabular}{llll}
\hline\hline
Species & 11523 & 12028 & 11579   \\
\hline
Global &  $750\pm2$   &  $756\pm3$     &  $768\pm3$   \\
\hline
\vspace{0.07cm}
C\1     &  ...     &  ...     & ...   \\
\vspace{0.07cm}
C\1*     &  ...   & ...  &  ...   \\
\vspace{0.07cm}
C\2      &  $755^{+5}_{-7}$   &    $763^{+6}_{-4}$ & $769^{+5}_{-7}$   \\
\vspace{0.07cm}
C\2*      & $739^{+6}_{-2}$   &   $749^{+11}_{-8}$ & $748^{+7}_{-4}$  \\
\vspace{0.07cm}
N\1      &  $735^{+10}_{-3}$   &  $737^{+8}_{-6}$ & $769^{+5}_{-5}$  \\
\vspace{0.07cm}
O\1      &  $761^{+14}_{-16}$  & ...      & ...    \\
\vspace{0.07cm}
Al\2       &  $748^{+7}_{-5}$   & $753^{+6}_{-7}$  & ...   \\
\vspace{0.07cm}
Si\2      & $757^{+2}_{-6}$   &  $763^{+5}_{-3}$ & $765^{+4}_{-5}$  \\
\vspace{0.07cm}
Si\2*      & ...   & ...      & ...      \\
\vspace{0.07cm}
P\1     &  ...  &  ... &  ...  \\
\vspace{0.07cm}
P\2     &  ...      &  ...     & ...   \\
\vspace{0.07cm}
S\1      & ...   & ...    &...    \\
\vspace{0.07cm}
S\2     &  $747^{+11}_{-9}$ & $749^{+10}_{-10}$  & $763^{+6}_{-7}$    \\
\vspace{0.07cm}
Cl\1  &  ...       &  ...       & ...     \\
\vspace{0.07cm}
Mn\2     &  ...   &  $721^{+10}_{-14}$ & $765^{+8}_{-6}$   \\
\vspace{0.07cm}
Fe\2    &  $749^{+4}_{-3}$ &   $759^{+8}_{-5}$ & $770^{+4}_{-6}$ \\
\vspace{0.07cm}
Ni\2     & $733^{+12}_{-4}$    &  $743^{+12}_{-7}$  &  $769^{+14}_{-17}$     \\
 \hline
 \end{tabular}\\
 \tablefoot{The first row gives the radial velocity of the neutral component, as derived from the simultaneous fit all the species in the table (assuming that the species share the same radial and turbulent velocities; Sect.\,\ref{sec:abs_sys}). 
 For each species, we also provide the radial velocity calculated by isolating the species in its own component, i.e., the radial velocity and the turbulent velocity are not constrained by other species.}
\end{table}

\subsection{Line broadening}\label{sec:broadening}

The observed line width is a combination of the intrinsic broadening (thermal and turbulent) with the instrumental line spread function. The latter is a critical parameter upon which the $b$ value determination relies. Systematic errors on the line spread function could lead to wrong diagnostics on the line saturation and, therefore, to unreliable column density determinations. 

In the case of a point-source well-centered in the aperture, the instrumental full width at half maximum (FWHM) is $\approx0.08$\AA. If the source has extended emission or if it is a collection of point-sources, the spectral resolution is significantly degraded\footnote{COS instrument handbook, \textit{http://www.stsci.edu/hst/cos}, Sect.\,5.9. }. Depending on the source distribution in the aperture, the line profiles might be shifted and/or skewed, but we found the lines to be Gaussian and symmetric, suggesting a rather uniformly distributed light in the aperture. 

We estimated the effective instrumental spectral resolution in two ways. First, we measured the one-dimensional flux profile of stars in the NW component as derived from a FUV image obtained by \cite{Brown02}. The profile looks roughly gaussian with a FWHM$=1.25\arcsec$, i.e., half the size of the COS aperture. The flat-topped profile of geocoronal Ly$\alpha$ emission, which uniformly fills the COS aperture, has a FWHM$=0.84$\AA, so the expected instrumental profile is expected to have a FWHM$=0.4$\AA. 
The second estimate of the spectral resolution makes use of the data itself. For this, we considered the sharpest lines in the COS spectrum and also the line blends for which the 2 lines are barely separated (in particular at $\approx1134$\AA\ and $\approx1260$\AA). These profiles give a reliable constraint on the instrumental FWHM, $\approx0.45-0.50$\AA. 
We use $0.50$\AA\ ($\approx90$\kms) for the line fitting of the 11523 and 12028 datasets (Sect.\,\ref{sec:abs_sys}). A slightly larger FWHM was required for the 11579 dataset, $0.55$\AA, which could be due to pointing jitter over the long exposure time. For comparison, the point-like source FWHM with the FUSE telescope is $20$\kms\ \citep{Hebrard02,Wood02}. 

We used the instrumental resolution above to calculate the intrinsic line broadening.
For the latter, we could not distinguish between the thermal and turbulent components. We assumed that turbulence is the dominant source of broadening by setting the temperature to $1000$\,K, corresponding to a thermal broadening of $\lesssim1$\kms.
The turbulent velocity value we determine for I\,Zw\,18 is $b\approx17.8\pm3$\kms\ for the 12028 dataset, and $18.2\pm8$\kms\ for the 11579 dataset, in good agreement with the value $17.3\pm5$\kms\ found by L04. The 11523 dataset did not converge on a well constrained $b$ value, which is likely due to the relatively low S/N ratio of this observation and also to the presence of spurious features. 
We forced the $b$ value in the 11523 dataset to be $\approx17.8\pm3$\kms.

\subsection{Heavy elements}\label{sec:cds_metals}

\begin{table*}
\caption{Column densities. \label{tab:cds}}
\centering
\begin{tabular}{l | lll | l | l l | l }
\hline\hline
            & \multicolumn{4}{c|}{COS}   & \multicolumn{2}{c|}{FUSE} & Model\tablefootmark{a} \\
           & 11523 &   12028 &    11579   & Adopted  & L04 & A03 ($\pm1\sigma$) &   \\
\hline
H\1    &  $21.34$    &               $21.34\pm0.04$          &               $21.34$    &     $21.34\pm0.04$   & $21.30\pm0.10$     & $21.35\pm0.10$     &   $21.34$          \\
\hline
\vspace{0.07cm}
C\1    &  ...               & $12.85^{+0.22}_{-0.46}$          & $<13.42$   & $<13.40$  & ...   & ...      &    ...       \\
\vspace{0.07cm}
C\1*    &   $<12.91$   &  $<13.18$        & $<12.97$  &  $<13.00$   & ...  & ...       &    ...     \\
\vspace{0.07cm}
C\2      &   ($15.46^{+0.41}_{-0.06}$)            & ($15.38^{+0.96}_{-0.36}$)      & ($15.37^{+0.36}_{-0.19}$)  & $15.35\pm0.30$    &  ...  &  ...     &    $15.49\  (-0.05)$    \\
\vspace{0.07cm}
C\2*     & $13.84^{+0.08}_{-0.05}$               &  $13.83^{+0.14}_{-0.07}$       & $13.76^{+0.09}_{-0.08}$    & $13.80\pm0.10$   &  ...  & ...       &   $13.53\  (-0.37)$      \\
\vspace{0.07cm}
N\1    &  ($14.55^{+0.02}_{-0.07}$)            & ($14.45^{+0.13}_{-0.06}$)    &  ($14.59^{+0.07}_{-0.03}$)   & $14.55\pm0.10$   & $14.22^{+0.14}_{-0.17}$   &  $14.44\pm0.04$     &    $14.54\ (-0.01)$     \\
\vspace{0.07cm}
O\1    & ($16.05^{+0.15}_{-0.75}$)              &  $<18.08$             & ($16.00^{+0.50}_{-2.00}$)  & $\approx16.37$\tablefootmark{b}  & $16.60^{+0.80}_{-0.55}$  &   $15.98\pm0.26$    &    ...      \\
O\1*    & ($<13.31$)              &  ...           & ($<13.87$)  & ($<13.90$)  & ...  & ...      &    ...      \\
\vspace{0.07cm}
Al\2      &   ($13.32^{+0.14}_{-0.07}$)         & ($13.25^{+0.28}_{-0.07}$)      & -\tablefootmark{c}      &  $13.32\pm0.30$   &  ... & ...      &  $13.32\  (-0.19)$    \\
\vspace{0.07cm}
Si\2    &  ($14.76^{+0.09}_{-0.13}$)              & ($14.53^{+0.42}_{-0.26}$)       & ($14.51^{+0.22}_{-0.24}$)  & $14.52\pm0.30$  & $14.80^{+0.25}_{-0.25}$  &   $14.81\pm0.07$      &  $14.86\  (-0.00)$      \\
\vspace{0.07cm}
Si\2*     &  $<12.31$       & $<11.69$  &  $<12.06$  & $<11.90$  & ...  &    ...      &   ...   \\
\vspace{0.07cm}
P\1      &   $<12.10$    &  $<12.51$         & $<12.18$    &  $<12.50$   &  ...  &   ...        &    ...    \\
\vspace{0.07cm}
P\2      & $12.91^{+0.21}_{-0.49}$    &  $12.86^{+0.26}_{-0.58}$          & $12.49^{+0.31}_{-1.03}$     & $12.92\pm0.30$  &  ...  &  $<-13.60$       &   ...       \\
\vspace{0.07cm}
S\1    & $12.93^{+0.18}_{-0.24}$       & $<12.86$              & $<12.30$      & $<12.30$   &  ...  & ...     &   ...    \\
\vspace{0.07cm}
S\2      & $14.75^{+0.09}_{-0.06}$            &  $14.71^{+0.07}_{-0.07}$        & ($14.56^{+0.08}_{-0.10}$)\tablefootmark{d}   & $14.73\pm0.10$   &...  & ...     &    $14.74\  (-0.00)$       \\
\vspace{0.07cm}
Cl\1      &     ...        &  $<12.95$                     &  $<13.05$   & $<13.05$   &  ... &   ...      &    ...      \\
\vspace{0.07cm}
Mn\2     & $13.10^{+0.05}_{-0.40}$       &  $13.23^{+0.14}_{-0.27}$             & $<13.30$      & $13.10\pm0.50$   &   ...  &    ...      &  $13.12\  (-0.00)$    \\
\vspace{0.07cm}
Fe\2    &  ($14.71^{+0.08}_{-0.05}$)     & ($14.49^{+0.20}_{-0.06}$)       & ($14.76^{+0.06}_{-0.07}$) & $14.75\pm0.20$  & $14.60^{+0.12}_{-0.13}$  &  $15.09\pm0.06$    &   $14.66\  (-0.00)$      \\
\vspace{0.07cm}
Ni\2      & $13.37^{+0.11}_{-0.10}$        &  $13.43^{+0.08}_{-0.10}$          & $13.42^{+0.04}_{-0.22}$     &  $13.42\pm0.10$  & ...  &   ...       &    $13.33\  (-0.01)$     \\
 \hline
 \end{tabular}\\
 \tablefoot{Error bars are $2\sigma$ error bars unless noted otherwise. Note that the error bars for the 11523 dataset are likely underestimated because of the tight constraint on $b$ for this dataset (Sect.\,\ref{sec:broadening}). 
 The values in parentheses could be affected by systematic uncertainties on the $b$ value or due to spurious features (see text). The O\1\ column density in dataset 12028 is an upper limit since it is determined using the undetected $\lambda1355$ line.  }
 \tablefoottext{a}{Column density calculated by the \texttt{Cloudspec} model (Sect.\,\ref{sec:nebular}). The value in parentheses gives the correction when the emission component is accounted for (Sect.\,\ref{sec:emission}). }
 \tablefoottext{b}{Estimated using S\2\ column density as a tracer (Sect.\,\ref{sec:alpha}).}
 \tablefoottext{c}{No coverage.}
 \tablefoottext{d}{Noisy profile.}
\end{table*}

The fitting of the absorption line profiles is shown in the Appendix. Final column densities are given in Table\,\ref{tab:cds}. The error bars were calculated by using the $\Delta\chi^2$ method, which gives the variation of $\chi^2-\chi^2_{\rm min}$, where $\chi^2_{\rm min}$ is the lowest $\chi^2$ value corresponding to the global minimum over the entire parameter space. Since we use the information from all available lines in the wavelength range, the error bar can be significantly smaller than when considering a single line. We refer to L04 and \cite{Lebouteiller06,Lebouteiller09} for more details. The comparison between the column densities we derive and the FUSE results asks for caution. The large FUSE aperture ($30\arcsec\times30\arcsec$) includes the entire main body of I\,Zw\,18 while COS only observed the NW component. However, most of the UV-bright stars are located toward NW \citep{Brown02}, and the corresponding lines of sight presumably dominated the FUSE spectral continuum. Therefore, we do not expect large differences between the column densities derived with COS and with FUSE. In the following we provide individual comments for each species.

The C\2\ column density is not well constrained because the only C\2\ line, $\lambda1334.4$, is heavily saturated. Our determination relies on the confidence that the $b$ value is well constrained by other lines from other species included in the fitting and on the fact that hidden saturation effects are negligible (Sect.\,\ref{sec:abs_sys}). The column density of C\2* is much better determined. 
C\1\ and C\1* lines could be present in our spectra but they are either barely detected or not detected. The contribution from  C\1\ and C\1* to the total C column density is negligible. 

The N\1\  $\lambda1134$ triplet, observed in the 11579 dataset, is not saturated (see also L04), while the $\lambda1200$ triplet is close to saturation. 
The N\1\ column density determination from the 3 datasets are in good agreement. They fall somewhat above the value from L04 and agree well with that from A03.  

O\1\ is observed in the $\lambda1302.2$ line (oscillator strength $f=0.048$) in the 11523 and 11579 datasets and in the $\lambda1355.6$ line (oscillator strength $f=10^{-6}$) in all datasets. The former line is contaminated by geocoronal O\1\ emission (Sect.\,\ref{sec:datasets}), and, although the contaminated exposures were ignored, we cannot exclude that a weak contamination remains in the other exposures. The $\lambda1355.6$ line is never detected and provides a large upper limit to the O\1\ column density. L04 and A03 measured different O\1\ column densities from the same spectrum, mostly because of the set of O\1\ lines used for the fitting. L04 ignored the $\lambda1039$ line  because of contamination by a terrestrial airglow and because of the line saturation. Our O\1\ column density determination is closer to that of A03, but we consider our value as highly uncertain because of saturation. 

Aluminum has a single strong line in the COS range at $1670.8$\AA. The line is unfortunately saturated, and the Al\2\ column density determination relies on the constraints on the $b$ value. We adopt an average column density from the datasets 11523 and 12028.

All the Si\2\ lines are saturated. The Si\2\ column determination is thus relatively uncertain since it depends strongly on the $b$ value determination and on the assumption that absorption arises from a single broad component (Sect.\,\ref{sec:abs_sys}). We use both 12028 and 11579 datasets for the Si\2\ column density determination, and ignore the 11523 dataset because of its relatively lower S/N. 
Our adopted column density is somewhat lower than the FUSE measurements by L04 and A03. We thus keep in mind that our Si\2\ column density might be underestimated and adopt a large error bar that reconciles our value with the FUSE values. 

The strongest P\2\ line ($\lambda1152.8$) is barely detected. Our column density determination falls well below the upper limit from A03. The strongest P\1\ line at $1381.5$\AA\ is not detected, and the contribution from P\1\ to the total P column density is negligible. 

S\2\ is observed in the $\lambda1250.8$, $\lambda1253.8$, and $\lambda1259.5$ lines. The main constraint on the S\2\ column density is given by the weakest line, $\lambda1253.8$. It is well detected in the 12028 dataset, while the profile is more noisy in the 2 other datasets, especially in 11579. We adopt an average column density between the 11523 and 12028 datasets. 

Chlorine is expected as Cl\1\ and Cl\2\ in the neutral phase. Cl\2\ has no lines in the COS range and it was not detected in the FUSE spectra. Cl\1\ has numerous lines below $\sim1400$\AA\, with the strongest at $1347.24$\AA. The latter is however not detected in any of the datasets.

Manganese is observed via several Mn\2\ lines with strong oscillator strengths. Unfortunately all the lines are heavily blended with saturated (or close to being saturated) N\1\ lines from I\,Zw\,18 and from the Milky Way. Therefore, we consider that the Mn\2\ column density is highly uncertain and may be affected by systematic uncertainties. 

Iron is observed via several Fe\2\ lines spanning a wide range of oscillator strengths. The presence of weak lines should result in a reliable Fe\2\ column density determination. Our column density determination falls between the value from L04 and A03. It agrees slightly better with L04 while it is not compatible within errors with the value from A03. The difference between L04 and A03 is due to the lower $b$ value ($<8$\kms) assumed by A03 for the Fe\2\ lines.

Nickel is observed via many weak Ni\2\ lines, most of which are not detected. We stress that the column density determination of Ni\2\ is better than $5\sigma$ because it uses the information from all the lines we included in the fitting. The column density determination from the 3 datasets are in remarkable agreement.

\subsection{Hydrogen}\label{sec:hicd}

The column density of H\1\ is an important quantity since metal abundances are tied to it. However, it is a difficult parameter to measure correctly. The blue wing of the Ly$\alpha$ profile is contaminated by numerous metal lines and by Ly$\alpha$  absorption from the Galaxy and from a high-velocity cloud along the line of sight to I\,Zw\,18 (Sect.\,\ref{sec:abs_sys}). Furthermore, the red wing of Ly$\alpha$ is contaminated by nebular continuum emission (2-photon and Balmer continuum) arising in the H\2\ region of I\,Zw\,18  and by the N\5\ $\lambda1238$, $\lambda1240$ wind lines from the ionizing stars in I\,Zw\,18. Finally, the continuum level is uncertain, because Ly$\alpha$ is such a broad line. In principle, both red and blue wings can be contaminated by stellar Ly$\alpha$ absorption, but this contamination turns out to be insignificant (see also \citealt{Lebouteiller08}). 
 
We worked with the COS spectrum processed through step 3 (i.e., corrected for Ly$\alpha$ in the Milky Way and in the high-velocity cloud; Sect.\,\ref{sec:datasets}). We measured the Ly$\alpha$ column density in I\,Zw\,18 by multiplying the observed spectrum by $e^{\tau}$, with $\tau=N$(H\1)$ \times 10^{-19.37} \times (\lambda-\lambda_0)^{-2}$ \citep{Jenkins71}. The column density that best canceled the observed absorption is $N$(H\1)=$2.2\times10^{21}$\,cm$^{-2}$.
We find a similar value for the 3 datasets. We list in Table\,\ref{tab:hicd} the various H\1\ column density determinations in the literature, either derived in absorption from the Lyman series or in emission using the hyperfine-structure 21\,cm line. Note that the comparison between observations in absorption and in emission should account for the simple geometrical effect that the absorption only probes the gas in the foreground of the UV-bright stars.

The H\1\ spatial distribution observed by \cite{Lelli12} is particularly informative. The high-spatial resolution ($2\arcsec$ beam) reveals that the NW region lies in a H\1\ hole (see also Fig.\,\ref{fig:acs}). The NW region is located between the southeast H\1\ peak, with $N$(H\1)$\approx1.5\times10^{22}$\,cm$^{-2}$, and a region to the north with a rather flat column density distribution ($\approx6\times10^{21}$\,cm$^{-2}$). Interpolating between these 2 values, we would expect an H\1\ column density of $\sim10^{22}$\,cm$^{-2}$ toward NW, which is significantly larger than our observed value. As discussed in \cite{Lelli12}, the young stellar cluster in NW likely created an H\1\ hole, either due to ionization or to a blow out of the ISM. Assuming the H\1\ hole is due to ionization, the column density of ionized hydrogen should be on first approximation $10^{22}-2.2\times10^{21}=7.8\times10^{21}$\,cm$^{-2}$.
We discuss further the distribution of ionized gas in Sect.\,\ref{sec:nebular}. 

\begin{table}
\caption{H\1\ column density determinations.\label{tab:hicd}}
\centering
\begin{tabular}{llll}
\hline\hline
Value  & Method  & Instrument$^{\rm a}$ & Ref. \\
\hline
      & Absorption  & &  \\
\hline
$21.34\pm0.04$ & Ly$\alpha$ (NW) & HST/COS ($\approx2.5\arcsec$) & (1) \\
$\approx21.32$ & Ly$\beta$  & FUSE ($30\arcsec\times30\arcsec$) & (2) \\
$21.54\pm0.06$ & Ly$\alpha$ (NW)  & IUE \& HST/GHRS ($2\arcsec$) & (3) \\
$21.34\pm0.07$ & Lyman series  &FUSE ($30\arcsec\times30\arcsec$) & (4) \\
$21.32\pm0.09$ & Lyman series  &FUSE ($30\arcsec\times30\arcsec$) & (5) \\
$22.3$ & Ly$\alpha$ (peak) & HST/STIS ($52\arcsec\times0.5\arcsec$) & (6) \\
\hline
      & Emission  & &  \\
\hline
$\approx21.2$ &  21\,cm (NW)  & VLA ($8\arcsec$) & (7)\\
$\approx21.20$ &  21\,cm   & VLA ($20\arcsec$) & (8)\\
$21.53\pm0.15$ & 21\,cm (NW)  & VLA ($5\arcsec$) & (8)\\
$\approx21.48$ & 21\,cm (NW) & VLA ($2\arcsec$) & (8) \\
 \hline
 \end{tabular}\\
 \tablebib{(1) This study; (2) \cite{Vidal00}; (3) \cite{Kunth94}; (4) A03; (5) L04; (6) \cite{Brown02}; (7) \cite{vanZee98}; (8) \cite{Lelli12}.
 }
 \tablefoottext{a}{The value in parentheses indicate either the aperture (absorption measurements) or the beam size (emission measurements). }
\end{table}

\section{Nebular models}\label{sec:nebular}

\subsection{Geometry and physical conditions}\label{sec:geometry}

Models were constructed to compute the ionization structure of the nebula and the relative emission and absorption strength of the intervening medium along the line of sight to UV bright stars. 
We made the following simplifying assumptions: (1) the ionizing stars of NW can be treated as a single, central source; (2) nebular emission arises over the whole volume of H\2\ region, but nebular absorption arises only in material along the line of sight to the central source; and (3) there is a clear-cut boundary between the H\2\ and H\1\ regions. The latter assumption is problematic.  The photoionization model of I\,Zw\,18 by \cite{Pequignot08} indicates that the covering fraction of the H\2\ region is only $40$\%, so the bulk of the photoionizing radiation must escape into the H\1\ region, and deep observations of I\,Zw\,18 show H$\alpha$ emission in arcs outside the star-forming region and in a diffuse halo $3$\,kpc across \citep{Dufour90}.

Spectral analysis of the H\1\ cloud of I\,Zw\,18  is much more complicated than for normal (i.e., star-less) damped Ly$\alpha$ systems (DLAs). For example, the spectral profile of the C\2\ 1334-5 resonance doublet is formed by absorption by relatively lower-mass and/or cooler young stars, and by emission from the outer parts of the H\2\ region, as well as by absorption in the H\1\ cloud. Even Ly$\alpha$, nominally formed in the H\1\ cloud is not immune to contamination. Balmer continuum emission and 2-photon emission from the H\2\ region affect the long-wavelength side of the line, and absorption by H$^0$ atoms in the atmospheres of cool stars can affect both the short- and long-wavelength wings of Ly$\alpha$ (Sect.\,\ref{sec:hicd}). 

We used the software package, \texttt{cloudspec} \citep{Hubeny00b}, which combines the photoionization code \texttt{Cloudy} \citep{Ferland98} with the spectral-synthesis code \texttt{synspec} \citep{Hubeny95} to compute the emission and absorption spectrum of an object. The physical model is a medium photoionized by a central source and the cosmic background. \texttt{Cloudy} computes the ionization structure and emission spectrum (both lines and continuum) of the medium. The \texttt{synspec} part of \texttt{cloudspec} then calculates the level populations and the attenuation spectrum produced by such a medium along the line of sight towards the central ionizing source. 

\begin{figure}
\centering
\includegraphics[angle=0,width=9.0cm,clip=true]{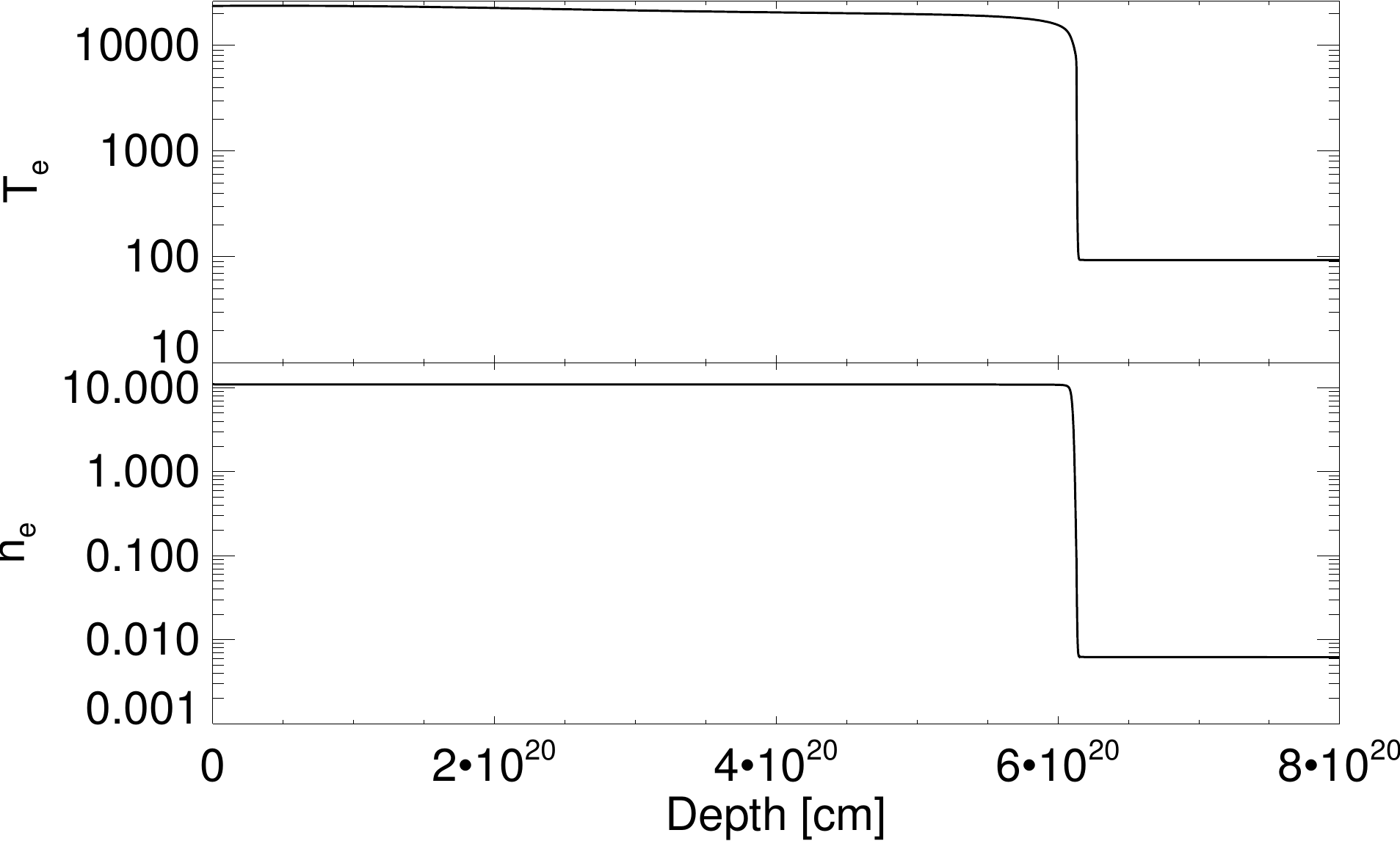}
\caption{Electron temperature (\textit{top}) and density (\textit{bottom}) profile of the simple model approximation of I\,Zw\,18. The abscissa is the radial distance from the ionizing stars in cm.}
\label{fig:simple0}
\end{figure}

We ran \texttt{cloudspec} through the H\2\ region and into the H\1\ region until the H\1\ column density reached its observed value toward the NW component, $2.2\times10^{21}$\,cm$^{-2}$ (Sect.\,\ref{sec:hicd}). Figure\,\ref{fig:simple0} shows the radial profile of the electron temperature and density and the sharp change in the physical conditions between the H\2\ and the H\1\ region. Close in to the ionizing stars, the nebula is quite hot, due to the low abundance of oxygen, the main element for collisional cooling in the ionized gas. The electron temperature decreases outward from the ionizing stars with an abrupt drop near the edge of the H\2\ region, and outside the Str{\"o}mgren sphere, the temperature reaches down to $100$\,K. The electron density drops by 3 orders of magnitude at the edge of the Str{\"o}mgren sphere, because most of the free electrons supplied by hydrogen are no longer available.

The NW component is dominated by the H\2\ region which sits an H\1\ hole either due to hydrogen ionization or to a blow out of the ISM \citep{Lelli12}. In our  model, we find that the column density of ionized hydrogen should be $\log N$(H\1)$=21.79$, i.e., almost a factor $3$ larger than the observed H\1\ column density. 
Following the hypothesis that the total (ionized+neutral) hydrogen column density is flat across the main body of I\,Zw\,18, we would expect the ionized hydrogen column density toward NW to be around $\log N$(H\1)$\sim21.89$ in order to compensate for the relatively lower H\1\ column density (Sect.\,\ref{sec:hicd}). Our models are in good agreement with this value and support the idea that the gas was not blown out but mainly ionized. 

In assuming spherical symmetry, the models are highly simplistic. The ionizing stars in I\,Zw\,18 are offset from H$\alpha$ emission shell \citep{Hunter95,Brown02}, and the H\2\ region has a knot of high density. Nevertheless, the models help us to understand important physical processes operating in I\,Zw\,18 (Sect.\,\ref{sec:ciiratio}) and to guide the abundance analysis of the H\1\ region (Sect.\,\ref{sec:abs_discuss}). 

\subsection{Spectral line formation in the H\2\ region}\label{sec:emission}

\begin{figure*}
\centering
\includegraphics[angle=0,width=17cm,height=7cm,clip=true]{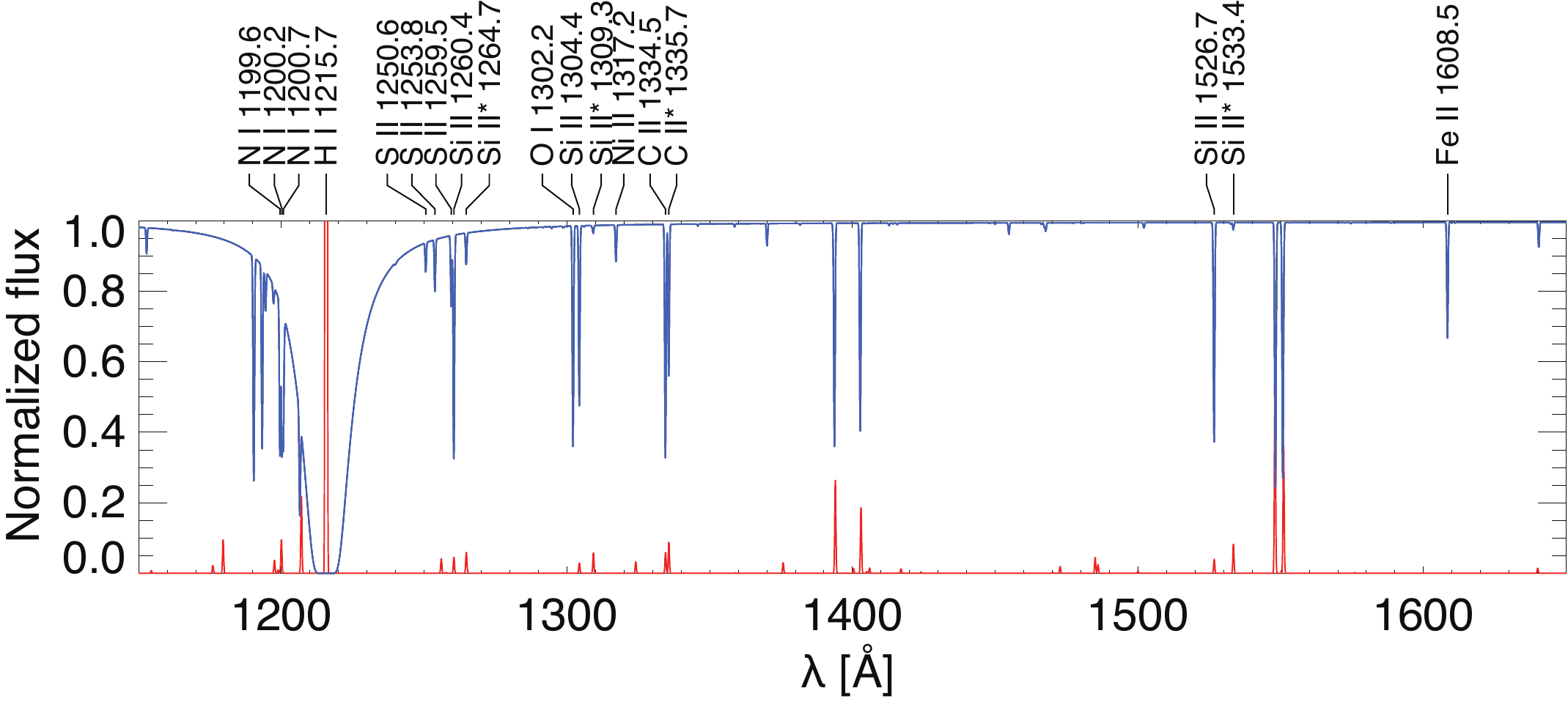}\\
\includegraphics[angle=0,width=18cm,height=10cm,clip=true]{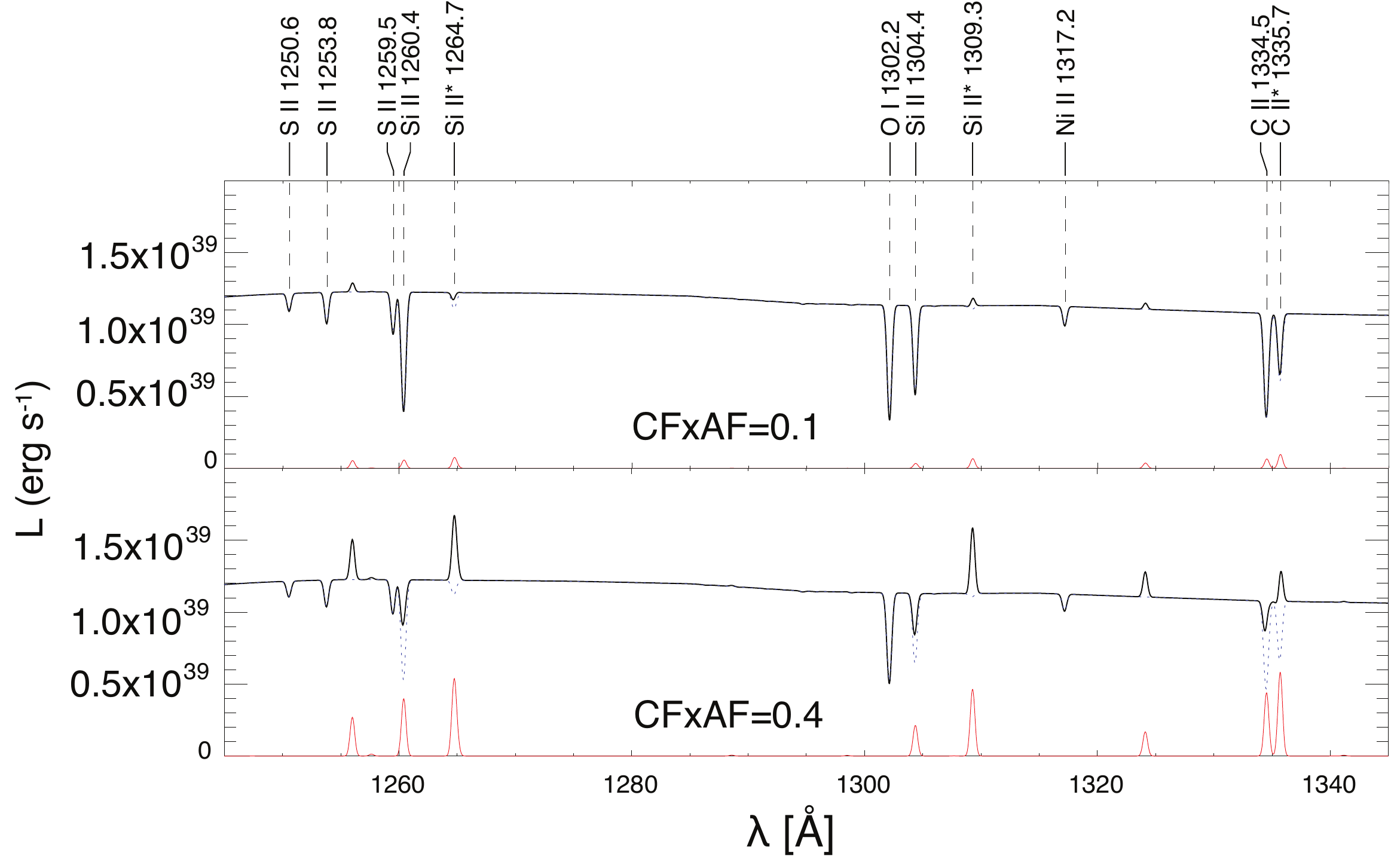}
\caption{\texttt{Cloudspec} model spectra. \textit{Top} $-$ Stellar and nebular (H\1\ and H\2\ region) component normalized to the stellar continuum (blue) and emission component (red). Both component spectra have been convolved with a gaussian line-spread function having a FWHM$=0.5$\AA\ to simulate the observations. Major absorption lines formed in the H\1\ region are labeled on top. Most other strong lines, e.g. the Si\4\ $\lambda1400$ doublet, C\4\ $\lambda1549$ doublet, are higher ionization lines formed in the H\2\ region. The absorption component of Ly$\alpha$ completely erases the Ly$\alpha$ emission formed in the H\2\ region. \textit{Bottom} $-$ Simple model spectrum for the case $CF \times AF=0.1$ (top) and for $CF \times AF=0.4$ (bottom), with $CF$ the covering fraction and $AF$ the aperture fraction (see text). The dotted line is the absorption spectrum, and the bold line is the net spectrum. The component and net spectra have been convolved with a gaussian line-spread function having a FWHM$=0.5$\AA. }
\label{fig:simple_lines2}
\end{figure*}

Singly ionized ions are formed by recombination of higher ionization stages inside the H\2\ region near the edge of the Str{\"o}mgren sphere. Recombination and subsequent cascade down to the ground state in the H\2\ region creates emission that works to reduce the apparent strength of absorption lines arising in the H\1\ region.

Figure\,\ref{fig:simple_lines2}a shows the emission and absorption components of the model line spectrum as derived by \texttt{cloudspec}. The observed spectrum is the net. As can be seen from the figure, the emission is strong relative to the absorption for C\2* $\lambda1335.7$.
How important is contamination by emission lines? The answer depends on two factors: the covering fraction of the H\2\ region, $CF$, and the fraction of the H\2\ emission region subtended by the entrance aperture to the spectrograph, $AF$. \cite{Pequignot08} estimates the covering fraction of the H\2\ region in I\,Zw\,18 at $CF=0.4$, based on the strength of H$\beta$ emission to Lyman continuum flux of the model best fitting the observations. The COS aperture fraction is not known, but is certainly much less than $1$. Figure\,\ref{fig:simple_lines2}b compares a segment of the model UV spectrum for two cases, $CF \times AF=0.1$ (top) and $CF \times AF=0.4$ (bottom). It is clear that the latter case disagrees with observation by generally overestimating the strength of emission, making the ground-state absorption lines too weak, and the fine-structure lines appearing in emission (particularly C\2* and Si\2*). The agreement is much better for the case where $CF \times AF=0.1$. In our abundance analysis, we have therefore neglected the effects of emission except for the C\2\ doublet at 1334-5\AA.

Contamination by emission cannot always be neglected. In fact, the spectra of star-forming galaxies at high-redshift often show emission in the fine-structure lines \citep{Pettini00,Shapley03}. This emission is easily identifiable because these galaxies have outflows. The H\1\ region is expanding with respect to the central, ionizing stars, so the absorption component, which arises along the line of sight to the stars, is blue-shifted. The emission component is formed over the outermost shell within the H\2\ region, so it is centered on the stellar radial velocity. Expansion of the H\2\ region simply broadens the line. Thus, the red half of the emission reaches the observer unscathed, but the blue half of the emission is at least partially nullified by the blue-shifted absorption component. The net profile looks like a P\,Cygni profile.
The emission spectrum of I\,Zw\,18 is as real as in high-redshift, star-forming galaxies. The difference is that there is no detectable velocity gradient in I\,Zw\,18, so the emission is not separable from the absorption component.

While the emission component in the H\2\ region is negligible for most lines, absorption can also arise from the H\2\ region. A few resonant lines originating in the ionized gas are seen in absorption in the FUV spectra of I\,Zw\,18 and provide an independent, direct, estimate of the ionization structure of the nebula. For instance, A03 measured the Fe\3\ $\lambda1122$ line in the FUSE spectrum of I\,Zw\,18 and set an upper limit to the fraction of Fe\2\ in the ionized gas of $5$\%. A03 also cautioned that a large fraction of Si\2\ absorption might arise in the ionized gas. We are now able to test this hypothesis, with the Si\3\ $\lambda1206.5$ line, which we detect at a velocity of $\sim755\pm50$\kms\ in the COS spectra. This line partly originates from stellar atmospheres and partly from the H\2\ region. The observed profile of the Si\3\ line is complex and it depends on the stellar contribution that was removed (Sect.\,\ref{sec:stellar}). 
The column density determination also depends on the $b$ value for which we have little information in the ionized gas. 
Assuming that the line is not saturated, we find a consistent Si\3\ column density in the 3 datasets with $\sim2\times10^{13}$\,cm$^{-2}$. 
What is the fraction of the Si\2\ column density we measured ($3.3\times10^{14}$\,cm$^{-2}$) arising in the ionized gas? In the warm ionized medium in the Milky Way, about $90$\%\ of  silicon is expected into Si\2\ and $10$\%\ into Si\3\ \citep{Sembach00}. Assuming this result is applicable to I\,Zw\,18, we find that only $\sim40$\%\ of the measured Si\2\ column density might arise from the neutral gas. However, the situation is likely different in I\,Zw\,18 where silicon in the ionized gas is expected mostly into higher ionization stages (Sect.\,\ref{sec:ionization}; see also A03).
We assume in the following that most of Si\2\ should lie in the H\1\ region gas. 

Unfortunately, no S\3\ line could be used to perform a similar analysis with sulfur. The brightest S\3\ line, $\lambda1190$, is blended with Si\2. Nevertheless, we consider that the ionization correction smaller than for Si and that most of S\2\ lies in the H\1\ region.

\section{Physical conditions and thermal equilibrium in the H\1\ region}\label{sec:ciiratio}

The C\2* column density represents only $\approx3$\%\ of the total carbon column density (Table\,\ref{tab:cds}), yet C\2* is key to understanding the physical conditions, in particular the thermal processes toward the NW region.  
The ground state of C$^+$ is split into the fine-structure levels $^2$P$_{1/2}$ and $^2$P$_{3/2}$. The absorption line C\2\ $\lambda1334.5$ originates from the $^2$P$_{1/2}$ level while C\2* $\lambda1335.7$ originates from $^2$P$_{3/2}$. The upper level is populated by collisions with H$^0$, H$_2$, and $e^-$. The spontaneous emission between the two fine-structure levels is observed in the far-infrared emission line [C\2] 157\mic\ and dominates over collisional de-excitation as long as the density is lower than the transition critical density.

\subsection{Electron fraction in the H\1\ region}\label{sec:eden}

The population of the C$^+$ fine-structure level constrains the electron density toward the NW region. Following \cite{Jenkins00b}, the volume-density ratio $n(C^{+*})/n(C^+)$ provides the electron density as long as collisions with electrons dominate the population of the fine-structure level. 
Collisions with H$_2$ should be unimportant in I\,Zw\,18 given the lack of diffuse H$_2$ \citep{Vidal00}. Furthermore, the collision strength with $e^-$ is larger than the collision strength with H atoms by a factor of $\approx150$ for $T=6000$\,K. Collisions with $e^-$ will therefore dominate as long as $n_e/n_{\rm H}$ remains larger than $\sim0.75$\%\ (e.g., \citealt{Lehner04}). Potential sources of free electrons are soft X-rays, cosmic rays, and ionization of C$^0$.
We approximate the number density ratio $n(C^{+*})/n(C^+)$ with the column density ratio $R_{\rm CII}=N(C^{+*})/N(C^+)$ and, by assuming that electrons are the dominant collision partners, we derive an upper limit on the electron density
\begin{equation}\label{eq:ne}
n_e < \frac{A_{21}}{\gamma_{21}(e)}\ \frac{R_{\rm CII}}{2 e^{-T_{\rm exc}/T}-R_{\rm CII}}\ {\rm cm}^{-3}, 
\end{equation}
where $A_{21}$ is the radiative decay probability from the fine-structure level from the $^2$P$_{3/2}$ level to the $^2$P$_{1/2}$, $T_{\rm exc}$ is the $^2$P$_{3/2}$ level excitation temperature ($\approx91$\,K), $T$ is the electron temperature and $\gamma_{21}(e)$ is the collisional deexcitation rate coefficient, which we take from \cite{Goldsmith12}. 

For a temperature range of [$1000-8000$]\,K, we find $n_e \lesssim [10-20]\ R_{\rm CII}$\,cm$^{-3}$. Using the observed column density ratio $R_{\rm CII}=2.8\times10^{-2}$, we calculate $n_e \lesssim 0.4$\cc. Since the C\2\ absorption-line is saturated, we also checked that the result holds when using S\2\ as a replacement for C\2. The S\2\ lines are not prone to saturation, and both S\2\ and C\2\ are the dominant ionization stages in the H\1\ region, so we expect the ionization fractions of S into S$^+$ and C into C$^+$ to be well correlated over a wide range of conditions. Carbon is more readily depleted on dust grains than sulfur \citep{Sofia97,Sofia98b}, but rather than assuming a depletion factor, we assume that the abundance ratio in the neutral gas equates that in the ionized gas (C/S)$_{\rm HII}\approx0.98$ \citep{Pequignot08}. The latter assumption also implies that C and S disperse and mix in the ISM over similar timescales (Sect.\,\ref{sec:carbon}). 
We can then approximate $R_{\rm CII}$ with
\begin{equation} 
R_{\rm CII} \approx \frac{N(C^{+*})}{N(S^+)}\ \frac{1}{(C/S)_{\rm HII}},
\end{equation}
from which we derive a somewhat lower upper limit of $n_e \lesssim 0.2$\cc.

On the one hand, the upper limit on the electron density we determine is significantly lower than the value measured in the ionized gas of the star-forming regions derived in Dufour \&\ Hester (1990) and in the H$\alpha$ arc derived by \cite{Izotov01a}, on the order of $10$\cc. This suggests that C\2* arises in the diffuse ionized gas or in the H\1\ region. 
On the other hand, the upper limit on $n_e$ is significantly larger than the electron density in the H\1\ region inferred from our \texttt{Cloudy} models ($\sim0.01$\cc; Sect.\,\ref{sec:nebular}). The models suggest that the C\2* emission component can be neglected and thus does not impact our estimate of $R_{\rm CII}$ (Sect.\,\ref{sec:emission}).
However, it is plausible that collisions partners other than $e^-$ contribute to populating the C$^+$ fine-structure level, in particular H atoms. 
We can use the same method as Eq.\,\ref{eq:ne} to derive an upper limit on the hydrogen density, i.e., assuming that hydrogen atoms are the dominant collision partners. For a temperature range of [$100-1000$]\,K, we find $n_{\rm H} \lesssim [30-40] $\cc.

If we assume that the H\1\ region has a density around or lower than $10$\cc, we must conclude that the electron fraction in the H\1\ region is likely significant, possibly reaching a few percents. Such values would lie many orders of magnitude above the fraction of electrons provided by ionization of C$^0$ alone (equal to the carbon abundance relative to hydrogen), suggesting that the H\1\ region is partially ionized, with electrons created by UV photons, soft X-rays, or cosmic rays originating from the star-forming regions. We investigate further the electron fraction in Sect.\,\ref{sec:cii_hi}.

\subsection{Theoretical cooling rate from the H\1\ envelope}\label{sec:cii_hi}

In order to examine further the physical conditions in the H\1\ region, we now compare the observed cooling rate to the theoretical rate expected from the H\1\ envelope, with H atoms and free electrons as collisions partners. 
The observed cooling rate per H atom is calculated following \cite{Pottasch79} and \cite{Gry92}, with
\begin{equation}\label{eq:coolperh}
L = A_{21}\ h\ \nu_{21}\ \frac{n(C^{+*})}{n_{\rm H}}\ \approx 2.89\times10^{-20}\ \frac{N(C^{+*})}{N(H^0)}\ {\rm erg\ s}^{-1}\ ({\rm H}^{-1})
\end{equation}
where $\nu_{21}$ is the transition frequency and where we have again approximated the number density ratio with the column density ratio. We find $L_{\rm COS}\approx8\times10^{-28}$\,erg\ s$^{-1}$\,(H$^{-1}$). We can rewrite the cooling rate per surface unit of the galaxy,
\begin{equation}
L' \approx 2.89\times10^{-20}\ N(C^{+*})\ {\rm erg\ s}^{-1}\ {\rm cm}^{-2},
\end{equation}
which leads to $L'_{\rm COS} \approx 1.8\times10^{-6}$\,erg\ s$^{-1}$\ cm$^{-2}$. Assuming the source is extended, the intensity per surface unit of the galaxy is the same as the intensity per surface unit on the detector, so we can directly relate $L'$ to the corresponding 157\mic\ emission intensity $I(157{\mu}{\rm m})$ in ${\rm erg\ s}^{-1}\ {\rm cm}^{-2}\ {\rm sr}^{-1}$ via:
\begin{equation}
I(157{\mu}{\rm m}) = L'\ \frac{\Phi}{4\pi},
\end{equation}
where $\Phi$ is the fraction of the beam filled with [C\2] emission. 
Assuming $\Phi=1$, we find $I_{\rm COS}(157{\mu}{\rm m})\sim1.5\times10^{-7}\,{\rm erg\ s}^{-1}\ {\rm cm}^{-2}\ {\rm sr}^{-1} = 1.5\times10^{-10}$\wmsr. This is the observed cooling rate toward the NW component. 

\begin{figure}
\includegraphics[angle=0,scale=0.46,clip=true]{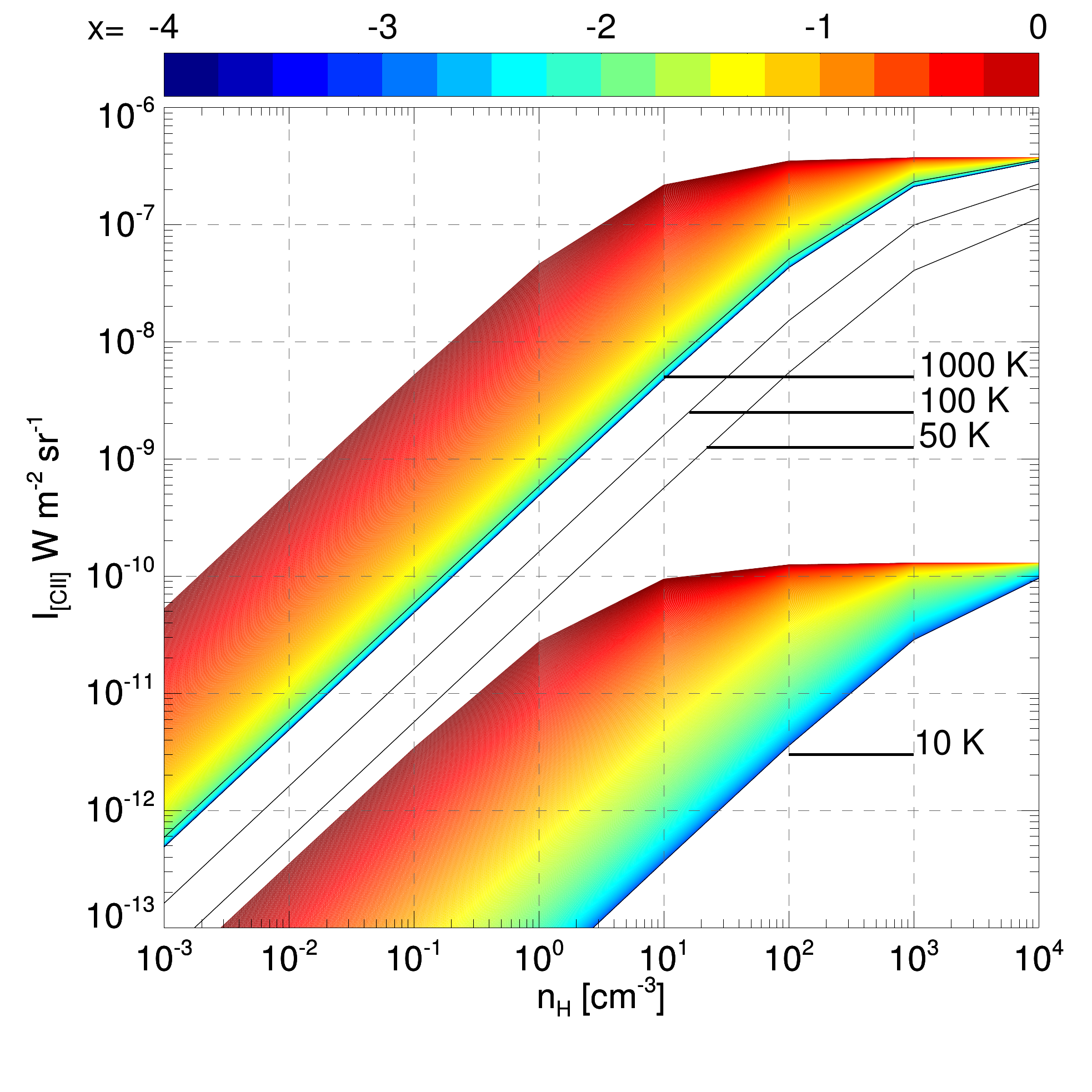}
\caption{The 157\mic\ cooling line intensity is calculated from Eq.\,\ref{eq:icii} and is plotted as a function of the hydrogen number density and of the electron fraction ($x=n_e/n_{\rm H}$). Note that the result should be scaled by the fraction of carbon in C$^+$. The result should also be scaled by the carbon abundance with respect to the solar value and by the column density of hydrogen nuclei in units of $10^{21}$\,cm$^{-2}$. For temperatures $50$\,K and $100$\,K, only the curve for $x=0$ is shown.
\label{fig:emcal_cii}}
\end{figure}

For the theoretical cooling rate in the H\1\ envelope, we consider a medium with most free electrons produced by cosmic rays, soft X-rays, or escaping UV photons (see Sect.\,\ref{sec:eden}).
Following \cite{Crawford85} and \cite{Madden93}, the cooling rate 
expressed in ${\rm W}\ {\rm m}^{-2}\ {\rm sr}^{-1}$ can be written as 
\begin{equation}\label{eq:icii}
I_{\rm th}(157{\mu}{\rm m}) = 5.8\times10^{-7}\ \frac{Z}{({\rm Z}_\odot)}\ \frac{N_{\rm H}}{(10^{21}\,{\rm cm}^{-2})}\ F_{{\rm C}^+}\ \Phi\ y ,
\end{equation}
where $Z$ is the abundance of carbon relative to the solar value (C/H$=2.69\times10^{-4}$; \citealt{Asplund09}), $N_{\rm H}$ is the hydrogen nuclei column density, $F_{{\rm C}^+}$ is the fraction of carbon into C$^+$ ions along the line of sight, $\Phi$ is the fraction of the beam filled by C$^+$ gas, and $y$ is the dimensionless factor defined as
\begin{equation}
y = \frac{ 2\ e^{-91/T} }{ 1 + 2\ e^{-91/T} + (n_e/n_{\rm cr,e}+n_{\rm H}/n_{\rm cr,H})^{-1}}, 
\end{equation}
where $n_{\rm cr,e}$ and $n_{\rm cr,H}$ are the critical densities for collisions with electrons and hydrogen (which we take from \citealt{Goldsmith12}). We assume in the following $F_{{\rm C}^+}=1$ and we define $x$ as the electron fraction $x=n_e/n_{\rm H}$. 
Figure\,\ref{fig:emcal_cii} provides $I_{\rm th}(157{\mu}{\rm m})$ as a function of density for various temperatures and electron fractions. When the ionization fraction is large, proton impact could contribute to the $^2$P$_{3/2}$ level excitation, but here we neglect this contribution since we focus on the H\1\ region. 
It can be seen that $I_{\rm th}(157{\mu}{\rm m})$ flattens for densities $>10$\cc\ for large $x$ values (medium significantly ionized), and $>10^3$\cc\ for low $x$ values (purely neutral medium), which is a consequence of the critical densities for collisions with electrons and hydrogen atoms. $I_{\rm th}(157{\mu}{\rm m})$ depends little on temperature for $T$ greater than a few hundreds K.

In I\,Zw\,18, the abundance of carbon relative to solar is $Z=10^{-2}$, where we have used sulfur abundance as a substitute and assumed the C/S ratio in the ionized gas (see Sect.\,\ref{sec:eden}).
We first consider a mostly neutral medium, with $T\sim1000$\,K and use the H\1\ column density derived in Sect.\,\ref{sec:hicd} ($10^{21.34}$\,cm$^{-2}$). From Fig.\,\ref{fig:emcal_cii}, we calculate that the observed cooling rate $I_{\rm COS}(157{\mu}{\rm m})$ can be reproduced by a wide range of conditions, from purely neutral with $n_{\rm H}\approx15$\cc, to a medium with a significant electron fraction with $n_{\rm H}\approx5$\cc\ and $x\approx2$\%. Assuming a homogenous medium with a low clumping factor, such number densities imply small H\1\ scales on the order of $\gtrsim100$\,pc, which is compatible with an H\1\ shell, left over after the NW star cluster blew out and ionized the pre-existing cloud (Sect.\,\ref{sec:emission}).

For larger electron fractions, the medium is diffuse and ionized and we need to account for the column density of H$^+$. Following the hypothesis that the total hydrogen column density toward NW is $\approx10^{22}$\,cm$^{-2}$ (Sect.\,\ref{sec:hicd}), we find that the cooling rate can be reproduced by an electron density of $n_e\approx0.04$\cc. However, such a low number density would result in an unrealistic nebula size ($>50$\,kpc).

In summary, although the lack of knowledge on the hydrogen number density in the H\1\ region prevents us from deriving a precise electron fraction, our results are compatible with most of the C\2* absorption arising in the H\1\ region with likely a significant electron fraction.

How does the cooling rate derived from C\2* compare with the predicted value from the \texttt{Cloudy} models?
The models (Sect.\,\ref{sec:nebular}) predict that radiative de-excitation of the C$^+$ fine-structure level dominates the cooling in the H\1\ region, with $L_{\rm model}\approx3.5\times10^{-28}$\,erg\,s$^{-1}$\,(H$^{-1}$), which is a factor of $\approx2$ lower than the observed value $L_{\rm COS}\approx8\times10^{-28}$\,erg\ s$^{-1}$\,(H$^{-1}$). While we assumed a constant hydrogen density of $10$\cc\ in the models, a lower density would result in a higher temperature, with other coolants contributing significantly to the total cooling rate, such as [O\1] 63\mic\ or [Si\2] 35\mic.
This would suggest that the total cooling rate in the models is likely underestimated by several factors.
Since our models assume a thermal equilibrium, the heating rate is also underestimated by the same amount. We discuss in Sect.\,\ref{sec:cii_heating} the heating mechanisms and how to reconcile the models with the observed C\2* cooling rate.

\subsection{Cooling rate and SFR}\label{sec:cooling}

The observed cooling rate derived from Eq.\,\ref{eq:coolperh} is $L_{\rm COS}\approx8\times10^{-28}$\,erg\ s$^{-1}$\,(H$^{-1}$). This value is significantly lower than in the diffuse medium of our galaxy ($\approx10^{-26}-10^{-25}$\,erg\ s$^{-1}$\,(H$^{-1}$); \citealt{Lehner04}) or in other nearby galaxies (e.g., \citealt{Madden93}), but it is close to the average cooling rate in DLAs ($\approx10^{-27}$\,erg\ s$^{-1}$\,(H$^{-1}$); \citealt{Wolfe03b}). 
Assuming that the photoelectric effect on dust is the dominant gas heating mechanism (see Sect.\,\ref{sec:cii_heating}), the cooling rate scales with the dust-to-gas ratio (D/G), the photoelectric efficiency, and the FUV field. Since the latter is a good gauge of the star-formation rate (SFR) per unit physical area, the similar cooling rate between I\,Zw\,18 and DLAs could in principle imply a similar surface SFR density \citep{Wolfe03a}. Following \cite{Bowen05}, we can estimate the SFR from C\2* by using the Milky Way (MW) as a reference
\begin{equation}
\Sigma_{\rm SFR} ({\rm I\,Zw\,18}) \approx [Z]^{-1}\ \frac{L_{\rm COS}}{L ({\rm MW})}\ \Sigma_{\rm SFR} ({\rm MW}),
\end{equation}
where $\Sigma_{\rm SFR}$ is the SFR per unit area and $[Z]$ represents the D/G normalized to the MW value. While $[Z]$ should in principle scale with the gas metallicity, we use instead the D/G as measured with the \textit{Herschel} Space Telescope ($2000^{+1000}_{-1300}$ times lower than the MW value; R{\'e}my et al.\ in prep). We assume $\Sigma_{\rm SFR} ({\rm MW})\approx4\times10^{-3}$\,M$_\odot$\,yr$^{-1}$\,kpc$^{-2}$ \citep{Rana91} and $L ({\rm MW})\approx2\times10^{-26}$\,erg\,s$^{-1}$\,(H)$^{-1}$ \citep{Lehner04}, and we finally obtain $\Sigma_{\rm SFR} ({\rm I\,Zw\,18})\sim0.3\pm0.2$\,M$_\odot$\,yr$^{-1}$\,kpc$^{-2}$. As an alternative measurement of the SFR, we use the empirical relation between the gas surface density $\Sigma_{\rm gas}$ and $\Sigma_{\rm SFR}$ (e.g., \citealt{Kennicutt98b}). The H\1\ column density is used as a tracer of $\Sigma_{\rm gas}$, with a correction factor of $\sim4$ due to the ionized gas along the line of sight (Sect.\,\ref{sec:geometry}). We find $\Sigma_{\rm gas}\approx87.6$\,M$_\odot$\,pc$^{-2}$ and $\Sigma_{\rm SFR} ({\rm I\,Zw\,18})\approx0.13^{+0.20}_{-0.08}$\,M$_\odot$\,yr$^{-1}$\,kpc$^{-2}$. Both these determinations are close to the values found by \cite{Aloisi99}, with $10^{-2}-10^{-1}$\,M$_\odot$\,yr$^{-1}$\,kpc$^{-2}$, and \cite{Petrosian97}, with $0.14$\,M$_\odot$\,yr$^{-1}$\,kpc$^{-2}$ (corrected to a distance of $19$\,Mpc). The slightly higher SFR found from the C\2* cooling rate, if genuine, might be due to an underestimated D/G because of the presence of metal-free gas on the line of sight (Sect.\,\ref{sec:disc}) and/or to a larger photoelectric efficiency in I\,Zw\,18 as compared to the MW (see also Sect.\,\ref{sec:cii_heating}).

\subsection{Heating mechanisms}\label{sec:cii_heating}

In the warm ionized medium, photoionization of H should dominate the heating (see \citealt{Pequignot08} for I\,Zw\,18), with likely extra heating from photoelectric effect on dust and from dissipation of interstellar turbulence (Reynolds et al.\ 1999). The heating mechanisms in the H\1\ region are comparatively much less constrained. The gas heating rate in the ISM of our Galaxy is dominated by photoelectrons ejected from polycyclic aromatic hydrocarbons (PAHs) and small dust grains \citep{Wolfire95,Weingartner01a}, but low metallicity environments could show fundamental differences, with for instance the low PAH abundance of PAHs (e.g., \citealt{wu06}). 

Ultraviolet photons escaping the H\2\ region play an important role in the H\1\ gas heating through photoionization and through the photoelectric effect.  \cite{Dufour90} discovered a faint halo of H$\alpha$ emission extending at least $2.7$\,kpc from the star-forming massive star clusters of I\,Zw\,18.
Furthermore, \cite{Pequignot08} found that only $30-40$\% of ionizing photons from the NW and SE massive star clusters is absorbed by surrounding H\2\ regions. The leakage of ionizing photons into the neutral medium should result in a heating effect, affecting the relative population of the ground and fine-structure levels of C\2, as well as the strength of the C\2* $\lambda1335$ fine-structure line. 

The total heating rate predicted by the \texttt{Cloudy} models in the H\1\ region (Sect.\,\ref{sec:nebular}) is dominated by cosmic rays, and amounts to $\Gamma \sim 3.5\times10^{-27}$\,erg\,s$^{-1}$\,cm$^{-3}$. Photoelectric effect heating is comparatively negligible, with $\sim2$\%\ of the total heating rate\footnote{If the D/G is simply scaled to the metallicity of I\,Zw\,18, the photoelectric effect heating rate becomes the dominant heating mechanism, contributing to half of the total heating rate, itself increased by a factor of $2$.}. 
However, our simple model has important limitations. First, the heating rate is likely underestimated by several factors (Sect.\,\ref{sec:cooling}), which could be due an underestimated photoelectric effect heating rate. 
The latter is highly dependent on the model geometry and parameters since it is proportional to the incident UV radiation field, the photoelectric efficiency, and  the D/G ratio (e.g., \citealt{Bakes94,Weingartner01a}). Apart from the remarkably low D/G in I\,Zw\,18 (R{\'e}my {\rm et al.}\ in prep.), the low photoelectric heating rate predicted by our models is due to the relatively weak radiation field in the H\1\ region ($G_0<100$, expressed in Habing units; \citealt{Habing68}), and to the low photoelectric efficiency, due to the large grain charging parameter ($G_0\sqrt{T_{\rm gas}}/n_e \sim 10^5$\,K$^{1/2}$\,cm$^{3}$).
The photoelectric effect might become the dominant heating mechanism in the models if a more realistic approach was taken. We list below some of the most important parameters,
\begin{itemize}
\item The radiation field in the H\1\ region depends on the ISM topology and in particular on the covering factor of the H\2\ region. The UV photons likely escape the star-forming region to reach and heat the gas on large spatial scales across the galaxy. 
\item The electron density in the models might be underestimated (Sect.\,\ref{sec:eden}). A larger electron density will increase the photoelectric efficiency, resulting in a larger heating rate. 
\item The grain size distribution could be biased toward the smallest sizes in low-metallicity dwarf galaxies (e.g., \citealt{Galliano05}), with likely a significant impact in the gas heating rate and efficiency. 
\item The D/G ratio is uncertain, between $700$ and $3000$ times lower than the MW value.
\end{itemize}
The study of the far-IR [C\2] line and its relation to the IR luminosity will be important to confirm the importance of dust in the gas heating process \citep{Lebouteiller13b}.

\begin{figure}
\centering
\includegraphics[angle=0,width=8.8cm,clip=false]{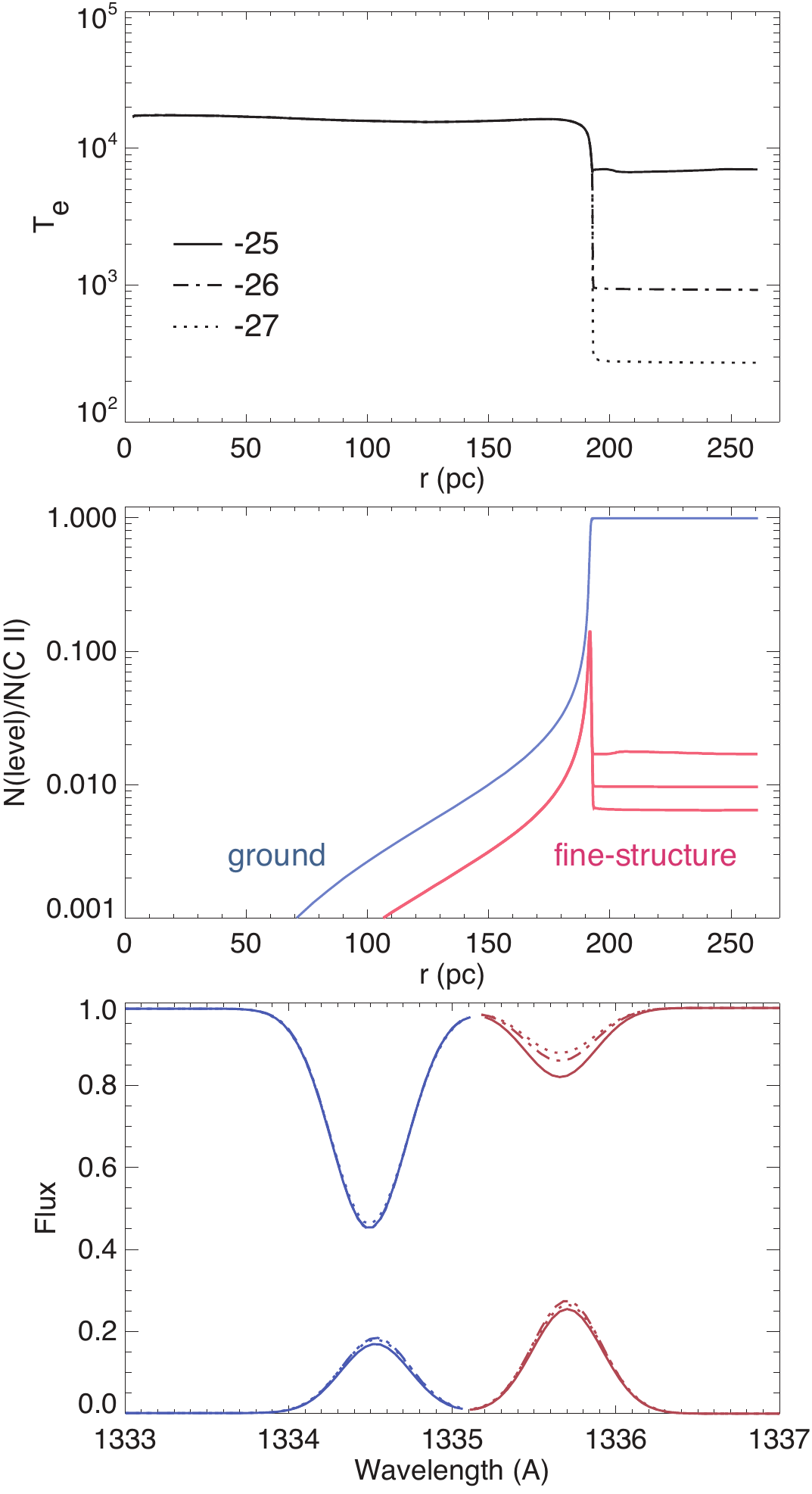}
\caption{Effect of extra heating in the H\1\ region. The electron temperature (\textit{top panel}) and the fraction of C\2\ and C\2* to total carbon (\textit{middle panel}) are shown as a function of radius to the ionizing stars. The predicted line strength is shown in the \textit{bottom panel}. The models assume unity for both the covering fraction of the H\2\ region and the fraction of the emission subtended by the COS aperture (see Sect.\,\ref{sec:emission}). 
}
\label{fig:HE_effect}
\end{figure}

Since our model appears to underestimate the heating rate (Sect.\,\ref{sec:cooling}), we explored further the heating of the H\1\ gas through the use of the ``extra heating" parameter in \texttt{Cloudy}. The extra heating represents all heating sources that are not well constrained or not modeled (e.g., escaping ionizing photons, compressional heating, dissipation of turbulence, mechanical energy transfer from stellar winds and supernov{\ae}...). We assumed no radial dependence for this parameter. 
Figure\,\ref{fig:HE_effect} shows that the electron temperature, initially at $100$\,K in absence of extra heating, jumps from $300$\,K to $7000$\,K when the extra heating rate is increased from $10^{-27}$ to $10^{-25}$\,erg\,s$^{-1}$\,cm$^{-3}$. The higher temperature brings increased collisional excitation of C$^+$ to the fine-structure level, and hence, stronger absorption of C\2* $\lambda1335.7$. As expected, the strengths of the emission components to the C\2\ doublet are hardly affected. The emission appears very strong $-$ enough to overwhelm the absorption component of C\2\ $\lambda1334.5$, but the COS aperture only subtended a small fraction of the H\2\ region, so the emission components are significantly weaker (see Sect.\ref{sec:emission}). Nevertheless, the unknown strength of the emission component adds to the uncertainty on the heating rate as derived from the observed C\2* profile. 

Assuming an extra heating rate of $\sim10^{-25.8}$\,erg\,s$^{-1}$\,cm$^{-3}$, we are able to reconcile our models with the cooling rate derived from C\2* (Sect.\,\ref{sec:cooling}). The electron temperature $T_e$ in the H\1\ region with such extra heating increases from $100$\,K to $1000$\,K. The electron temperature is unfortunately difficult to measure observationally, and, although the column density ratio $N$(Si$^{+*}$)/$N$(C$^{+*}$) provides in principle the gas kinetic temperature \citep{Howk05}, our upper limit on the Si\2*\ column density is too large to provide any constraint. 

By including the extra heating, the main IR cooling lines predicted by the models are [C\2] ($\sim40$\%\ of the total cooling rate), [O\1] 63\mic\ ($\sim30$\%), and [Si\2] 35\mic\ ($\sim26$\%). 
We wish to recall here that these values correspond to the diffuse medium, and that far-IR emission-line intensities might include other components (e.g., photodissociation regions). The O\1* $\lambda1304.9$ absorption-line, arising from the $^3$P$_1$ fine-structure level of O$^0$, is unfortunately undetected. Following Eq.\,\ref{eq:coolperh}, we calculate an upper limit on the [O\1] cooling of $<10^{-26}$\,erg\,s$^{-1}$\,(H$^{-1}$), which is significantly larger than the cooling by [C\2] inferred from C\2*\ (Sect.\,\ref{sec:cooling}). Therefore, we cannot examine in detail the effect of extra heating on the relative cooling line contributions. At best, the large upper limit on the cooling by [O\1] is compatible with the extra heating.

\section{Chemical abundances}\label{sec:abs_discuss}

\subsection{Dominant ionization stages in the H\1\ region}\label{sec:ionization}

In order to calculate the abundances relative to atomic hydrogen, all the ionization stages in the neutral phase should be accounted for. Fortunately, the COS range gives access to the lines of the dominant ionization stages in the neutral gas, namely C\2, N\1, O\1, Mg\2, Si\2, P\2, S\2, Mn\2, Fe\2, and Ni\2. Only chlorine is expected about equally between Cl\1\ and Cl\2. Most studies of DLAs assume that singly ionized metals such as C, Al, Si, S, Fe, and Ni reside exclusively in the H\1\ region so that elemental abundances can be obtained from the ratio of the ion column density to the H\1\ column density (e.g., \citealt{Pettini00}). However, this assumption is not always correct for H\1\ clouds with embedded star-forming regions. To illustrate, we have constructed a simple \texttt{Cloudy} model of the gaseous medium surrounding the ionizing stars (Sect.\,\ref{sec:nebular}). The model is simplistic in having a constant total hydrogen density, 10\cc, but in other respects $-$ metallicity, ionizing radiation field, star-formation rate, and H\1\ column density $-$ the model is appropriate for I\,Zw\,18.

\begin{figure*}
\centering
\includegraphics[angle=0,width=18cm, height=11cm,clip=true]{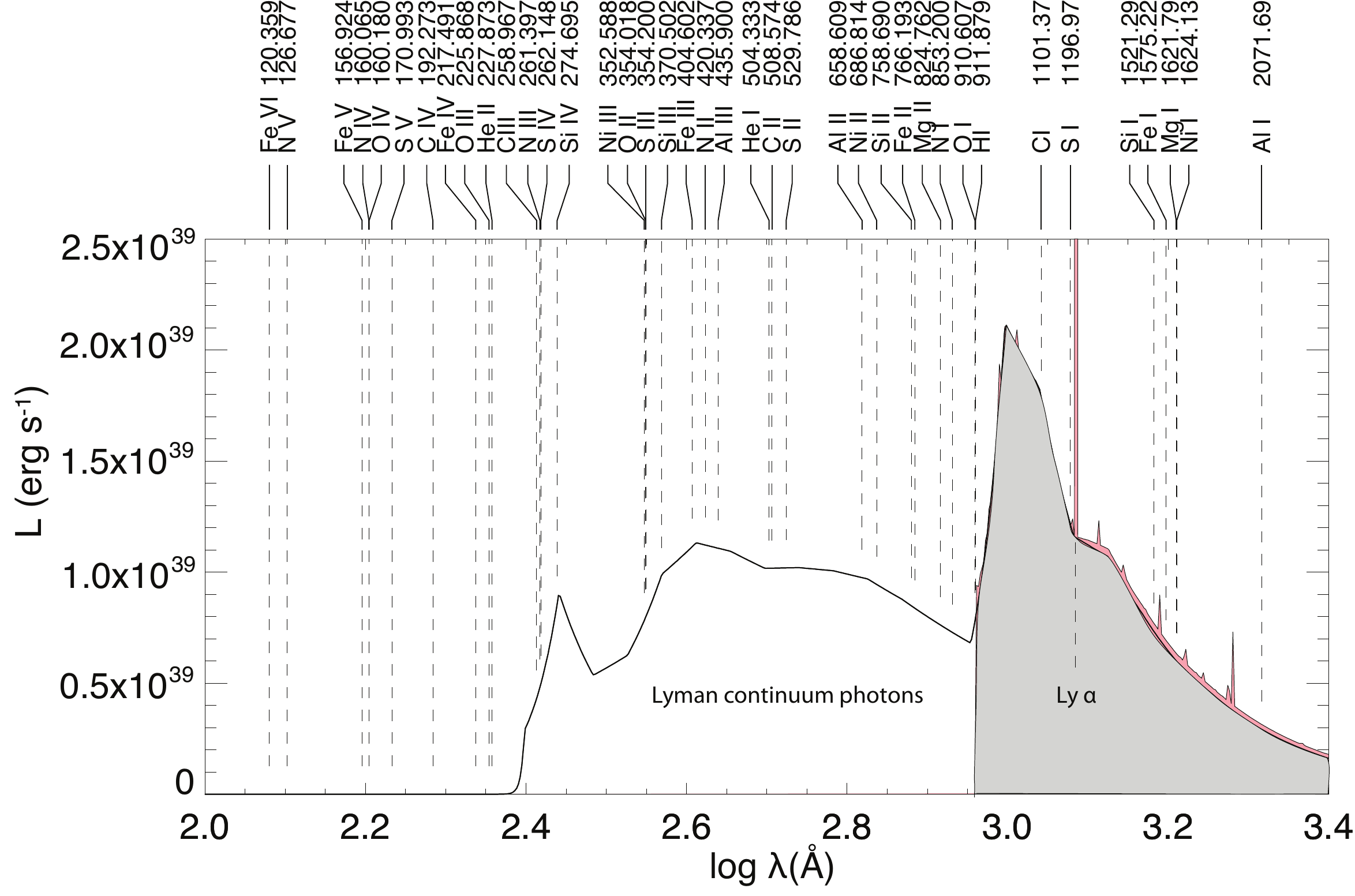}
\caption{\texttt{Cloudy} model spectral energy distribution of I\,Zw\,18. The flux distribution of the stellar population incident on the inner edge of the H\2\ region is shown in white, the stellar flux transmitted to the H\1\ region in gray. The wavelength of the ionization edge of relevant species is given along the top of the figure. The H\2\ region absorbs virtually all the H\1\ Lyman continuum and converts it to H\1\ Ly$\alpha$, nebular continuum, and line emission (all shown in red).}
\label{fig:wedge}
\end{figure*}

Figure\,\ref{fig:wedge} compares the model spectral energy distribution of the ionizing stars incident on the surrounding gaseous medium and after passage through the H\2\ region. No H-, N-, or O-ionizing radiation gets through to the H\1\ region, so oxygen and nitrogen are neutral in the H\1\ region. However, stellar radiation longward of the H\1\ Lyman limit, although diluted by distance from ionizing stars, has basically the same relative flux distribution except for the addition of very strong Ly$\alpha$ and other, much weaker nebular features. Hence, other elements such as C, Al, Ni, Mg, Fe, Si, S, and Cl will be ionized in the H\1\ region.

\begin{figure*}
\centering
\includegraphics[angle=0,width=9.0cm,height=15cm,clip=true]{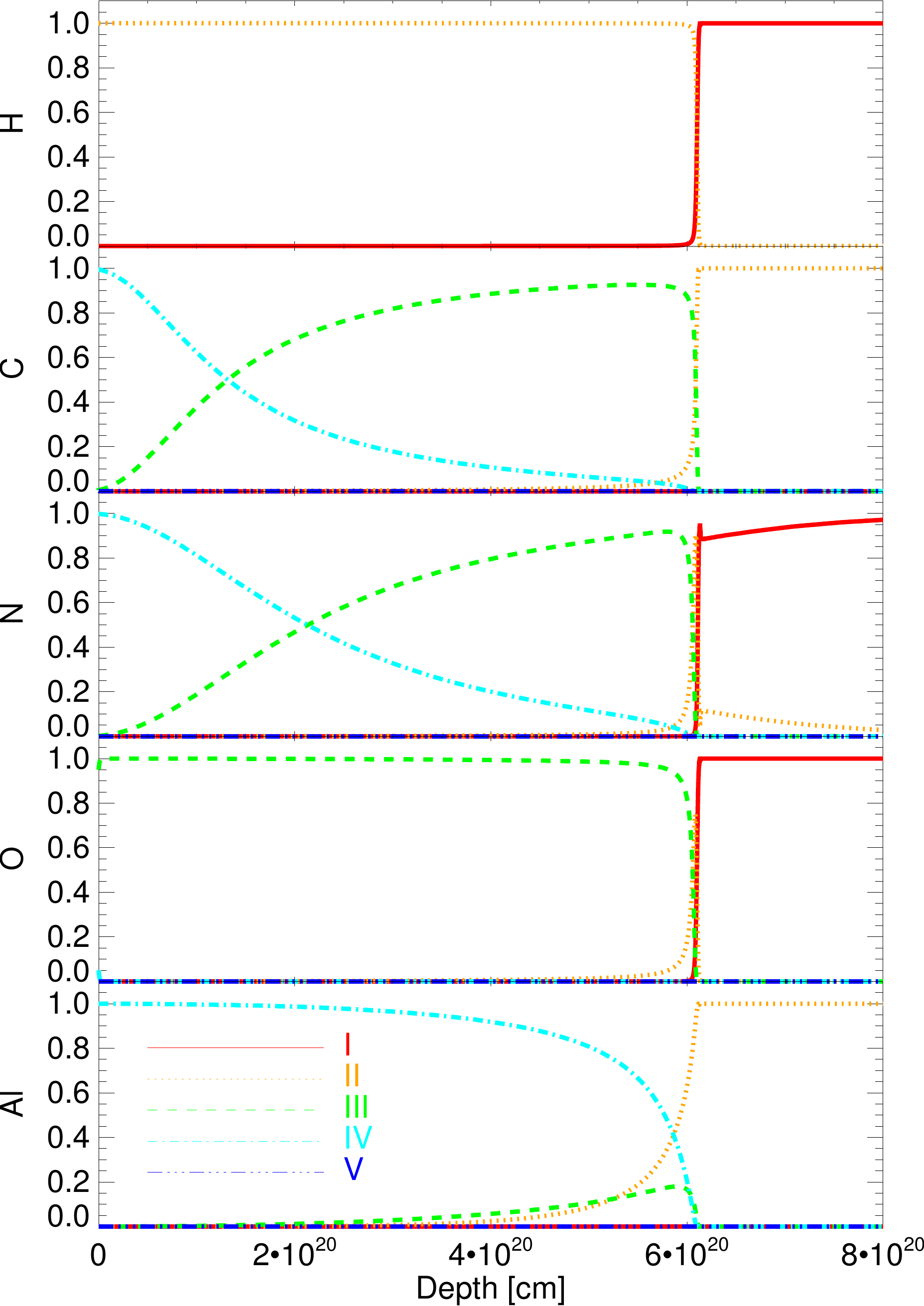}
\includegraphics[angle=0,width=9.0cm,height=15cm,clip=true]{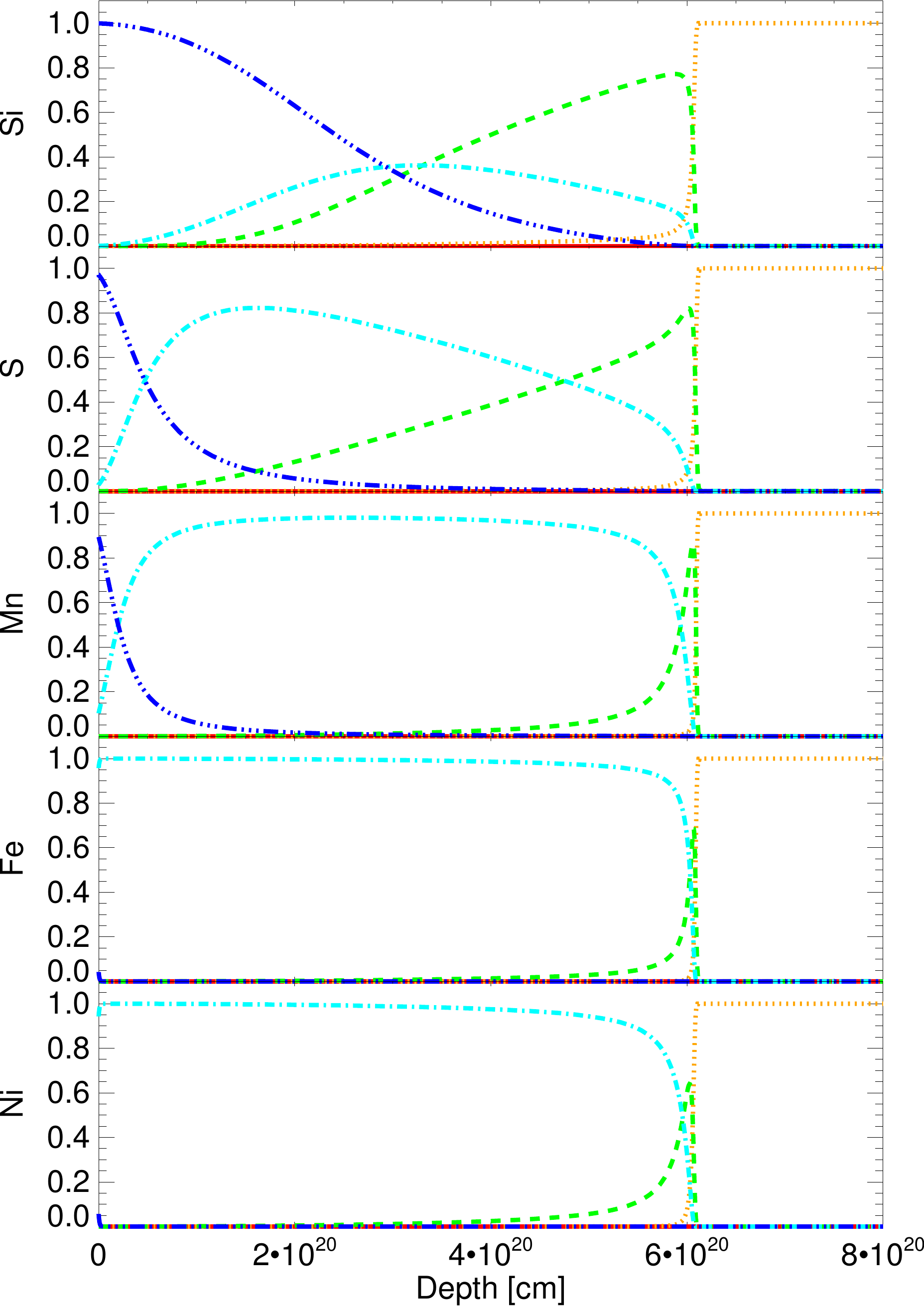}
\caption{Ionization structure of the simple model approximation of I\,Zw\,18 NW. The abscissa is the radial distance from the ionizing stars in cm. Ionization stages are identified by their color: dark blue (V), cyan (IV), green (III), orange (II), red (I). See Fig.\,\ref{fig:simple0} for the corresponding electron temperature and density profiles. }
\label{fig:simple1}
\end{figure*}

From Fig.\,\ref{fig:simple1}, we can see that in the H\1\ region, the dominant ionization state of metals with ionization edges longward of the H\1\ Lyman limit is singly ionized. We use the column densities of the corresponding species to determine the total abundance. A fraction of these species also arises in the H\2\ region where it can lead to emission (recombination of higher stages; Sect.\,\ref{sec:emission}) and absorption. 

A caveat should be that partially-ionized regions might exist on the lines of sight that our model cannot account for. The existence of such regions might affect the ionization structure of species with a large photoionization cross-sections (PiCS). Even if the ionization potential is greater than $13.6$\,eV, the PiCS can be large enough for ionization to occur while hydrogen remains atomic. This effect is unimportant for thick clouds well shielded from the FUV radiation, but becomes significant for thin clouds or clouds embedded in a diffuse UV field. Among the species observed in absorption in galaxies, Ar\1\ is the species the most affected because of its large PiCS (e.g., \citealt{Sofia98a,Jenkins13}). More relevant to our study, N\1\ and O\1\ also have a PiCS larger than that of H\1\ \citep{Verner96}, but also lower than that of Ar\1\ by a factor of $\gtrsim3$, so we expect a much smaller correction than for Ar. Most importantly, the ionization fraction of N and O are coupled to the ionization fraction of H via resonant charge exchange reactions (see also \citealt{Jenkins00a}). For all the other species\footnote{Except Mn\2, which is the only species for which we could not find the PiCS. }, the first ionization potential is lower than $13.6$\,eV, and the second ionization potential is well above, so we always expect them as single-charged ions in the neutral phase even in partially-ionized regions.

In the following, we compare our abundances to the solar values from \cite{Asplund09} except for argon for which we choose \cite{Lodders03}. For the H\2\ region abundances, we use the results of \cite{Pequignot08} and adopt error bars that encompass the determinations by \cite{Izotov99b}. Following \cite{Lebouteiller04}, we define the abundance difference between the H\2\ and H\1\ region as:
\begin{equation}
\delta_{\rm HI} ({\rm X}) = \log ({\rm X/H})_{\rm HII} - \log ({\rm X/H})_{\rm HI},
\end{equation}
where $({\rm X/H})_{\rm HII}$ and $({\rm X/H})_{\rm HI}$ are the abundances of the element X in the H\2\ region and H\1\ region respectively. We also define the abundance relative to the solar value as:
\begin{equation}
[{\rm X/Y}] = \log ({\rm X}/{\rm Y}) - \log ({\rm X}/{\rm Y})_\odot,
\end{equation}
where $({\rm X}/{\rm Y})_\odot$ is the solar abundance reference.

\begin{figure}
\centering
\includegraphics[angle=0,width=9.cm, height=6.8cm,clip=true]{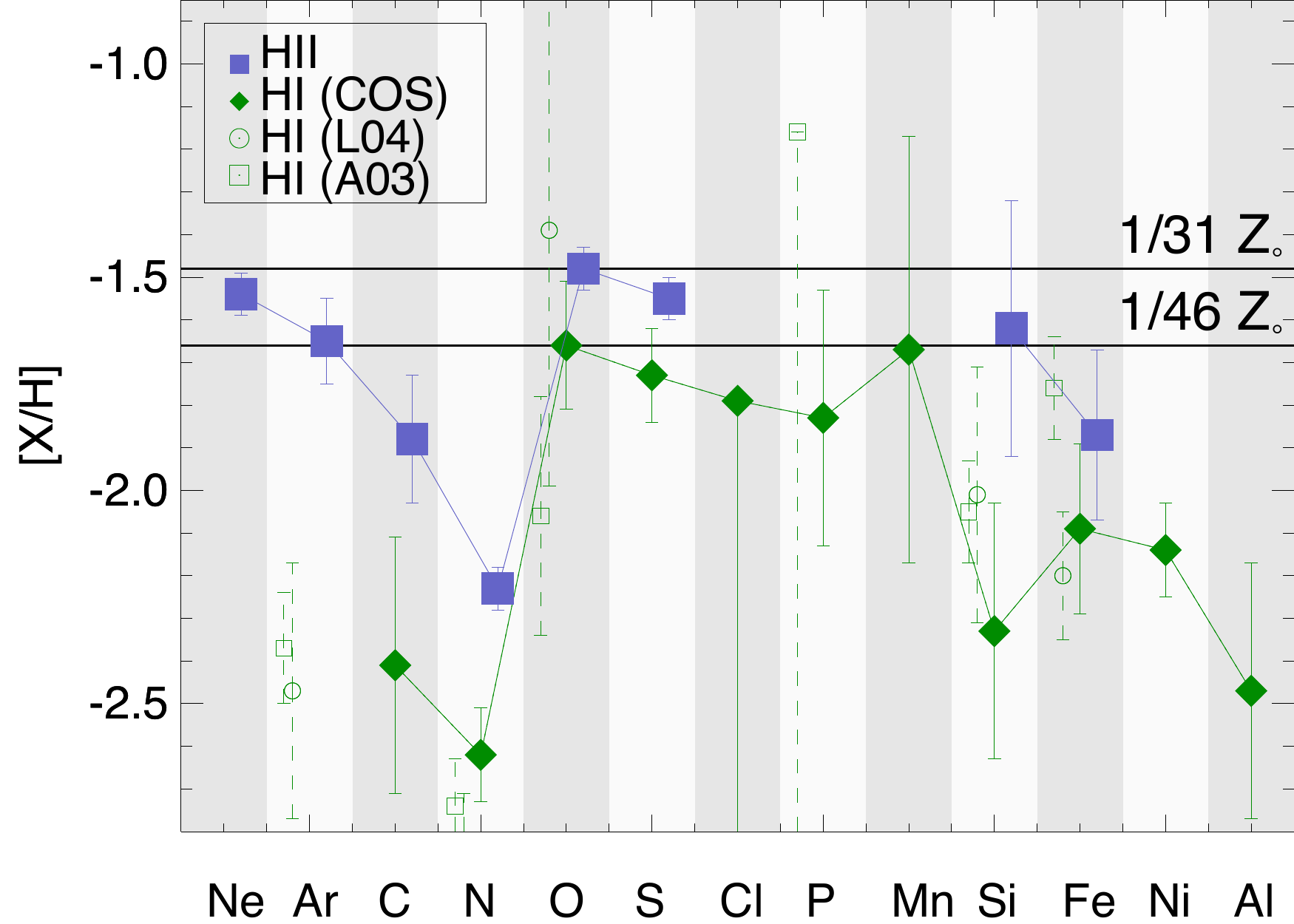}
\caption{The abundance of each element is plotted in the H\1\ region and in the H\2\ region. Elements are ordered by increasing condensation temperature (temperature for which $50$\%\ of an element is removed from the gas phase). The horizontal lines show the metallicity for each phase based on [O/H]. For the H\1\ region, S is used to estimate O (Sect.\,\ref{sec:alpha}).
}
\label{fig:abs}
\end{figure}

Figure\,\ref{fig:abs} shows the abundances in the neutral gas (see Table\,\ref{tab:abs} and Sect.\,\ref{sec:abs_discuss}) as compared those in the ionized gas. The line fits are shown in the appendix, including the line profiles as they would appear if the H\1\ region abundances would be similar to those in the H\2\ region.

\begin{table*}
\caption{Chemical abundances.\label{tab:abs}}
\centering
\begin{tabular}{l | l  ll | l l}
\hline\hline
X  & This study & L04 (FUSE) & A03 (FUSE)  & H\2\ region & Adopted (X/H)$_{\rm HII}$   \\
 \hline
 C  & $6.02\pm0.30$  & ... &... &  $6.30\pm0.10$ (3), $6.55$ (5), $6.54\pm0.10$ (6), $6.47\pm0.15$ (7) & $6.55\pm0.15$ \\
 N  & $5.21\pm0.11$ & $4.92\pm0.20$ & $5.09\pm0.11$ &  $5.58\pm0.60$ (1), $5.61\pm0.03$ (4), $5.60$ (5) & $5.60\pm0.05$ \\
 O  & $7.05\pm0.15$\tablefootmark{a} & $7.30^{+0.85}_{-0.60}$ & $6.63\pm0.28$  &  $7.13\pm0.05$ (1), $7.17\pm0.03$ (2),  Ê$7.22\pm0.01$ (4), $7.21$ (5)  & $7.21\pm0.05$ \\
 Ne & ...  & ...   & ...  &   $6.38\pm0.02$ (1,2), $6.45\pm0.02$ (4), $6.39$ (5)     & $6.39\pm0.05$  \\
 Al  & $3.98\pm0.30$ & ... & ...&  ... &  ... \\
 Si  & $5.18\pm0.30$ &$5.50\pm0.30$ & $5.46\pm0.12$ & $5.61\pm0.23$ (3), $5.89$ (5), $5.65\pm0.22$ (7) & $5.89\pm0.30$  \\
 P  & $3.58\pm0.30$ & ... & $<4.25$ & ... & ... \\
 S  & $5.39\pm0.11$ & ... & ... &  $5.57\pm0.40$ (1), $5.75\pm0.06$ (2), $5.59\pm0.05$ (4), $5.57$ (5) &$5.57\pm0.05$   \\
 Cl  & $<3.71$  & ... &...   &...  & ... \\
 Ar  & ... &$4.15\pm0.30$ &$4.25\pm0.13$ & $4.87\pm0.15$ (1), $5.10\pm0.04$ (2), $4.92\pm0.02$ (4), $4.97$ (5) & $4.97\pm0.10$ \\
 Mn  & $3.76\pm0.50$ &  ... &...  & ... & ... \\
 Fe  & $5.41\pm0.20$ & $5.30\pm0.15$ & $5.74\pm0.12$ & $5.95\pm0.90$ (1), $5.41\pm0.09$ (4), $5.63$ (5)   & $5.63\pm0.20$  \\
 Ni  & $4.08\pm0.11$ &  ... & ...&  ... & ...  \\
 \hline
 \end{tabular}\\
 \tablebib{(1) (09 34 02.1 +55 14 25 J2000) \cite{Thuan05b}; (2) (NW) \cite{Izotov99b}; (3) (NW) \cite{Izotov99a}; (4) (09 34 02.2 +55 14 25.06) \cite{Izotov97b}, (5) \cite{Pequignot08}, (6) \cite{Garnett97}, (7) \cite{Garnett95}. }
 \tablefoot{Abundances are given as $12+\log$\ (X/H). 
  }
\tablefoottext{a}{Calculated from S\2\ column density and assuming S/O=$-1.64$ (ionized gas value from \citealt{Pequignot08}).}
\end{table*}

L04 and A03 examined the same FUSE spectrum and the differences are due to a different method of measuring the column densities and also to the choice of lines (Sect.\,\ref{sec:cds_metals}). Our method is similar to L04's method, i.e., a simultaneous fit of the lines. The differences in the abundances between our study and L04 and A03 directly reflect the differences in the metal column density determinations since our H\1\ column density is similar to their's and since no ionization correction was used in our study nor in L04 and A03.

\begin{table}
\caption{Comparison between abundances in the H\1\ and H\2\ region.\label{tab:compa}}
\centering
\begin{tabular}{lllll}
\hline\hline
X  & (X/H)$_\odot$\tablefootmark{a} & [X/H]$_{\rm HI}$ & [X/H]$_{\rm HII}$ & $\delta_{\rm HI}$(X) \\
\hline
  C   &  $-3.57\pm0.05$  &   $-2.41\pm0.30$     &   $-1.88\pm0.16$      &  $0.53\pm0.34$    \\
  N   &  $-4.17\pm0.05$  &   $-2.62\pm0.12$     &   $-2.23\pm0.07$      &  $0.39\pm0.14$    \\
  O   &  $-3.31\pm0.03$  &   $-1.66\pm0.15$\tablefootmark{b}     &   $-1.48\pm0.06$      &  $0.18\pm0.16$   \\
  Ne   &  $-4.07\pm0.10$  &   ...     &   $-1.54\pm0.11$      &  ...    \\
  Al   &  $-5.55\pm0.03$  &   $-2.47\pm0.30$     &   ...      &  ...    \\
  Si   &  $-4.49\pm0.03$  &   $-2.33\pm0.30$     &   $-1.62\pm0.30$      &  $0.71\pm0.42$    \\
  P   &  $-6.59\pm0.03$  &   $-1.83\pm0.30$     &   ...      &  ...    \\
  S   &  $-4.88\pm0.03$  &   $-1.73\pm0.10$     &   $-1.55\pm0.06$      &  $0.18\pm0.11$    \\
  Cl   &  $-6.50\pm0.30$  &   $<-1.79$     &   ...      &  ...    \\
  Ar   &  $-5.38\pm0.08$  &   $-2.42\pm0.17$\tablefootmark{c}     &   $-1.65\pm0.13$      &  $0.77\pm0.21$    \\
  Mn   &  $-6.57\pm0.04$  &   $-1.77\pm0.50$     &   ...     &  ...    \\
  Fe   &  $-4.50\pm0.04$  &   $-2.09\pm0.20$     &   $-1.85\pm0.20$      &  $0.24\pm0.28$    \\
  Ni   &  $-5.78\pm0.04$  &   $-2.14\pm0.11$     &   ...      &  ...    \\
\hline
 \end{tabular}\\
\tablefoottext{a}{Abundances in the present-day solar photosphere from \cite{Asplund09} except for Ar for which we take Lodders (2003).}
\tablefoottext{b}{Calculated from S\2\ column density and assuming S/O=$-1.64$ (ionized gas value from \citealt{Pequignot08}).}
\tablefoottext{c}{Argon abundance is taken from A03 and L04, with $12+\log({\rm Ar/H})=4.20\pm0.15$ (Table\,\ref{tab:cds}).}
\end{table}

\subsection{O/H and the $\alpha$-elements}\label{sec:alpha}

$\alpha$-elements are produced in short-lived high-mass stars and are often used to trace the metallicity of galaxies. The majority of oxygen is produced by He and Ne burning during the hydrostatic burning phase with negligible contribution from the explosive stages. Silicon is produced by O burning (hydrostatic and explosive) and Ne burning, while sulfur is produced by O burning. Since the $\alpha$-elements are made in the same massive stars that make oxygen, the abundance ratio between two $\alpha$-elements should depend  little on metallicity (see \citealt{Izotov99b}).

The sulfur abundance is tightly correlated with that of oxygen in BCDs \citep{Izotov99b}. 
The S/O ratio in the ionized gas of I\,Zw\,18 is identical the solar ratio, with [S/O]$=-0.07\pm0.08$. Within the error bars, we find that S is barely underabundant in the neutral phase with $\delta_{\rm HI} ({\rm S})\approx0.18\pm0.11$.
Figure\,\ref{fig:fits2} shows that the line profiles are not consistent with the H\2\ region abundance. 
Sulfur is not expected to be depleted on dust  (except in molecular clouds) and ionization corrections are negligible (Sect.\,\ref{sec:ionization}), so we consider that this result implies a genuine underabundance of sulfur in the neutral gas. In other words, the metallicity of the neutral gas is between $51$\%\ and $85$\%\ that of the H\2\ region. 

\begin{figure}
\centering
\includegraphics[angle=0,width=9.cm, height=16cm,clip=true]{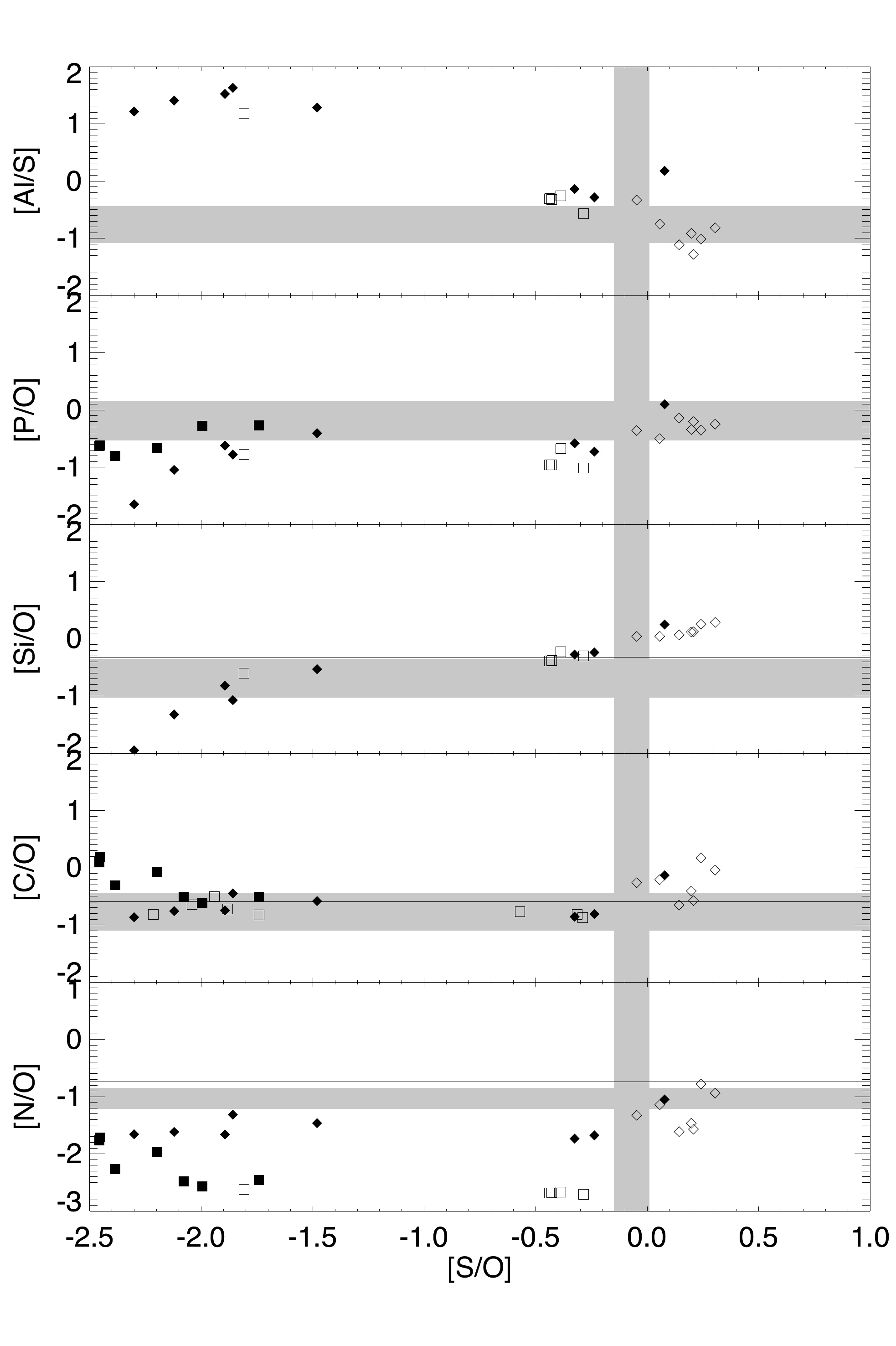}
\caption{Abundance ratios calculated from the stellar yields of \cite{Woosley95}. Yields are plotted for two mass ranges, $11-25$\,M$_\odot$ (open symbols) and $30-40$\,M$_\odot$ (filled symbols), and two metallicities, $0.1$\,Z$_\odot$ (diamonds) and $0.01$\,Z$_\odot$ (squares). The grey stripe shows the observed values. The horizontal black line shows the average ratio observed in the H\2\ region of BCDs \citep{Izotov99b}.  }
\label{fig:yields}
\end{figure}

The silicon abundance determination is fairly uncertain in the H\2\ region, with the values from \cite{Pequignot08} and \cite{Izotov99b} differing by almost a factor $2$. 
Taking at face the value from \cite{Pequignot08}, Si/O in the ionized gas is in good agreement with the solar ratio, with [Si/O]$=-0.14\pm0.33$. Considering the similar nucleosynthesis process of these elements, this indicates little or no depletion of Si on dust grains in the H\2\ region. It must be noted that \cite{Peimbert10} used the relatively smaller silicon abundance found by \cite{Garnett95} to conclude that $\sim39$\%\ of Si might be locked in dust in I\,Zw\,18. 
The silicon abundance in the H\1\ region is uncertain because of the Si\2\ line saturation; our determination is somewhat lower than the column density derived by both L04 and A03 (Table\,\ref{tab:cds}). 
Within errors, we find that silicon is underabundant in the neutral gas with respect to the ionized gas, although with a large uncertainty, with $\delta_{\rm HI} ({\rm Si})\approx0.71\pm0.42$. Figures\,\ref{fig:fits2},\ref{fig:fits3} show that the line profiles are not consistent with the H\2\ region abundance. 
Silicon can be affected by depletion on dust in the neutral phase, but this turns out to be a negligible effect (Sect.\,\ref{sec:depletion}). Furthermore, although a fraction of Si\2\ could originate in the diffuse ionized gas (Sect.\,\ref{sec:emission}), it would only increase the discrepancy between the silicon abundance in the two phases. 
The Si/S ratio in the ionized gas is solar ([Si/S]$=-0.08\pm0.30$), while it is somewhat sub-solar in the neutral gas ([Si/S]$=-0.60\pm0.32$). Assuming that [Si/S] should also be $\approx0$ in the neutral gas, we consider that our silicon abundance might be somewhat underestimated. This hypothesis seems to be confirmed in Fig.\,\ref{fig:yields}, which shows that our observed [Si/O] lies below the predicted value from massive star yields \citep{Woosley95}.

The oxygen abundance is widely used as a metallicity tracer. However the O\1\ lines in the COS spectra are either too weak to provide a useful upper limit or too strong to provide a reliable column density determination (Sect.\,\ref{sec:cds_metals}). To circumvent this issue, we can use silicon or sulfur as a replacement. Sulfur is a better candidate because the S\2\ lines are not prone to saturation. \cite{Bowen05} find that S\2\ traces well H\1\ in a diffuse slab over a wide range of ionization parameter (see also Sect.\,\ref{sec:ionization}). Sulfur is thus a reliable substitute for oxygen and a useful metallicity tracer. Assuming that the S/O ratio in the ionized phase of I\,Zw\,18 ($-1.64$) also holds in the neutral phase, we find (O$_{\rm S}$/H)$\approx7.03\pm0.15$. This value is barely in agreement with the O/H ratio measured in the ionized gas, leading to $\delta_{\rm HI} ({\rm {\rm O}_{\rm S}})\approx0.18\pm0.16$. 
Our indirect determination of the oxygen abundance lies between the FUSE value of L04, who found similar O/H between the two phases (although with large error bars), and the value of A03, who, despite using the same spectrum as L04, found $\delta_{\rm HI} ({\rm O})\approx0.54$ (Sect.\,\ref{sec:cds_metals}).

\subsection{The odd-Z elements}\label{sec:odd}

Although Al and P are not $\alpha$-elements, they are thought to be produced the same massive stars that make oxygen. No significant amount should be produced in the explosive phases and the yields are expected to be metallicity dependent because of the odd-even effect \citep{Woosley95}. Aluminum is produced by C and Ne burning while phosphorus is produced during Ne burning. \cite{Lebouteiller05} found that P\2\ and O\1\ column densities trace well each other over a wide range of environments as long as ionization corrections are not a major issue. Depletion on dust grains affects P more than O (Jenkins 2009), so part of the dispersion observed by \cite{Lebouteiller05} could also be due to a differential depletion factor.

In this section, we discuss the abundances of P and Al relative to $\alpha$-elements since there is no measurement of P or Al abundance in the ionized gas. Assuming a solar P/O ratio we estimate (O$_{\rm P}$/H)$\sim6.86\pm0.33$. This value agrees within errors with the oxygen abundance (O$_{\rm S}$/H)$\approx7.05\pm0.15$ (Sect.\,\ref{sec:alpha}). Figure\,\ref{fig:yields} shows that the expected yields from massive stars are also close to the solar P/O ratio.

The Al/S ratio in the neutral gas is remarkably lower than the solar value, with [Al/S]$=-0.74\pm0.32$. Aluminum abundance is determined using a single Al\2\ line that is unfortunately saturated. However, the saturation of the Al\2\ line is not as strong as C\2\ or Si\2\ lines, so we consider that saturation cannot explain the entire difference. It is also unlikely that there is a significant contribution from Al\1\ given its extremely low ionization potential of $5.99$\,eV. The low [Al/S] ratio could be due to a relatively stronger depletion of Al on dust grains as compared to S. The depletion of Al in the Local Interstellar Cloud is $\sim0.9$\,dex relative to O in the WNM \citep{Kimura03}. Such a factor would be more than enough to reconcile our Al/S ratio with the solar ratio. However, we investigate further the hypothesis of dust depletion in Sect.\,\ref{sec:depletion} and find that depletion alone cannot explain the low [Al/S] ratio. 
The Al/S ratio in the neutral gas, although sub-solar, is in fact consistent with expected yields from massive stars (Fig.\,\ref{fig:yields}). We therefore conclude that aluminum in the H\1\ region is not deficient with respect to the $\alpha$-elements.

\subsection{Carbon}\label{sec:carbon}

Carbon is produced in intermediate- and high-mass stars by He burning. During the early stages of a galaxy evolution, carbon is released mostly by the same young stars that make oxygen, so it is expected that C/O traces the yield ratio from massive stars. As galaxies evolve, older stellar population will contribute to the total carbon abundance, and a rise in C/O is expected (e.g., \citealt{Izotov99a}). 

In the ionized gas of I\,Zw\,18, [C/O]$=-0.40\pm0.17$, which is somewhat lower than more metal-rich BCDs, suggesting that C mostly comes from massive stars, as proposed in \cite{Izotov99a}. Note that the C/O measured by \cite{Izotov99b}, $-0.77\pm0.10$, is slightly lower than both the value in \cite{Garnett97}, $-0.63\pm0.10$, and than the value in \cite{Pequignot08}, $\approx-0.66$. \cite{Garnett97} found that C/O in I\,Zw\,18 is larger than expected from massive star nucleosynthesis and attribute the discrepancy to C enrichment by a lower mass stellar population.

The carbon abundance in the H\1\ region is somewhat lower than in the H\2\ region. 
Using sulfur to trace oxygen in the neutral gas (Sect.\,\ref{sec:alpha}), we find [C/O$_{\rm S}$]$=-0.77\pm0.33$, which falls somewhat below, although compatible with, the ionized gas value. We also note that [C/O$_{\rm S}$] appears to be compatible with expected massive stars yields (Fig.\,\ref{fig:yields}).

\subsection{Nitrogen}\label{sec:nitrogen}

The nucleosynthesis of nitrogen is not as well understood as for carbon or oxygen. Nitrogen can be produced in the CNO cycle by any stars more massive than $1.5$\,M$_\odot$. The observation of a N/O plateau at low-metallicity in BCDs (e.g., \citealt{Izotov99a,Nava06}) leads to the conclusion that nitrogen is produced as a primary element, i.e., using the C and O produced in situ (since little C is available when the stars form). At higher metallicities ($12+\log({\rm O/H})\gtrsim8.3$), secondary nitrogen production (from C and O from older generations) dominates. The origin of the N/O scatter at low-metallicity is still debated. If it is dominated by measurement errors, then nitrogen, like oxygen, is likely produced in short-lived massive stars \citep{Nava06}. The spread could also be intrinsic, due to the delayed injection of N from intermediate-mass stars. The observed scatter is theoretically explained either by effective yield variations and/or different star-formation history \citep{Henry00,Henry06,Meynet02b}. The results of Henry et al.\ also show that nitrogen production in massive stars is possible. 

The N/O ratio in the ionized gas of I\,Zw\,18, $-1.61\pm0.06$, is compatible with the low-metallicity BCD plateau located at $\approx-1.60$ \citep{Izotov99a}. In the neutral gas, nitrogen is somewhat deficient, with $\delta_{\rm HI} ({\rm N})\approx0.39\pm0.11$. This translates into a N/O ratio of N/O$_{\rm S}=-1.84\pm0.19$ which is only somewhat lower than the ionized gas ratio. Following the N/O plateau paradigm, we conclude that the ISM was enriched on large scales by the same stellar population. Assuming the lower N/O in the H\1\ region is genuine, it could be that intermediate-mass stars have contributed to the nitrogen enrichment in the H\1\ region, or that nitrogen abundance was enhanced by massive stars in the H\2\ region (see also Sect.\,\ref{sec:sample}).

\subsection{The iron group elements}\label{sec:irongroup}

The iron group elements are particularly important because they could be released in the ISM on longer timescales than $\alpha$-elements. Iron is produced partly in massive stars and in SNe\,Ia. A useful quantity to disentangle the two origins is [$\alpha$/Fe]. The [$\alpha$/Fe] ratio should increase with the progenitor mass \citep{Woosley95}. If star formation is stopped, SNe\,II contribute no more to the Fe enrichment and SNe\,Ia could become important. 
Our Fe abundance is the same as the value determined by L04 using FUSE, and both values hint at a lower Fe abundance in the neutral gas as compared to the ionized gas. However, our value remains compatible with the ionized gas value within errors, with $\delta_{\rm HI} ({\rm Fe})\approx0.24\pm0.28$. We find that [O$_{\rm S}$/Fe] in the neutral gas is $0.45\pm0.25$, as compared to $0.37\pm0.21$ in the ionized gas. It is thus likely that the enrichment from SNe\,Ia, if any, is identical in the H\1\ phase and in the H\2\ phase. This value of [O$_{\rm S}$/Fe] lies between $0$ (solar O/Fe ratio) and the value in Galactic halo stars \citep{Barbuy88} and in metal-poor stars in the solar neighborhood \citep{Jonsell05}, suggesting an early enrichment by high-mass stars (see \citealt{Izotov99a}). 

Nickel and manganese both belong to the iron group. The depletion strength of Ni is similar to that of Fe while Mn is less depleted than Fe by a factor $\lesssim0.5$\,dex \citep{Jenkins09}. Manganese is mostly made in explosive Si burning while nickel is produced mostly by neutron capture on the iron-peak elements in SNe\,Ia. It should be noticed that Ni and Fe are among the most heavily depleted elements in the ISM (e.g., \citealt{Savage96}). 
We discuss the influence of depletion on dust grains in Sect.\,\ref{sec:depletion}.
The Ni/Fe ratio in the neutral gas of I\,Zw\,18 is solar, with [Ni/Fe]$=-0.05\pm0.23$. We note that [Ni/Fe] remains essentially null over a wide range of metallicities in nearby stars (also in DLAs; \citealt{Izotov01b}) while at the same time [$\alpha$/Fe] increases at low-metallicity \citep{Jonsell05}. This implies that Ni and Fe evolve in parallel, and it is therefore not surprising that we also find a solar ratio in I\,Zw\,18. Since we already suggested that iron might be produced mainly in massive stars, the same result should apply to nickel. 
Concerning manganese, its abundance is uncertain, and we measure a Mn/Fe ratio slightly above the solar ratio, with [Mn/Fe]$=0.32\pm0.51$.

\subsection{Depletion on dust grains}\label{sec:depletion}

We observe a small abundance discontinuity between the H\1\ and the H\2\ region (Sect.\,\ref{sec:abs_discuss}). 
Could part of the discontinuity be explained by a differential depletion factor on dust grains?
Volatile elements such as sulfur and chlorine provide us with a way of estimating the fraction of the underabundance in the neutral gas that is \textit{not} due to depletion on dust. Unfortunately, Cl is expected both as Cl\1\ and Cl\2\ in the neutral phase so our upper limit on Cl abundance is not useful. On the other hand, sulfur is our best metallicity estimator since little or no S is expected to be locked in dust. We found that S/H is at least $\sim15$\%\ lower than in the ionized gas, implying that depletion alone cannot explain the lower metallicity of the neutral gas.

We then examine whether depletion could explain the abundance discontinuity observed for the refractory elements.
For this, we calculate the mass of each element in the H\1\ region locked in dust that would be required to explain the observed abundance discontinuity between the H\1\ and the H\2\ region. Given the small mass fraction of ionized gas in I\,Zw\,18 (e.g., \citealt{Petrosian97}), we expect most of the dust mass to lie in the H\1\ region and we therefore ignore the depletion in the H\2\ region.
Dust extinction toward I\,Zw\,18 is mostly Galactic, and the dust mass in I\,Zw\,18 is exceptionally low. Using data from the \textit{Spitzer} Space Telescope, \cite{Galliano08b} found an upper limit of $<2000$\,M$_\odot$, while \textit{Herschel} data constrains for the first time the dust mass, with $300^{+700}_{-100}$\,M$_\odot$ (R{\'e}my et al.\ in prep.). We assume that the metals are well mixed in the H\1\ region and express the mass locked in dust $m_{\rm d} ({\rm X})$ as:
\begin{equation}\label{eq:depl}
m_{\rm d} ({\rm X}) = (f-1)\ m_{\rm g} ({\rm X}) \sim (f-1)\ A({\rm X})\ ({\rm X/H})_{\rm obs}\ M_{\rm HI},
\end{equation}
where $m_{\rm d} ({\rm X})$ is the mass locked into dust, $m_{\rm g} ({\rm X})$ is the observed abundance in the neutral gas, $f$ is the total mass (dust+gas) divided by the mass in the gas, $A({\rm X})$ is the atomic mass, $({\rm X/H})_{\rm obs}$ is the observed abundance ratio, and $M_{\rm HI}$ is the H\1\ mass. The total H\1\ mass in I\,Zw\,18 is $M_{\rm HI}=6.9\times10^7$\,M$_\odot$, $38$\%\ of which ($2.6\times10^7$\,M$_\odot$) coming from the main H\1\ component associated with the star-forming regions \citep{Lequeux80,vanZee98}. 

\begin{table}
\caption{Mass estimates of elements locked in dust.\label{tab:mass}}
\centering
\begin{tabular}{lll}
\hline\hline
Element &    $m_{\rm g} ({\rm X})$\tablefootmark{a}  & $\Delta m$\tablefootmark{b}   \\
      &   (M$_\odot$)  &   (M$_\odot$)  \\
\hline
C  &  $325^{+325}_{-160}$  & $357-776$    \\ 
N  & $59^{+17}_{-13}$    & $59-86$       \\ 
O$_{\rm S}$  & $4670^{+1920}_{-1365}$  & $0-2100$     \\ 
Al &  $6.7^{+6.7}_{-3.3}$ &    $46-56$     \\ 
Si  &  $110^{+110}_{-55}$ &     $210-450$   \\ 
P   &  $3^{+3}_{-1.5}$    &   $0.4-3.4$  \\ 
S  &    $214^{+61}_{-48}$    & $33-95$     \\ 
Cl  & $<4.7$    &  $<4.4$     \\  
Fe  &  $374^{+220}_{-138}$ &     $0-275$  \\ 
\hline
 \end{tabular}\\
\tablefoottext{a}{Mass of element in the gas phase of the H\1\ region (observed).}
\tablefoottext{b}{Mass required in dust to reach the H\2\ region abundance, i.e., $\delta_{\rm HI} ({\rm X})=0$. For Al, P, and Cl, for which there is no H\2\ region reference, we provide the mass required to reach the metallicity calculated from O, [X/H]$=-1.64$.}
\end{table}

The mass of each element in the gas phase is tabulated in Table\,\ref{tab:mass}. We also list the mass required in the dust phase to reach the corresponding abundance in the H\2\ region. For Al, P, and Cl, whose abundances are not constrained in the ionized gas, we estimate the mass required to obtain a solar proportion with oxygen, i.e., [X/O]$=0$. 
The mass of elements locked into the dust can be then compared to the dust mass measured by \textit{Herschel}. Since we ignore the dust in the H\2\ region, the dust mass derived from \textit{Herschel} is an upper limit to the available dust in the H\1\ region.
We find that depletion alone cannot explain the underabundance of C and Si in the neutral gas. Even though the required mass of N, Al, and S into dust is below the observed dust mass, altogether they represent at least half of the available dust, requiring unlikely dust compounds composition. We thus conclude that depletion on dust cannot explain the observed underabundances in the neutral gas.

Since there is little dust available to remove elements from the gas phase, we also find that the abundance \textit{ratios} are not affected by depletion on dust. As an illustration, there is not enough dust to reconcile O/Fe with the solar ratio (Sect.\,\ref{sec:irongroup}). Similarly, we find that the low Al/S and Si/S ratio (Sects.\,\ref{sec:odd}, \ref{sec:alpha}) cannot be explained by aluminum and silicon locked in dust.

\section{Discussion}\label{sec:discussion}

The H\1\ region of I\,Zw\,18 is not pristine; it has already been enriched in metals. We also find that all the elemental abundances, in particular for Al, S, P, Fe, and Ni,  are consistent with production in massive stars (Sect.\,\ref{sec:abs_discuss}). 
Our best tracer of the enrichment in the H\1\ region by massive stars is sulfur (Sect.\,\ref{sec:alpha}). By using the solar sulfur abundance from Asplund et al.\ (2009), we estimate the metallicity of the neutral phase to be $\approx1/46$\,Z$_\odot$, as compared to $\approx1/31$\,Z$_\odot$ for the H\2\ region. Depletion on dust grains should have a negligible impact in both phases considering the low dust mass in the galaxy (Sect.\,\ref{sec:depletion}). We now discuss the enrichment of the H\1\ and H\2\ regions.

\subsection{Abundance discontinuity H\1-H\2}\label{sec:disc}

The ionized gas in I\,Zw\,18 is more metal-rich than the neutral gas, with abundances higher by $\approx0.18-0.77$\,dex depending on the element.
Our results are a priori compatible with either (1) the H\2\ region being enriched by the present or recent star-formation episode(s) or (2) the presence of extremely low-metallicity clouds in the H\1\ region diluting the observed abundances.

The elements synthesized and released by the current burst are expected to lie at first in a hot phase ($10^6$\,K) and mixing timescales could be as long as $10^9$\,yr. Hot gas outflows have been observed through the O\6\ FUV doublet in starburst galaxies, including a few dwarf galaxies (e.g., \citealt{Grimes09}). However, there is no evidence that I\,Zw\,18 developed such an outflow, even though the non-detection of the O\6\ doublet could be due to the low oxygen abundance in this object \citep{Heckman02}. 
Furthermore,  {\it Chandra} X-ray observations of SBS\,0335-052W, SBS\,0335-052E, and I\,Zw\,18 also failed to reveal any sign of hot gas breaking out from the stellar body implying that the galactic fountain scenario may not be valid in BCDs (\citealt{Thuan04b}; see also Sect.\,\ref{sec:hienrich}). 

As first hypothesized by \cite{Kunth86}, the H\2\ region could be self-enriched, with a fraction of the newly produced heavy elements dispersing on small spatial scales and mixing on short timescales (few Myrs). This would naturally explain the H\1-H\2\ region abundance discontinuity. 
\cite{Legrand00} find that abundances are remarkably uniform in the ionized gas of I\,Zw\,18, with a dispersion of $\lesssim0.1$\,dex on scales of several hundreds of parsecs, which seems to go against the fast and local enrichment hypothesis in the H\2\ regions.  
We recall that our H\1\ region abundances are measured toward the NW region, and compared to the H\2\ region abundances in that region. The NW region is where the youngest stars are located, which in principle provides us with the best chance of observing a local enrichment from the recent star-formation episode(s). 
The most significant abundance discontinuities H\1-H\2\ we calculate are for N and Si, with $\delta_{\rm HI}$(N)$=0.39\pm0.14$ and $\delta_{\rm HI}$(Si)$=0.71\pm0.42$ respectively (Sects.\,\ref{sec:nitrogen}, \ref{sec:alpha}), which is significantly larger than the abundance dispersion across the galaxy in its ionized gas. Although nitrogen abundance might be enhanced by very massive, rotating, stars in the H\2\ region (e.g., \citealt{Meynet02a,Hunter08,Maeder12}), local enrichment by massive stars does not explain the abundance enhancement of the other elements in the H\2\ region.

\begin{figure}
\centering
\includegraphics[angle=0,width=9.5cm, height=7cm,clip=true]{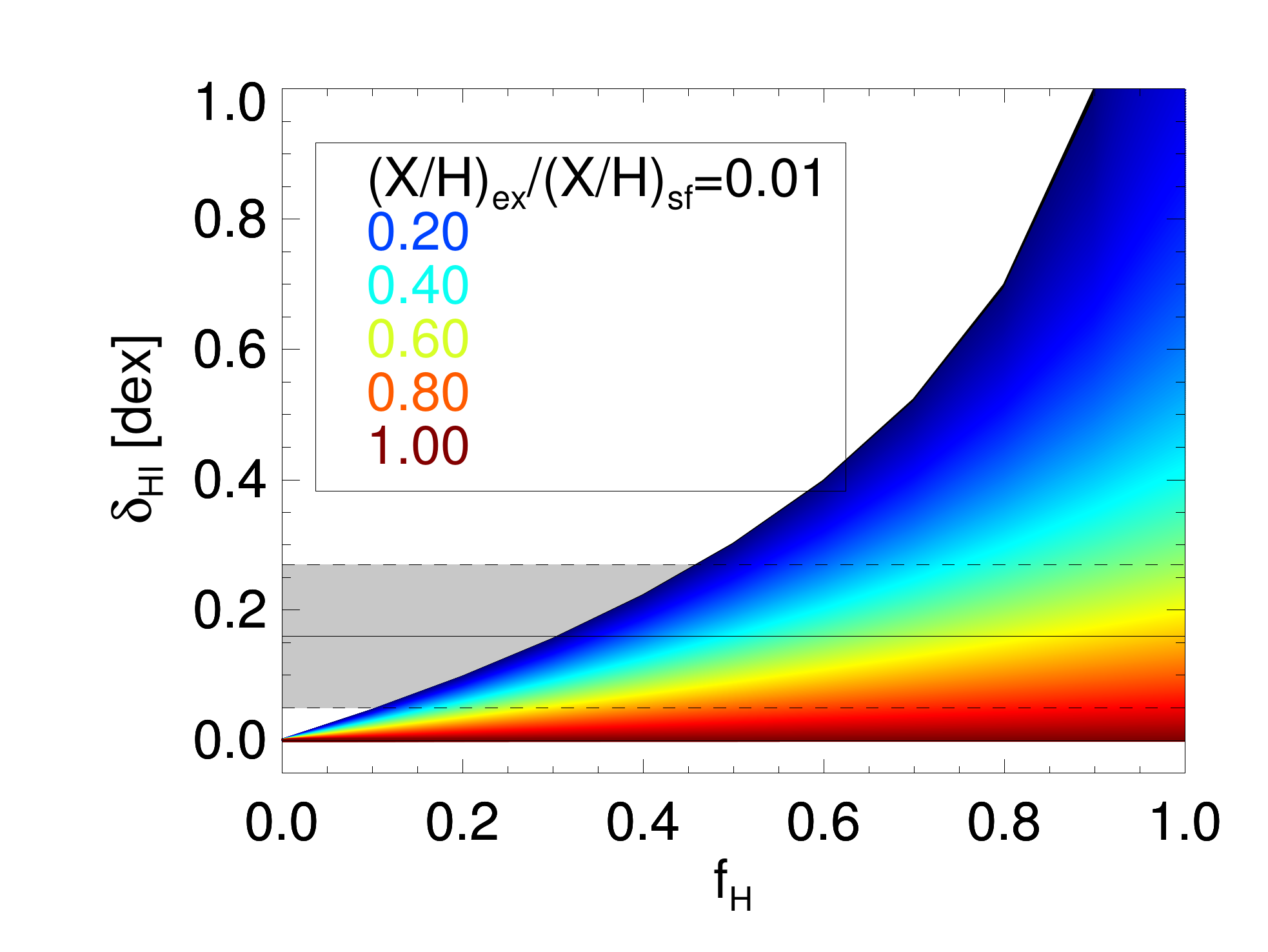}
\caption{The metal deficiency in the H\1\ region $\delta_{\rm HI}$ is plotted as a function of the fraction of hydrogen present in the \textit{HI-ex} component defined as the extended H\1\ component that may be less enriched in metals as compared to \textit{HI-sf} (the H\1\ component associated to the star-forming regions). We assume that the \textit{HI-sf} component has the same metallicity as the H\2\ region and calculate $\delta_{\rm HI}$ for several values of the sulfur abundance ratio between \textit{HI-ex} and \textit{HI-sf}. The horizontal dashed line and the grey stripe indicates illustrate the sulfur deficiency, $\delta_{\rm HI} ({\rm S})=0.18\pm0.11$.   }
\label{fig:phases}
\end{figure}

Another possibility to explain a metal deficiency in the neutral gas is to invoke the presence of extremely metal-poor regions along the lines of sight, such as the low-metallicity high-velocity clouds infalling on the Milky Way. The H\1\ envelope in I\,Zw\,18 extends as much as $\sim13$\,kpc away from the star-forming regions \citep{Lelli12}, and it is possible that the $-$ presumably lower density $-$ outskirts are not as enriched in heavy elements as the immediate surroundings of the H\2\ regions. 
In order to estimate the mass of this relatively metal-poor H\1\ gas, we tentatively assume that the H\1\ medium can be separated into two components, one (hereafter \textit{HI-sf}) is spatially associated with the star-forming region and another component (\textit{HI-ex}) is more extended and possibly less enriched. In the following, we neglect the size and mass of the H\2\ region. We further assume that the mixing of metals is uniform in the vicinity of the H\2\ region, i.e., abundances are similar in the H\2\ region and in  \textit{HI-sf}. The metal abundance in the H\1\ region can then be written as
\begin{equation}
({\rm X/H})_{\rm HI} = \frac{ 1 + f_{\rm X}\ f_{\rm H,ex} } { 1 + f_{\rm H} }\ ({\rm X/H})_{\rm sf}
\end{equation}
And the underabundance $\delta_{\rm HI} ({\rm X})$ in the neutral gas becomes:
\begin{equation}
\delta_{\rm HI} ({\rm X})= -\log\ [ (1-f_{\rm H}) + f_{\rm X}\ f_{\rm H,ex} ],
\end{equation}
where $f_{\rm X}=({\rm X/H})_{\rm ex}/({\rm X/H})_{\rm sf}$ is the relative abundance ratio of a given element between \textit{HI-ex} and \textit{HI-sf} and $f_{\rm H}$ is the fraction of the total H\1\ along the line of sight in the \textit{HI-ex} component. 
Figure\,\ref{fig:phases} shows the variations of $\delta_{\rm HI} ({\rm X})$ with $f_{\rm H}$ and $f_{\rm X}$. 

Taking sulfur as a metallicity tracer ($X=$S), it can be seen that if \textit{HI-ex} is pristine ($f_{\rm S}=0$), the observed sulfur deficiency is compatible with $\approx10-45$\%\ of hydrogen into \textit{HI-ex} along the line of sight. Assuming a spherical distribution with constant hydrogen volume density, this represents a factor $[1 - (1-f_{\rm H})^3]$ in mass, translating to $\sim25-80$\%\ of the total H\1\ mass in \textit{HI-ex}. In other words, about half the mass of the H\1\ envelope in I\,Zw\,18 could be pristine. This scenario is therefore not compatible with a small number of low column density H\1\ clouds such as the high-velocity clouds observed in the Milky Way. If the observed H\1-H\2\ region abundance discontinuity is caused by the presence of pristine gas, the latter should represent a significant mass fraction of the galaxy (whose baryonic mass budget is dominated by H\1\ gas).

We can then assume that the \textit{HI-ex} component has also been somewhat enriched ($f_{\rm S}>0$). Considering that the H\1\ mass in \textit{HI-ex} is diluted over a larger volume, the aperture fraction is probably very small, and we can reasonably assume that the fraction of hydrogen in the \textit{HI-ex} component is less than $75$\%. For such a value of $f_{\rm H}$, the observed sulfur deficiency is compatible with the sulfur abundance in \textit{HI-ex} being $\approx40-80$\%\ of the value in \textit{HI-sf}, and thus also of the value in the H\2\ region. This would imply a metallicity as low as $\sim1/70$\,Z$_\odot$ in the outskirts of the galaxy.
These results could put galaxy evolution scenarios to the test, with in particular the hypothesis of a minimal enrichment level at a given redshift (Sect.\,\ref{sec:hienrich}). Does a metallicity of $\sim1/70$\,Z$_\odot$ correspond to the minimal enrichment at redshift $z=0$? 
Deep H\1\ surveys such as ALFALFA \citep{Giovanelli05}, and later SKA, should identify large numbers of optically-faint H\1\ clouds in the Local Universe that will test the hypothesis of a minimal enrichment level at $z=0$.  
We also wish to emphasize that the hypothetical presence of metal-free gas has important consequences on the interpretation of the integrated D/G ratio, in particular if the latter does not scale linearly with metallicity.

\subsection{Comparison with the sample of BCDs}\label{sec:sample}

\begin{figure}
\centering
\includegraphics[angle=0,width=9.4cm, height=7.5cm,clip=true,trim=50 0 0 0]{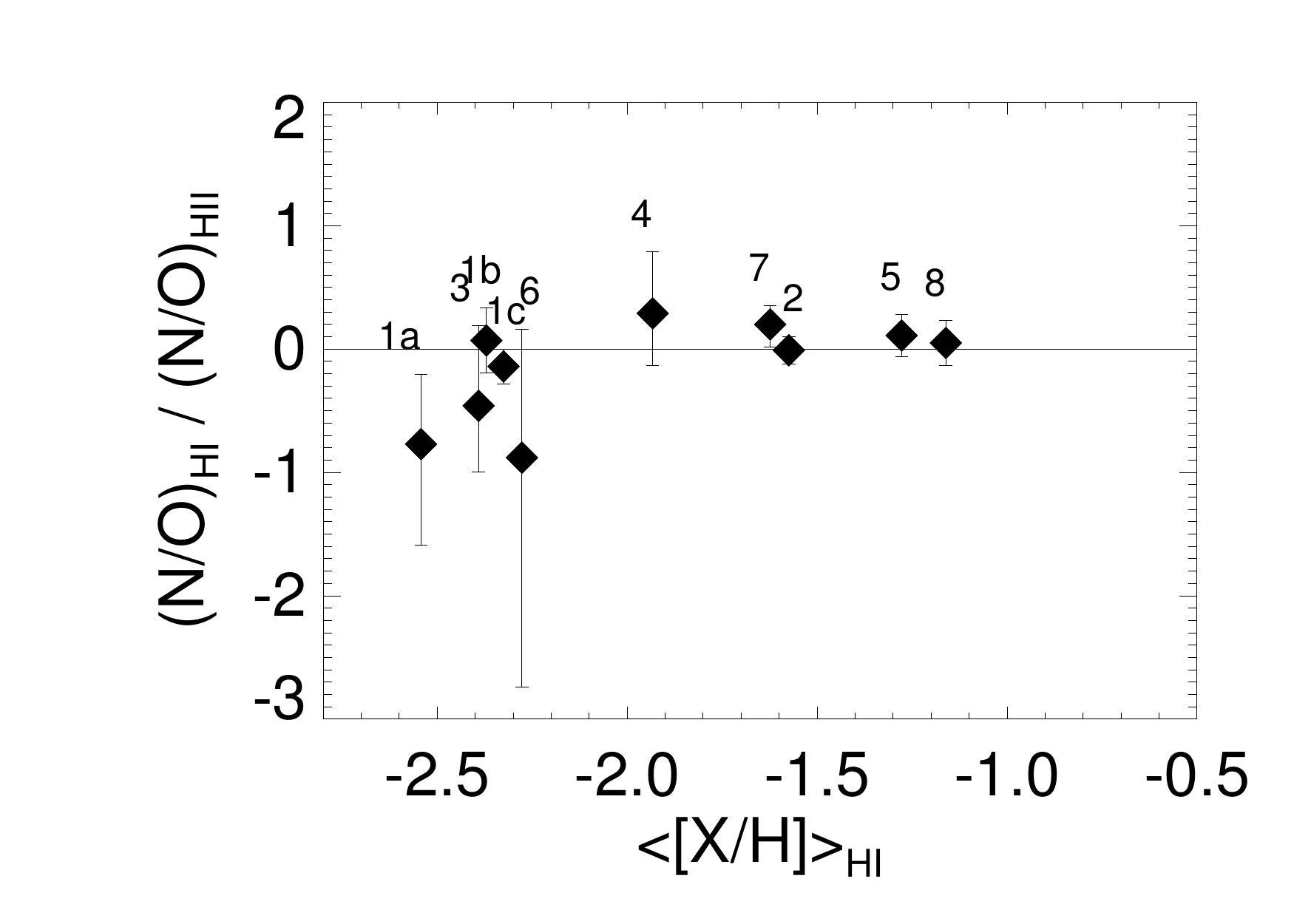}
\includegraphics[angle=0,width=9.4cm, height=7.5cm,clip=true,trim=30 0 0 0]{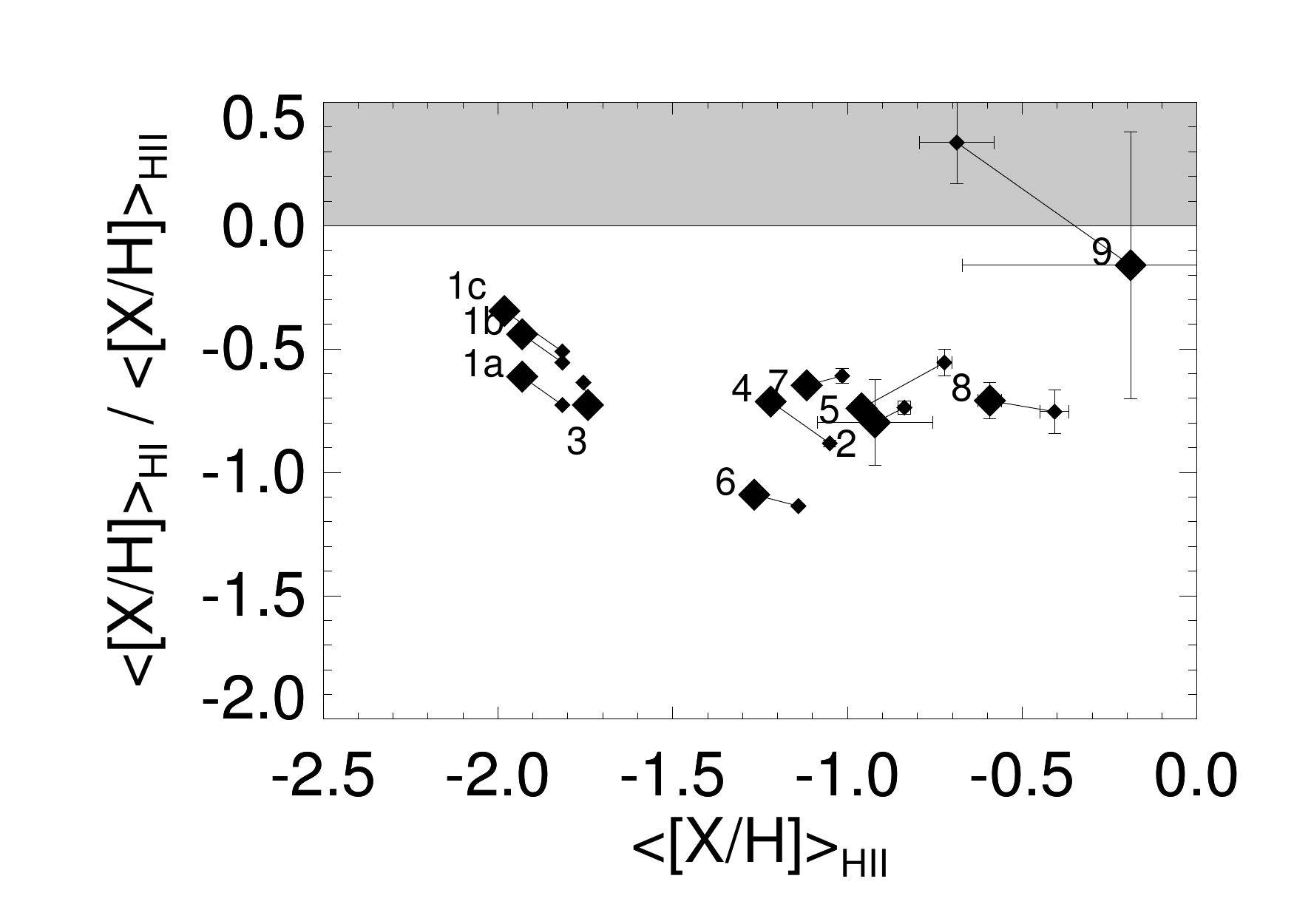}
\caption{\textit{Top} $-$ The N/O ratio is compared between the H\1\ and H\2\ region for the sample of BCDs studied with FUSE, and for I\,Zw\,18 with COS. The labels are defined as 1a: I\,Zw\,18 (L04), 1b: I\,Zw\,18 (A03), 1c: I\,Zw\,18 (this study, with S as a tracer of O; Sect.\,\ref{sec:alpha}), 2: NGC\,1705 \citep{Heckman01}, 3: SBS\,0335-052 \citep{Thuan05a}, 4: Mark\,59 \citep{Thuan02}, 5: NGC\,625 \citep{Cannon05}, 6: I\,Zw\,36 \citep{Lebouteiller04}, 7: Pox\,36 \citep{Lebouteiller09}, 8: NGC\,604 \citep{Lebouteiller06}, 9: SBS\,1543+593 \citep{Bowen05,Schulte04}. 
\textit{Bottom} $-$ Ratio of the average metal abundance $[{\rm M/H}]=<[{\rm X/H}]>$ between the H\1\ and H\2\ regions. [M/H] is calculated two ways, averaging all elements observed in each phase (small diamonds), and averaging only the elements observed in both phases (large diamonds).   }
\label{fig:sample}
\end{figure}

We now compare our results on I\,Zw\,18 to the sample of BCDs whose neutral gas has been investigated with the FUSE telescope. 
First, we examine the N/O abundance ratio as it potentially holds clues as to the stellar population responsible for enriching the ISM. Nitrogen can be synthesized in a wider range of stellar masses than oxygen, and rapidly rotating massive stars can potentially enrich the ISM on short timescales (Sect.\,\ref{sec:nitrogen}). In contrast, oxygen and the $\alpha$-elements are produced exclusively in massive stars and are released within a few Myrs through SNe (Sect.\,\ref{sec:alpha}). 
Figure\,\ref{fig:sample}a compares the N/O ratio in the H\1\ and in the H\2\ region of BCDs. The N/O ratio is similar in both phases for the highest metallicity objects. There might be a hint of a lower N/O in the H\1\ region for the lowest-metallicity objects (I\,Zw\,18, SBS\,0335-052, and I\,Zw\,36), but the result is inconclusive within error bars. We conclude that the stars producing N have the same mass range in both phases, which argues for an enrichment of the ISM before the present starburst episode. 

The hypothesis of relatively metal-poor outskirts (Sect.\,\ref{sec:disc}) can also be put to the test in the BCD sample. \cite{Lebouteiller09} summarized the FUSE results and showed that the metallicity in the H\1\ region in BCDs was equal (I\,Zw\,18, SBS\,0335-052) or lower (I\,Zw\,36, Mark\,59, Pox\,36, NGC\,625, NGC\,1705) than that in the H\2\ region. A trend might exist, with the discontinuity being essentially null for the lowest metallicity objects, and then constant around $\sim1$\,dex for metallicities $\gtrsim1/6$\,Z$_\odot$ (value in the ionized gas). However, if nitrogen is used as a metallicity tracer instead of oxygen, the results could also be interpreted as a constant discontinuity independent on metallicity. The lowest-metallicity object, I\,Zw\,18, 
is key to conclude on the ubiquity of the abundance discontinuity H\1-H\2\ region. While the result for I\,Zw\,18 was at the time inconclusive because of the discrepancy between A03 and L04, our study suggests that the abundance discontinuity in I\,Zw\,18 is rather small, smaller than what is observed in other BCDs. 
We now revisit the findings of \cite{Lebouteiller09} by including this new measurement, and by calculating the average metal abundance $[{\rm M/H}]=<[{\rm X/H}]>$ in the H\1\ and H\2\ regions (Fig.\,\ref{fig:sample}b). 
The choice of averaging the abundances is motivated by (1) possible systematic problems in the abundance determination of some elements (hidden saturation, stellar contamination, atomic data...), and (2) the fact that there is so far no clear evidence of a different nucleosynthesis origin of elements in both phases.
Figure\,\ref{fig:sample}b shows that the abundance discontinuity is around $0.7$\,dex for most objects, with no clear correlation as a function of metallicity. 
This implies that the chemical enrichment H\1\ and H\2\ region is disconnected in most objects, with the H\1 region systematically more metal-poor than the H\2\ region.  
Applying the results of Sect.\,\ref{sec:disc} to the sample of BCDs, Figure\,\ref{fig:phases} shows that for $\delta_{\rm HI}\sim0.7$\,dex, the \textit{HI-ex} component (extended H\1\ envelope) would have a metallicity between $0$\% and $30$\% that of the \textit{HI-sf} component (associated to the H\2\ region), and would represent more than $75$\% of the H\1\ mass. This result is relevant in particular for SBS\,0335-052, which, despite a metallicity close to I\,Zw\,18, shows a discontinuity of $\approx0.7$\,dex (Figure\,\ref{fig:sample}b). For SBS\,0335-052, the H\1\ envelope could be as metal-poor as $\sim1/150$\,Z$_\odot$. New abundance determinations in this object by COS should bring a definitive answer \citep{James13}.

\subsection{Enrichment of the H\1\ region}\label{sec:hienrich}

\subsubsection{Cosmic evolution}

Despite the weak discontinuity between the abundances of the H\1\ and H\2\ regions of I\,Zw\,18, the ISM is fairly well mixed on kiloparsec scales, with the metallicity of the neutral gas at least $\approx50$\%\ of that in the ionized gas (most likely value of $66$\%). 
How was the ISM, and in particular the H\1\ region, enriched? Considering the relatively uniform abundances, at least half of the metallicity in the H\1\ region was presumably set before the current starburst episode. A major uncertainty concerns the \textit{genuine} age of the galaxy as compared to its \textit{chemical} age. The latter is probed by the metallicity parameter, which is determined by a complex interplay between successive star-formation episodes and external gas flow exchanges (outflow of enriched material and infall of metal-poor gas).

Although the age of the oldest stars in I\,Zw\,18 is still subject to debate (see \citealt{Papaderos12}), stars as old as the Hubble time \citep{Aloisi07,Contreras11} might be present, suggesting that I\,Zw\,18 is not genuinely young, and therefore that its present-day metallicity is the result of the mechanisms mentioned above over cosmic times. For this reason, the comparison between I\,Zw\,18 and high-z objects is relevant and instructive. 
The intergalactic medium (IGM)  was already enriched to metallicities similar to that of the H\1\ region of I\,Zw\,18 at $z\sim1.6-2.9$ \citep{Telfer02}. Hence it is possible that I\,Zw\,18 was assembled from a collection of low column-density clouds in the IGM a few Gyrs after the Big Bang, and experienced little enrichment since then. The metallicity of the H\1\ envelope of I\,Zw\,18 is also typical of DLAs at $z>2$ \citep{Rafelski12}, again compatible with a slow chemical enrichment. 
The lack of strong constraints on the mass of DLAs makes the comparison with I\,Zw\,18 relatively uncertain within the chemical downsizing scenario in which galaxies with different mass have different star-formation efficiencies (e.g., \citealt{Brooks07}). Nevertheless, the lack of nearby galaxies as metal-poor as high-z DLAs ($\sim1/100$\,Z$_\odot$; e.g., \citealt{Prochaska99,Rafelski12}) implies that nearby BCDs, including I\,Zw\,18, have experienced a slow, minimal, enrichment over several Gyrs. \cite{Legrand00} suggested that I\,Zw\,18 went through a mild continuous star-formation rate (SFR), $\sim10^{-4}$\,M$_\odot$\,yr$^{-1}$, over several Gyrs as opposed to a series of violent short bursts separated by relatively longer quiescent eras.
Metal enrichment in BCDs through a series of violent star-formation episodes is an unlikely process as galaxies with different star-formation histories would result in different metallicities, thus including a class of nearby galaxies with no starburst episodes and an extremely low-metallicity, which is not confirmed by observations. 

The relation between stellar mass and metallicity in galaxies (e.g., \citealt{Tremonti04}) suggests that star formation proceeds slowly at low metallicity. However, the mass-metallicity relation can also be understood as a lower star-formation efficiency at low mass, with a relatively more efficient ejection of enriched material, or by a metallicity increase through star-formation triggered by infalling gas. \cite{Mannucci10} showed that a more fundamental relation exists between stellar mass, metallicity, and SFR, where downsizing, outflows, and infalls are the major actors.

Accretion of inflowing cold gas triggers a transient phase in which (1) metallicity is diluted by the injection of significant amounts of low-metallicity gas, (2) the SFR increases through the Schmidt-Kennicutt law, and (3) metallicity increases again as star formation proceeds.  
Although H\1\ companions are often observed around star-forming dwarf galaxies (e.g., \citealt{Taylor97,Ramya09,Pustilnik01b}), their accretion and the consequence on star formation is not definitive (e.g., \citealt{Telles00}). 
Gas inflow is expected to play a more important role at high-z. The cosmological inflow from the cosmic web could have been so high at high-$z$ that star-formation could not catch up in the least massive haloes like I\,Zw\,18 \citep{Krumholz12b}. This metallicity-dependent quenching would result in a growth of the H\1\ envelope, the latter still being observed today. Only at $z<2$ could a low-mass galaxy like I\,Zw\,18 start to build up its metallicity. Such a scenario remains compatible with the presence of old stars in I\,Zw\,18.

Alternatively (or possibly in parallel), galactic winds could expel most of the metals into the IGM through multiple supernovae explosions, effectively keeping the ISM metallicity to low levels (e.g., \citealt{Recchi04}). The most active star-forming regions would be responsible for those winds.
A major uncertainty remains however on the fate of the metals observed in galactic outflows, since the superbubble expansion might decelerate, and the outflowing material might cool down and return to the ISM instead of escaping the gravitational well (e.g., \citealt{Tenorio99,MacLow99,Rieschick03}). Despite the current starburst episode in I\,Zw\,18, there is no evidence of outflowing hot gas (Sect.\,\ref{sec:disc}). Furthermore, we note that outflows would play a more important role in galaxies whose star-formation history is dominated by a series of violent bursts, which has probably not been the case in I\,Zw\,18 \citep{Legrand00}. 

In summary, although gas flows play a major role in the cosmic evolution of galaxies, their effects in I\,Zw\,18 are not yet fully constrained and understood. A combination of metal-poor gas infall in the early stages of the galaxy, and a chemical downsizing could be responsible for the extremely low-metallicity observed today in the H\1\ and H\2\ regions. 
Finally, our results do not allow putting constraints on the reasons of the current starburst episode in I\,Zw\,18, which has been $\sim10-100$ times relatively stronger over the past $\sim50$\,Myr \citep{Aloisi99,Legrand00}. The sudden onset of star formation could have been triggered by gas compression from merging events. The disturbed H\1\ morphology suggests that several small structures might have been interacting (Lelli et al.\ 2012). Such structures could have been gas-rich dwarf galaxies or sub-DLAs. From the radial velocity of the neutral metals observed with COS, we did not find any evidence of infalling material on the NW region itself; the H\1\ gas is, if anything, outflowing from the H\2 region with a velocity of $\sim20$\kms\ (see also A03; L04). However, \cite{Lelli12} found a radial inflow/outflow motion on the order of $\sim15$\kms\ on larger scales and concluded that interactions or mergers are the most likely explanation for the current starburst episode. 
In general, the FUSE results for other BCDs show blue shifts of at most $10-20$\kms\ (e.g., \citealt{Lebouteiller09,Cannon05}), confirming that slow radial flows of H\1\ gas might be frequent.

\subsubsection{Local conditions for star formation}

We now discuss the star-formation process itself, and the influence of metallicity in the ability of clouds to form stars in an environment as metal-poor as I\,Zw\,18. Our present study puts strong constraints on the metallicity of the H\1\ region, which constitutes the bulk of the mass of the galaxy. The H\1\ region is thus key to understand the past star-formation efficiency. 

First, the fact that stars of all masses have formed for several Gyrs has some implications for the early metallicity evolution of the galaxy. \cite{Bromm03} showed that when the metallicity reaches more than $10^{-4}-10^{-3}$\,Z$_\odot$, the cooling by the far-IR fine-structure lines [C\2] 157\mic\ and [O\1] 63\mic\ enables cloud fragmentation and therefore low-mass star-formation. No dust or molecules were considered for this calculation and we note that H$_2$ could provide the most efficient cooling at even lower metallicity (e.g., \citealt{Omukai05}). However, the lack of diffuse H$_2$ in I\,Zw\,18 \citep{Vidal00} suggests that H$_2$ cooling is not significant in our practical study, except if present in small dense clumps where H$_2$ might be more abundant (see \citealt{Thuan05a}). We thus get the picture of I\,Zw\,18 forming stars of all masses only a few Gyrs after the Big Bang, when its metallicity reached $\gtrsim10^{-4}$\,Z$_\odot$. The oldest stars in I\,Zw\,18 could be the remnants of the star-formation episode following this early enrichment. 
The existence of a metallicity threshold for low-mass star-formation is however still debated and we note that the presence of dust to shield UV photons, and the D/G ratio, ultimately related to metallicity via the evolution of the dust-to-metal ratio, may be the most important parameters governing the star formation process \citep{Schneider12,Glover12a}. 

The importance of the ISM metallicity is fundamental for the ability of clouds to form stars, and it may well regulate the SFR as long as external gas flows are unimportant.
Although the observed SFR in BCDs is $\sim10^{-2}-10$\,M$_\odot$\,yr$^{-1}$, this corresponds to the current starburst episode, and a much milder SFR could have dominated beforehand. For example, in I\,Zw\,18 the past SFR could have been as low as $\sim10^{-4}$\,M$_\odot$\,yr$^{-1}$; e.g., \citealt{Legrand00}). It seems at first contradictory with \cite{Mannucci10} who finds that for a given stellar mass, the SFR is anti-correlated with metallicity. The fundamental metallicity relation found by \cite{Mannucci10} is, however, biased toward emission-line galaxies, and a ``pre-starburst'' I\,Zw\,18 would not have been selected by their survey. Furthermore, as discussed in the following, it is possible that the extremely low-metallicity of this galaxy has a major impact on the star-formation process itself. 

The temperature and density of the ISM are essential parameters for star formation. The latter can be inhibited by the inability of the gas to cool down enough for removing thermal pressure and for creating gravitational instabilities. 
The warm temperature in a metal-poor ISM is due to the paucity of efficient cooling mechanisms, such as molecules or the far-IR fine-structure line [C\2], which dominates the cooling in the diffuse atomic clouds. \cite{Krumholz12a} showed that the thermal equilibrium timescale is more important than the timescale for conversion of atomic to molecular gas in a low-metallicity gas. Star formation could proceed before the medium becomes molecular when the metallicity reaches below a few percents solar. The metallicity threshold above which thermal equilibrium is reached before a few free-fall timescales (once the gas becomes gravitationally unstable) thus depends mostly on the ability of the gas to cool down. The H\1\ region of I\,Zw\,18 is an ideal testbed for star-formation timescales in a low-metallicity atomic gas, since molecular gas has yet to be detected. 
The model in \cite{Krumholz12a} highly depends on the heating mechanisms and their metallicity dependence, and the uncertain heating mechanisms in the H\1\ region of galaxies such as I\,Zw\,18 (Sect.\,\ref{sec:cii_heating}) add to the uncertainty on the threshold metallicity and thermal equilibrium timescale. 
Following the method in \cite{Krumholz12a}, we use our determination of the heating rate in the H\1\ region, and an initial gas density and temperature of $\sim10$\,\cc\ and $\sim1000$\,K, respectively. We find that thermal equilibrium is reached before a few free-fall timescales only for a metallicity greater than $\sim1/100$\,Z$_\odot$. A lower threshold metallicity would be obtained if we consider a significant clumping factor. 
Hence we see that widespread star formation is possible in the H\1\ region of I\,Zw\,18, currently holding a metallicity of $\approx1/46$\,Z$_\odot$. If such physical conditions prevail in objects as metal-poor as $\lesssim1/100$\,Z$_\odot$, the thermal equilibrium timescales are rather long and the ability to form stars ultimately depends on the survival of the clouds. I\,Zw\,18 might have kept a low metallicity because its ISM remained warm on long timescales.  
A major uncertainty remains on the presence of small cold clumps  (cold neutral medium phase) in which star formation would be more efficient. The mixing factor of the warm neutral medium and cold dense clumps in the FUV absorption lines is difficult to quantify. Moreover the ISM pressure in I\,Zw\,18 might be too low to maintain a multi-phase ISM, so that the neutral gas we characterized in this study might be the only thermally-stable phase. The temperature from this phase could thus be a crucial parameter to understand the star-formation process on large spatial scales and over cosmic times.

\section{Conclusions}

We present a comprehensive investigation of the H\1\ envelope of I\,Zw\,18. The H\1\ region holds several fundamental clues to the chemical evolution of galaxies. The metallicity of the H\1\ region is set by the successive star-formation episodes over the galaxy history, and is modulated by outflows of supernovae products and hypothetical inflows of quasi-pristine gas from the cosmic web. The abundance of metals in the H\1\ region is also a critical parameter to understand the present and the future star-formation history, as metals provide most of the cooling in the H\1\ gas reservoir to facilitate and sustain the star-formation process. Our objectives are to characterize the H\1\ region by (1) its physical conditions and (2) its chemical abundances. We compare the abundances to those derived in the H\2\ region (from optical emission-lines), and discuss the enrichment history of the galaxy.

The H\1\ envelope of I\,Zw\,18 was observed toward the NW young stellar cluster in the FUV with the COS instrument on \textit{Hubble}. The massive stars provide the FUV continuum over which absorption-lines from neutral species are superimposed. We modeled the stellar absorption in order to remove its contribution and isolate the interstellar absorption. We derived the column density of H\1\ and of heavy elements (C\2, C\2*, N\1, O\1, Al\2, Si\2, P\2, S\2, Mn\2, Fe\2, Ni\2). Models were constructed to compute the ionization structure of the nebula. Our models predict a large ionized gas volume in the H\2\ region, suggesting that the H\1\ cavity observed at 21\,cm toward NW is due to ionization by the stellar cluster as opposed to blow-out by stellar winds and supernovae. 
We study potential emission components in the H\2\ region that could work out to reduce the apparent optical depth of absorption lines. 

\noindent The physical conditions in the H\1\ region are drawn from the study of the C\2* absorption-line. 
\begin{itemize}
\item The population of the fine-structure level of C$^+$ toward NW is compatible with collisions with H$^0$ atoms, implying that C\2* arises in the H\1\ envelope. 
\item We determine observationally that the electron fraction in the H\1\ region is less than $\sim2$\%; our models predict $\sim0.1$\%. 
\item From C\2* we derive a cooling rate that is larger than the models. An extra heating source is required in the models to reconcile them with the observations, leading to an elevated electron temperature in the H\1\ region of $\sim1000$\,K. 
\item The cooling rate inferred from C\2* is used to calculate the star-formation rate. Our value agrees with other determinations in the literature, although on the high-end. The D/G ratio used for this calculation is a factor of $\sim2000$ times lower than the Milky Way value, i.e., a factor of $\sim60$ times lower than what is expected if we simply scale D/G with  metallicity. 
\end{itemize}

\noindent We then infer the dominant ionization stages in the H\1\ region from the models and calculate the chemical abundances in the H\1\ region. 
\begin{itemize}
\item We find that sulfur is less abundant in the H\1\ region as compared to the H\2\ region, by a (log) factor of $0.18\pm0.11$.
\item Nitrogen, and possibly carbon, are also underabundant in the H\1\ region. 
\item The P/S ratio is solar while Al/S is consistent with yields from massive stars. The Fe abundance is the same in the 2 phases within errors. The [$\alpha$/Fe] ratio is consistent with production in massive stars. 
\item Depletion on dust grains is unlikely to explain the observed abundance discontinuity between the H\1\ and H\2\ regions. This is due to the low dust mass available.
\end{itemize}

\noindent Finally, we discuss possible scenarios concerning the abundance difference in the gas phases, and concerning the enrichment of the H\1\ region.
\begin{itemize}
\item Local enrichment in the H\2\ regions could explain the enhanced N abundance but fails to explain other results, especially for S. 
\item The presence of relatively lower metallicity outskirts in the H\1\ region is discussed. As much as $50$\%\ of the H\1\ mass could be pristine. 
\item The metallicity of the H\1\ region is not null, it is compatible with the IGM metallicity at $z=1.6-2.9$. I\,Zw\,18 could  be considered as an old DLA with little recent star formation. The presence of old stars confirm that I\,Zw\,18 is not a genuinely young galaxy. The inflow of quasi-pristine gas from the cosmic web for a few Gyrs after the Big Bang could explain the massive H\1\ envelope and the quenched star-formation in the early stages of the galaxy evolution. 
\item The current onset of star formation could be a due the merging of dwarfs or sub-DLAs, as suggested by the disrupted H\1\ morphology. 
\end{itemize}

The sample of BCDs whose neutral gas has been investigated by FUSE and COS shows that a systematic discontinuity exists between the H\1\ and H\2\ region abundances. I\,Zw\,18 is the object with the lowest discontinuity. Another BCD, SBS\,0335-052, despite having a metallicity similar to I\,Zw\,18 does show a significant discontinuity which could translate in the H\1\ region being as metal-poor as $1/150$\,Z$_\odot$.

\begin{acknowledgements}
V.L. is supported by a CEA/Marie Curie Eurotalents fellowship.
 We would like to thank Aur\'elie R\'emy for the \emph{Herschel} dust mass measurements. 
Based on observations obtained with the NASA/ESA Hubble Space Telescope, which is operated by the Association of Universities for Research in Astronomy, Inc., under NASA contract NAS5-26555. SRH and IH gratefully acknowledge support by the NASA Cosmic Origins Spectrograph program at the University of Colorado, Boulder.
\end{acknowledgements}

\bibliography{/Users/vianneylebouteiller/Documents/TeXstyle/mybib,/Users/vianneylebouteiller/Documents/TeXstyle/mybib_inprep}

\appendix

\section{Line fitting}\label{sec:profiles}

\begin{figure*}

\includegraphics[angle=90,width=9.3cm,height=7cm,clip=true]{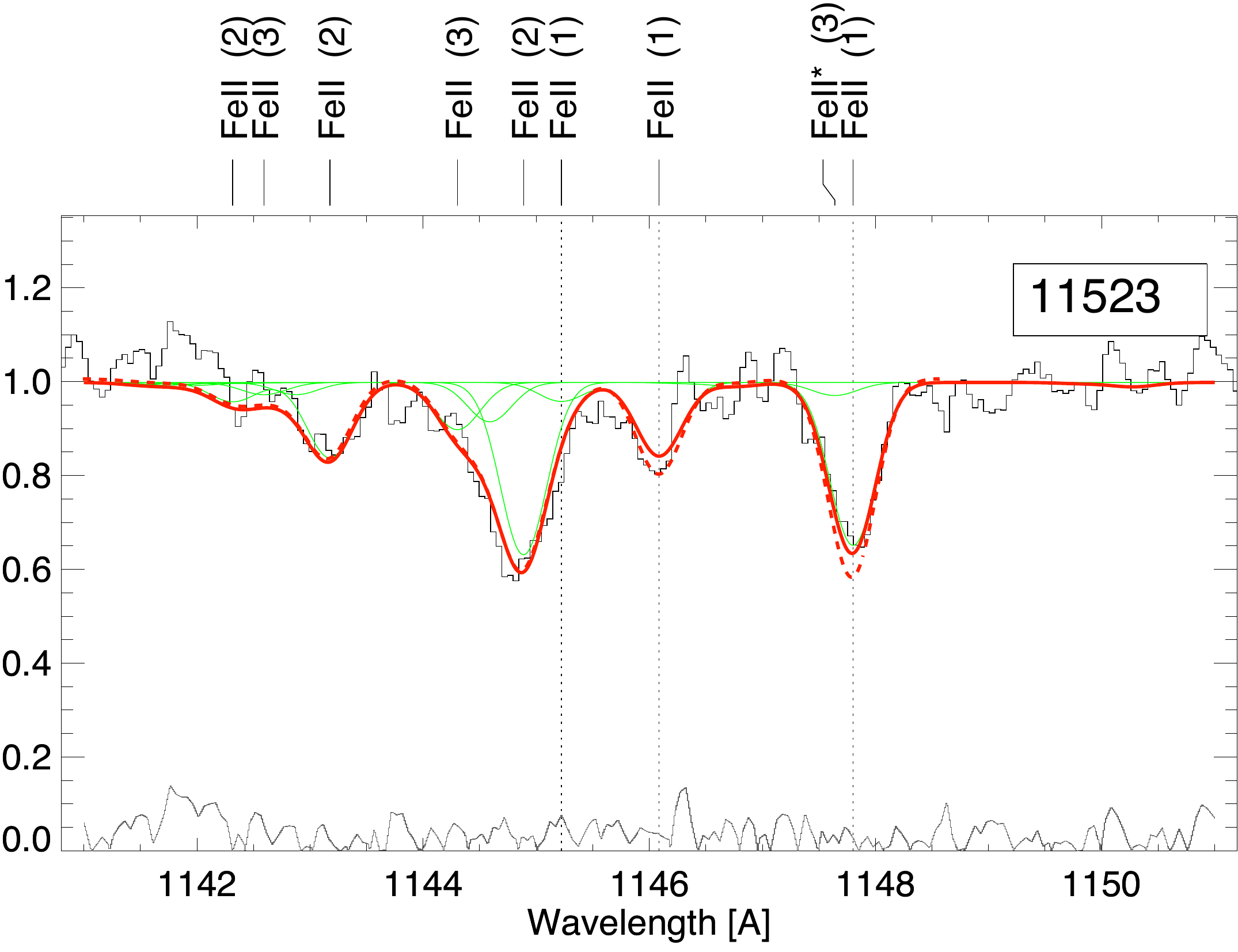}
\includegraphics[angle=90,width=9.3cm,height=7cm,clip=true]{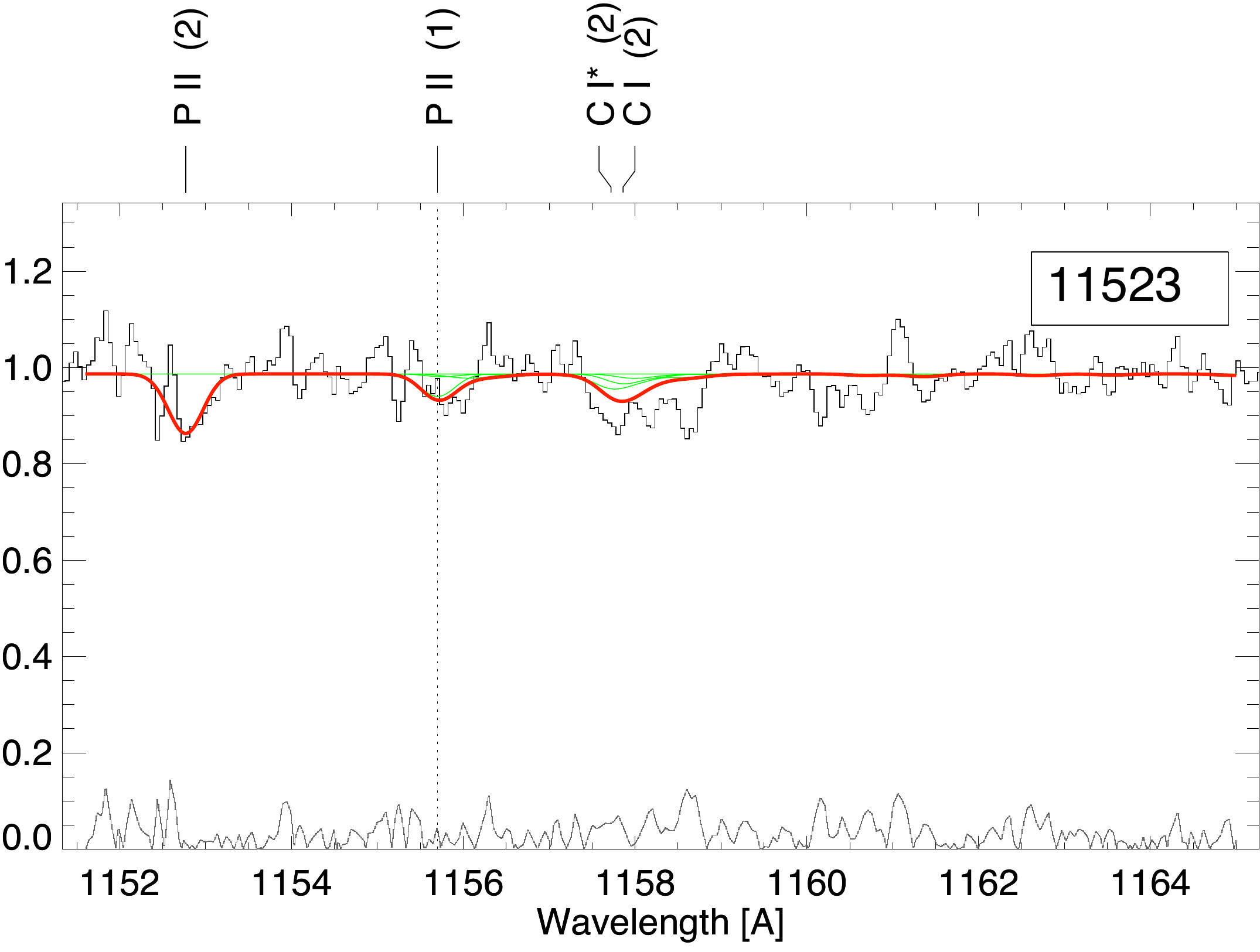}\\

\hspace{9.3cm}
\includegraphics[angle=90,width=9.3cm,height=7cm,clip=true]{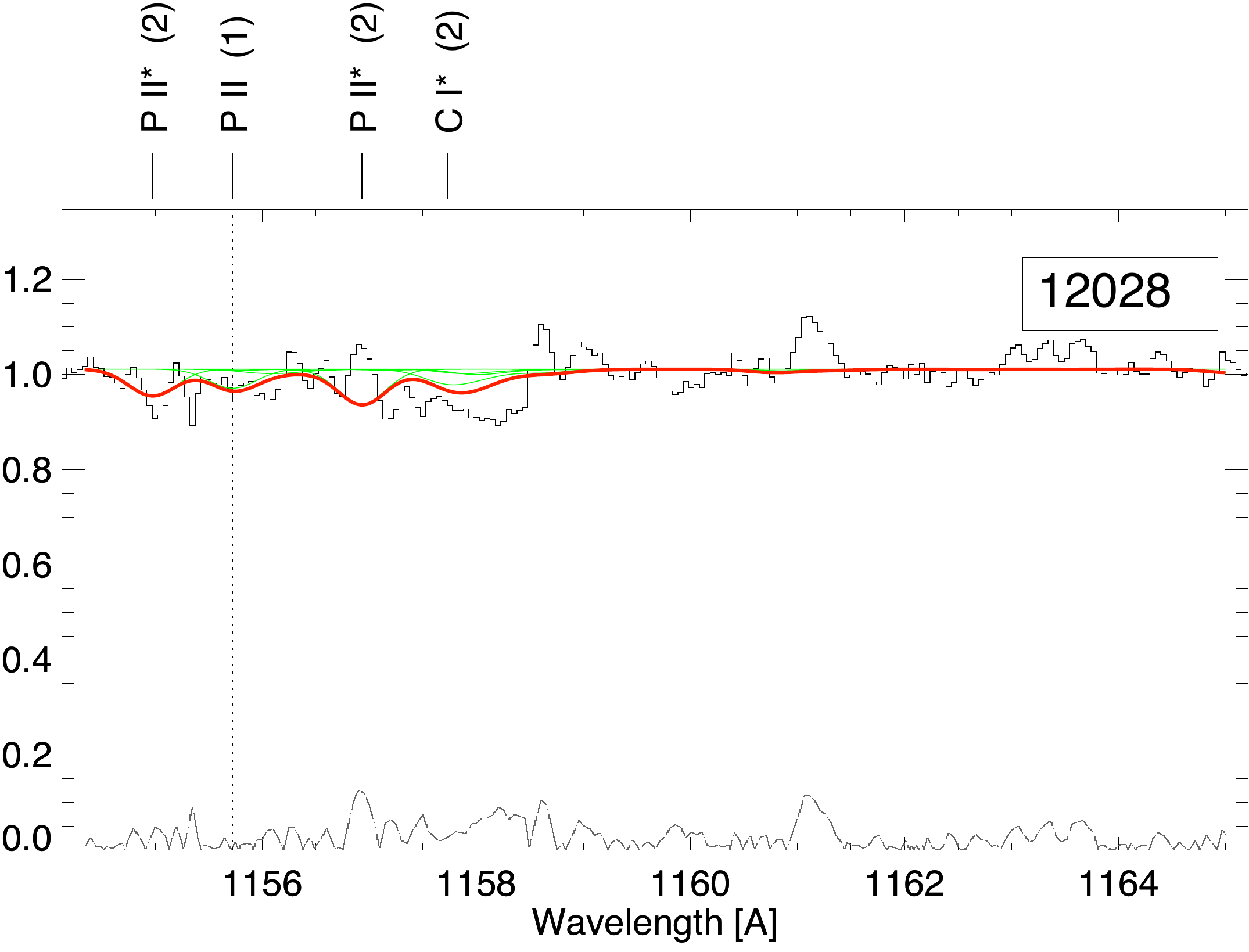}\\

\includegraphics[angle=90,width=9.3cm,height=7cm,clip=true]{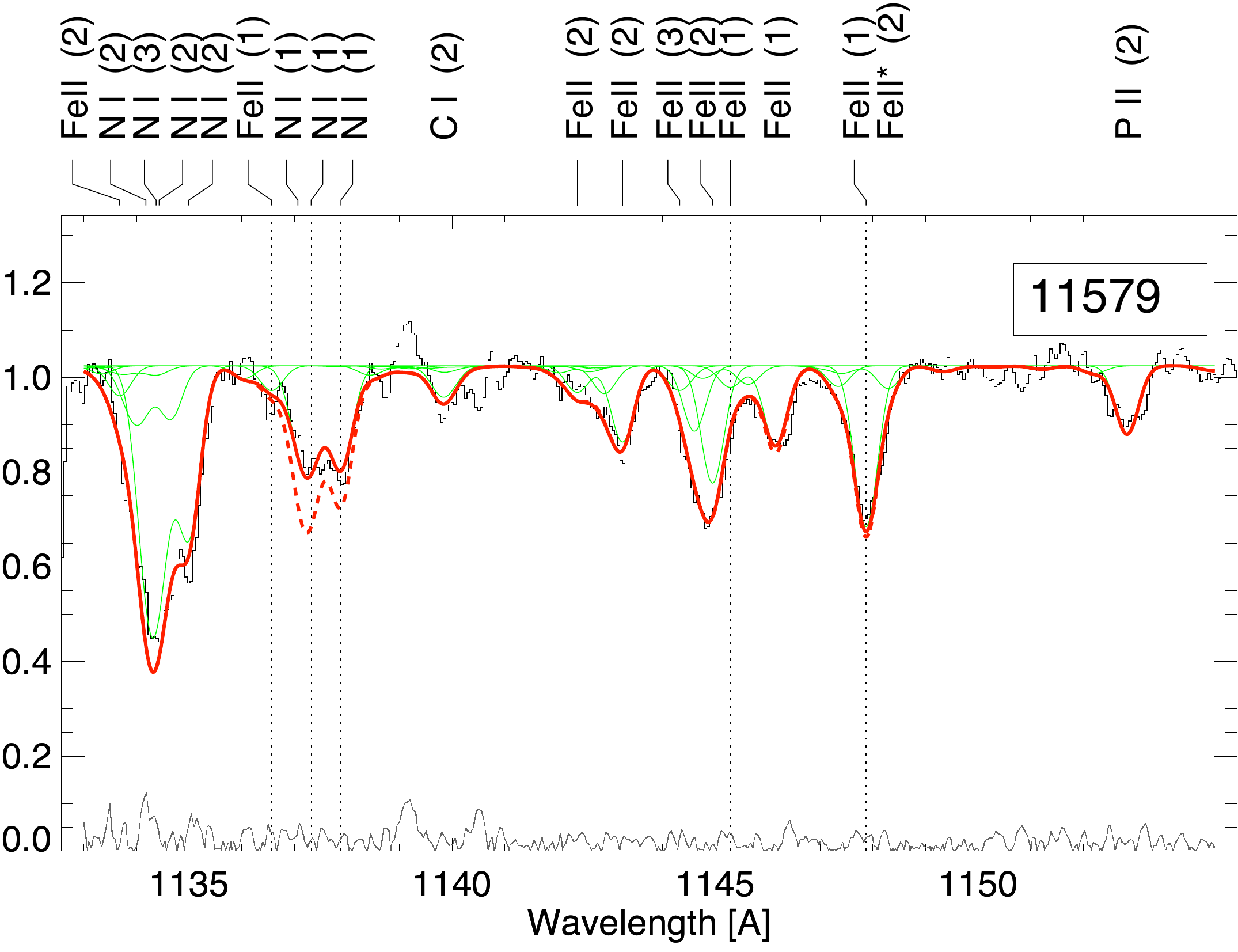}
\includegraphics[angle=90,width=9.3cm,height=7cm,clip=true]{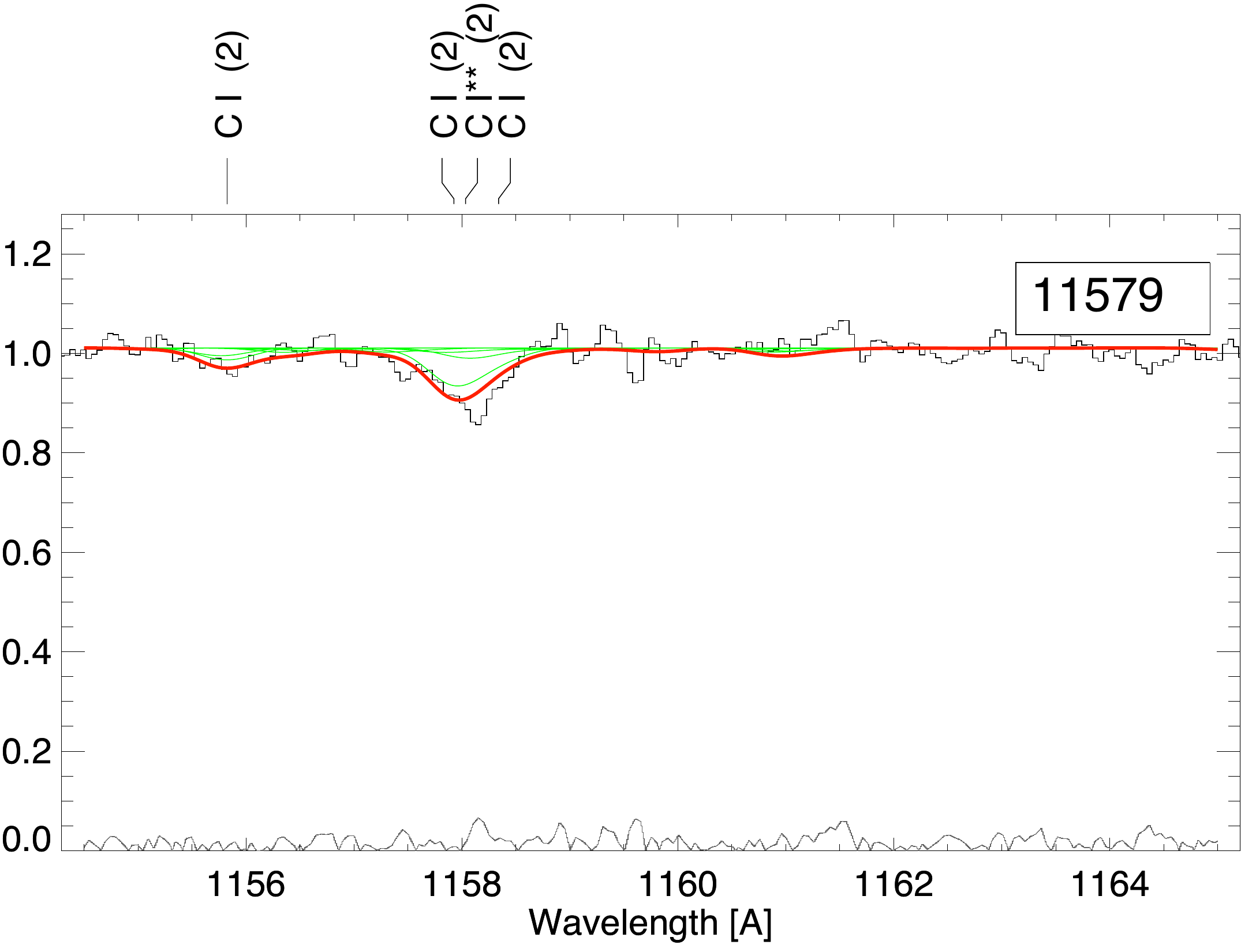}\\

\caption{Line profile fitting for the 11523 (top), 12028 (middle), and 11579 datasets (bottom). For display purposes, only the strongest lines from I\,Zw\,18 (component 1, also indicated by the vertical dotted lines), from the Milky Way (component 2), and from the high-velocity cloud (component 3) are labelled. The thin green line shows the individual line profiles. The solid thick red line shows the fit to the data while the dotted red line shows the line profiles assuming the H\1\ region abundances are equal to the H\2\ region abundances. }
\label{fig:fits1}
\end{figure*}

\begin{figure*}

\includegraphics[angle=90,width=9.3cm,height=7cm,clip=true]{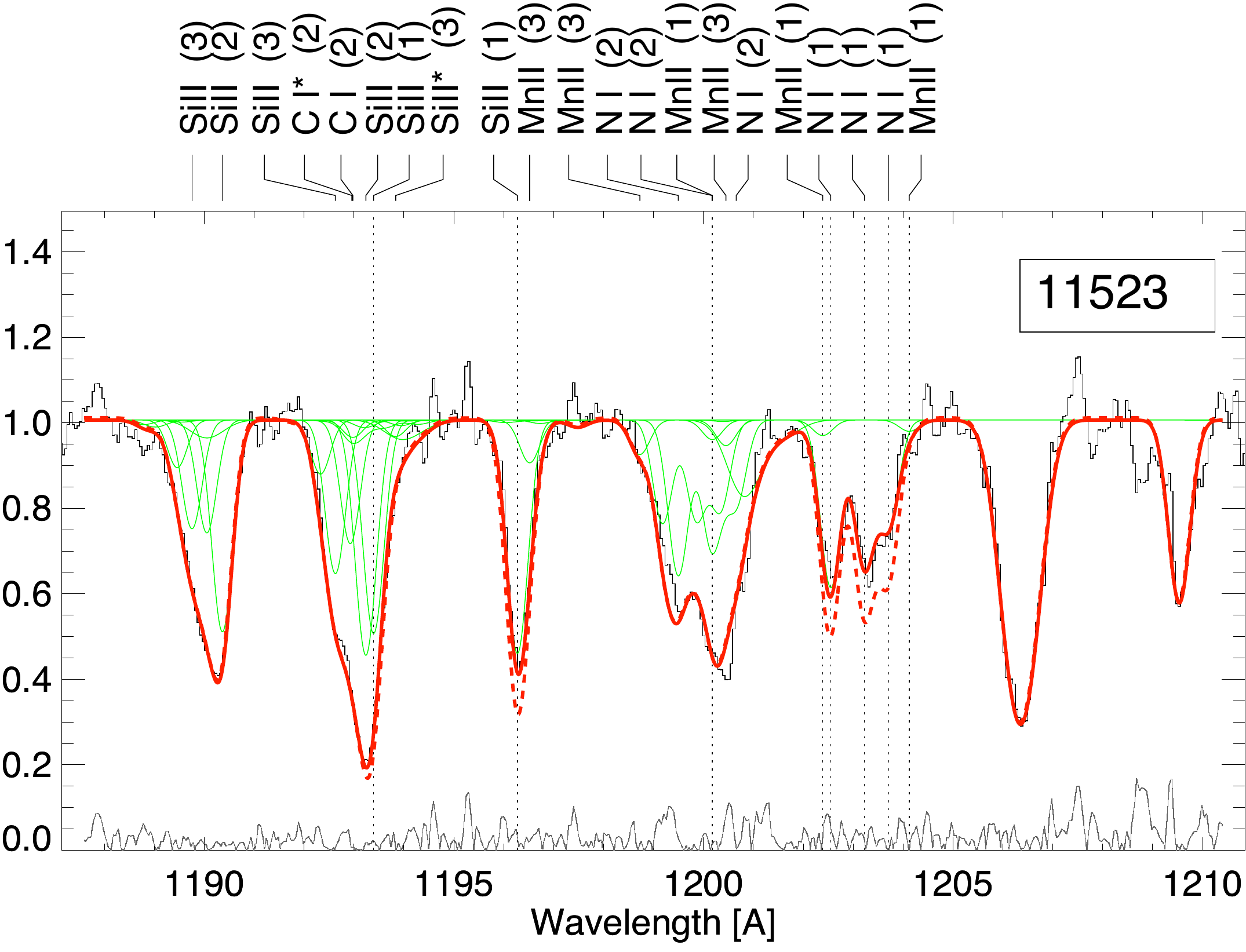}
\includegraphics[angle=90,width=9.3cm,height=7cm,clip=true]{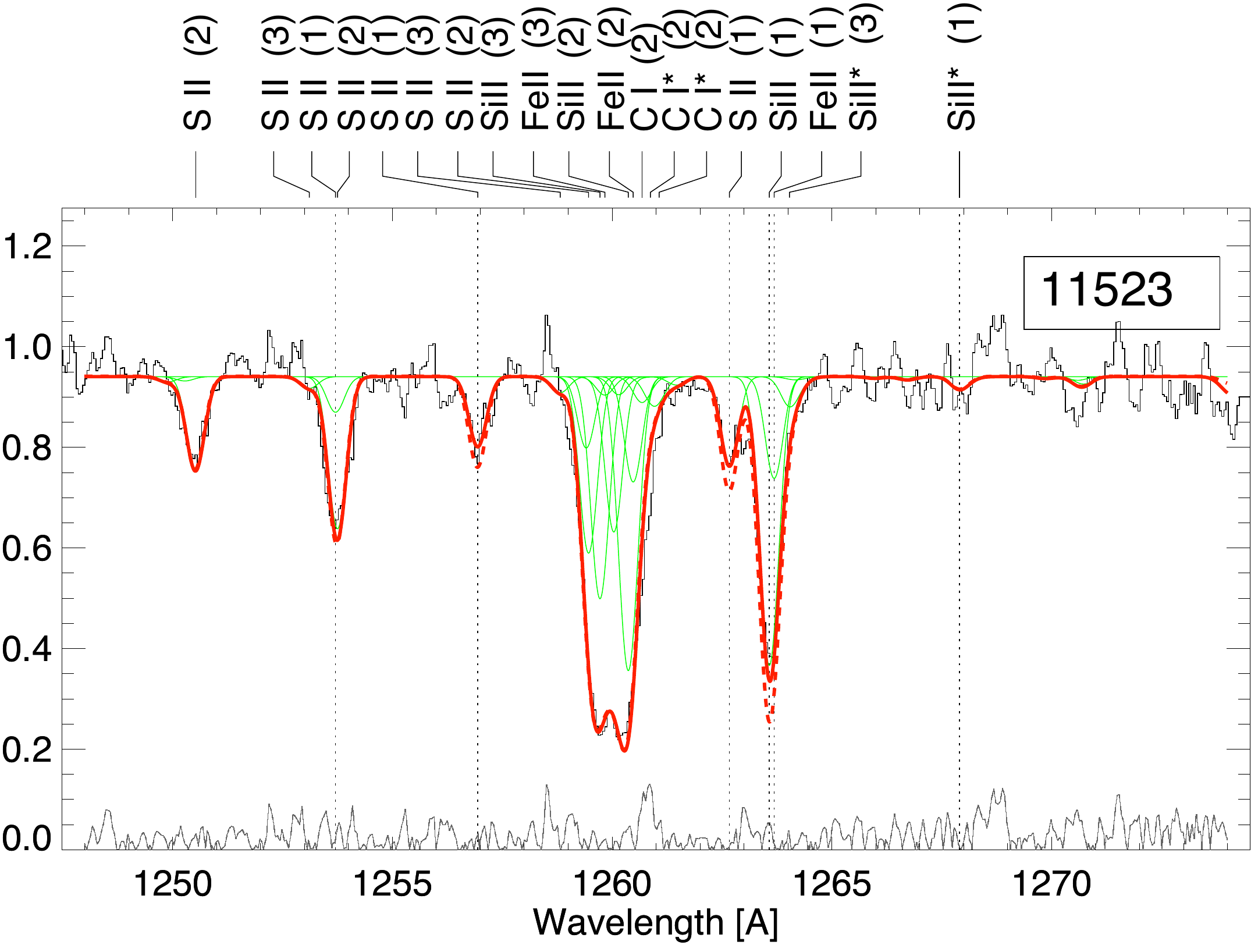}\\

\includegraphics[angle=90,width=9.3cm,height=7cm,clip=true]{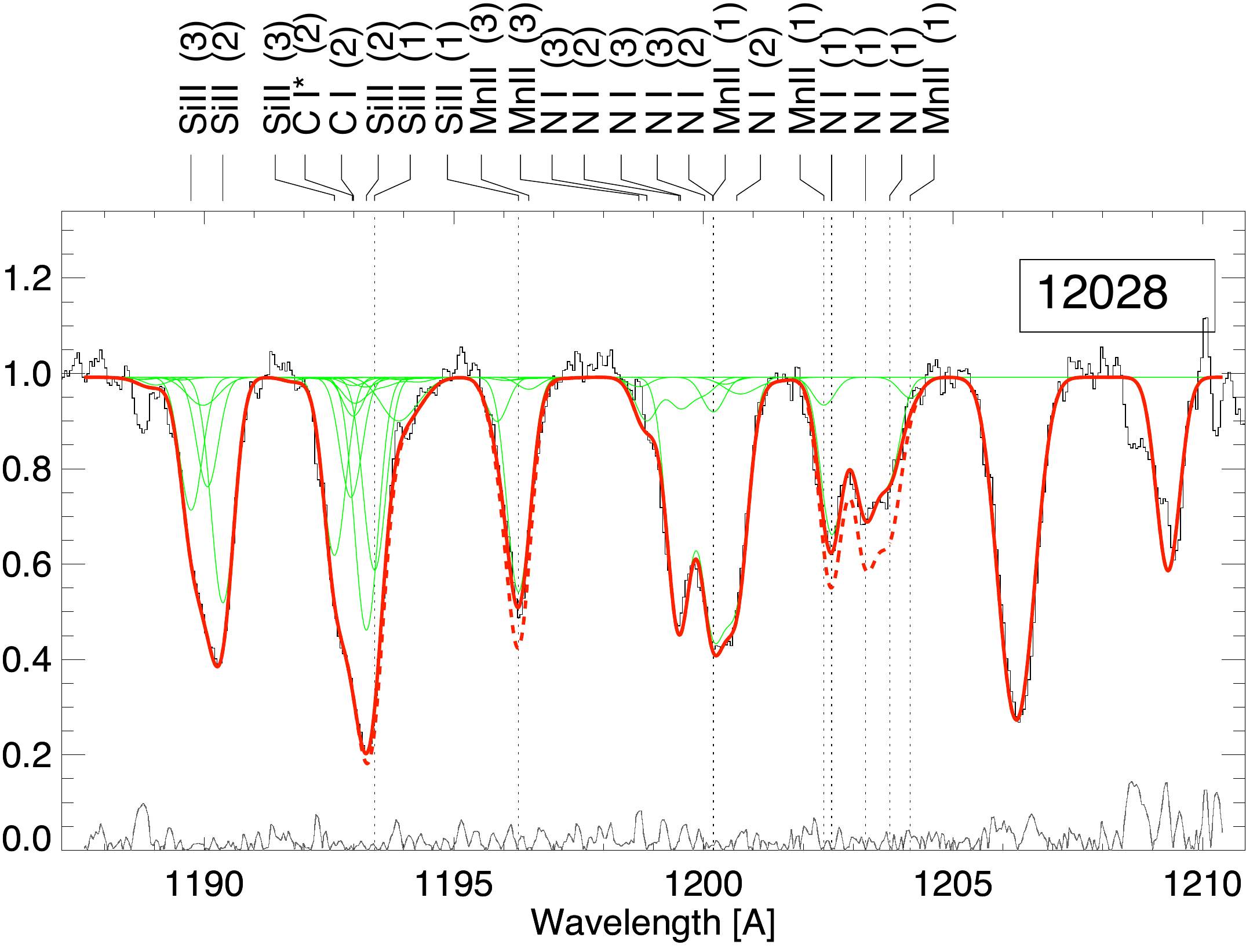}
\includegraphics[angle=90,width=9.3cm,height=7cm,clip=true]{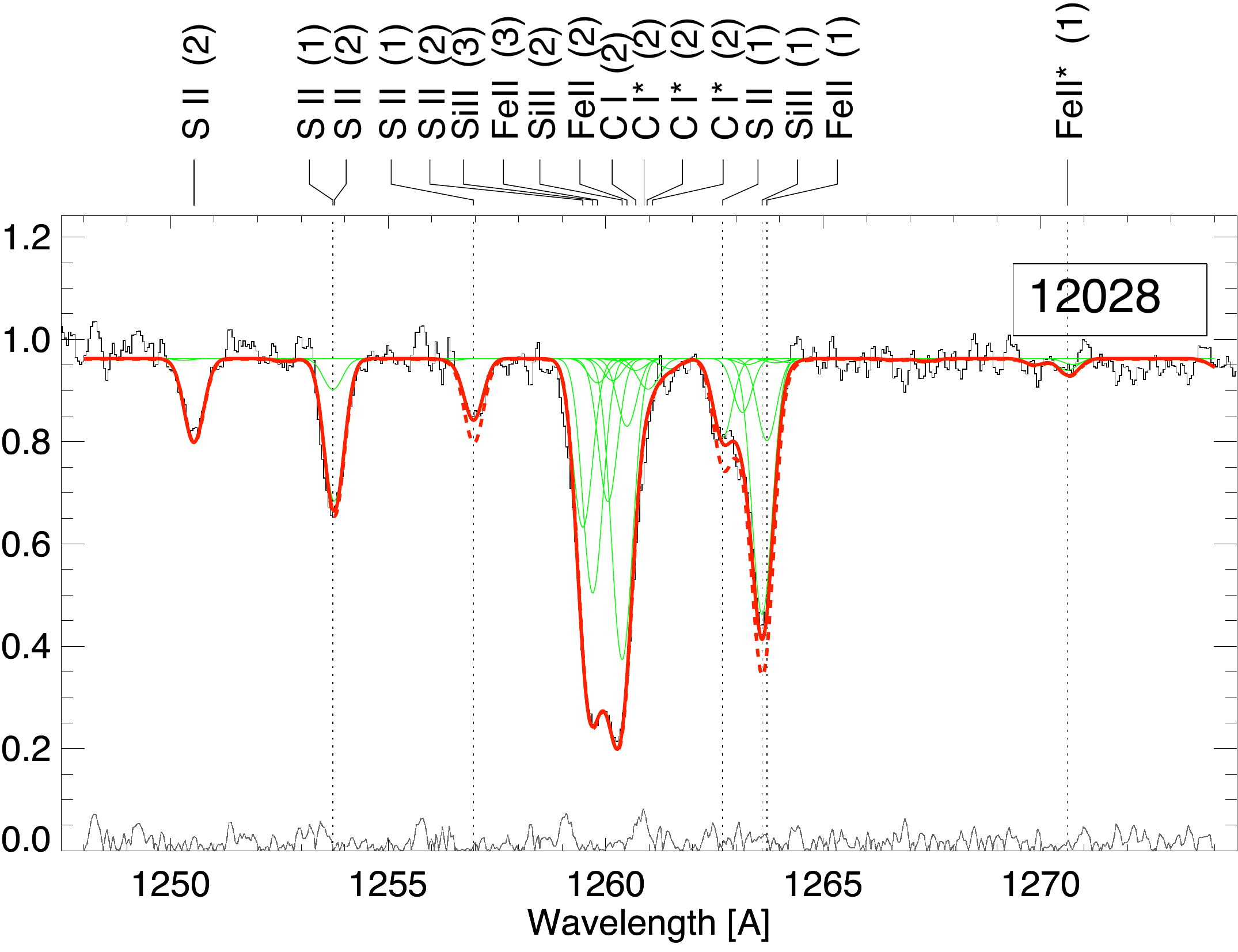}\\

\includegraphics[angle=90,width=9.3cm,height=7cm,clip=true]{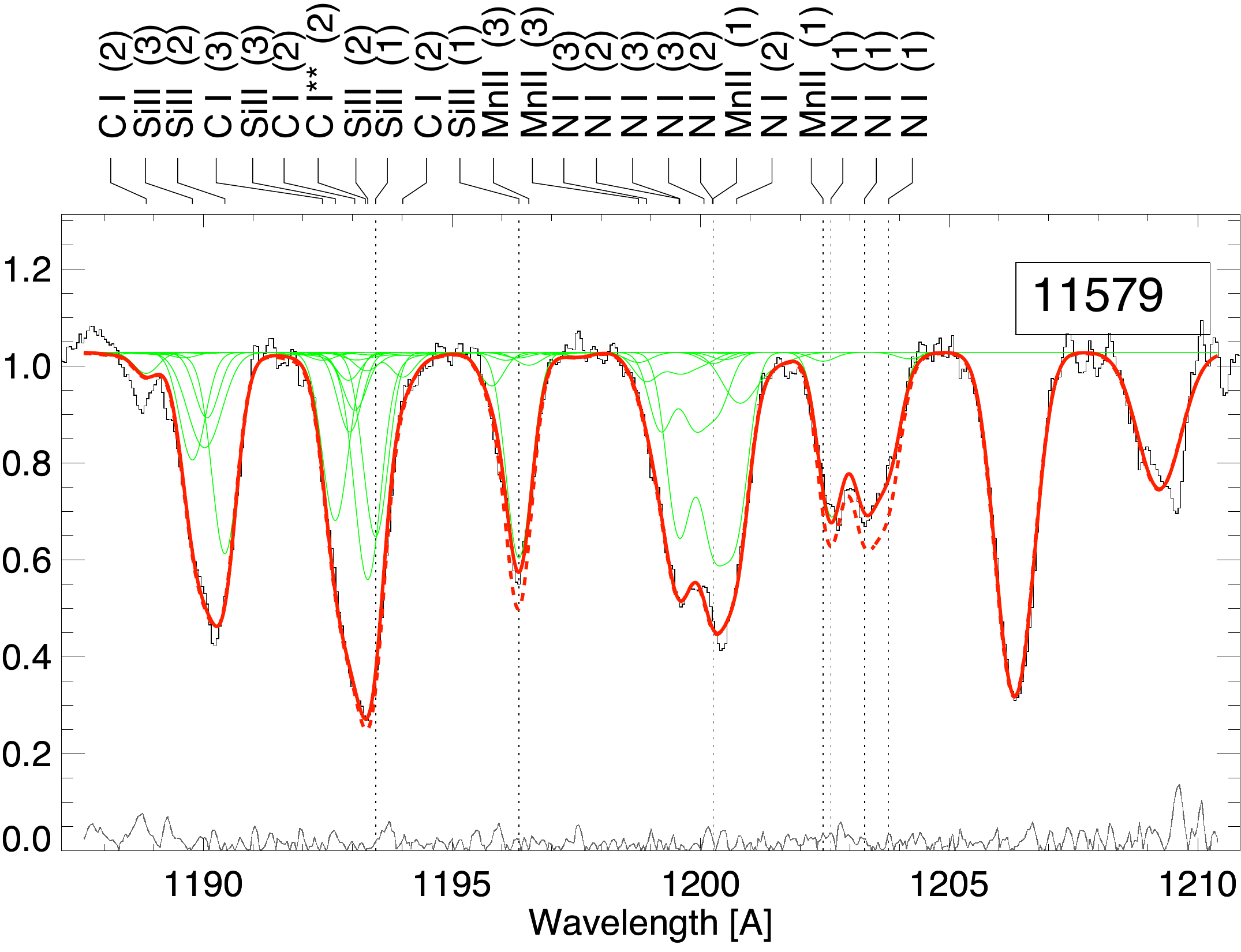}
\includegraphics[angle=90,width=9.3cm,height=7cm,clip=true]{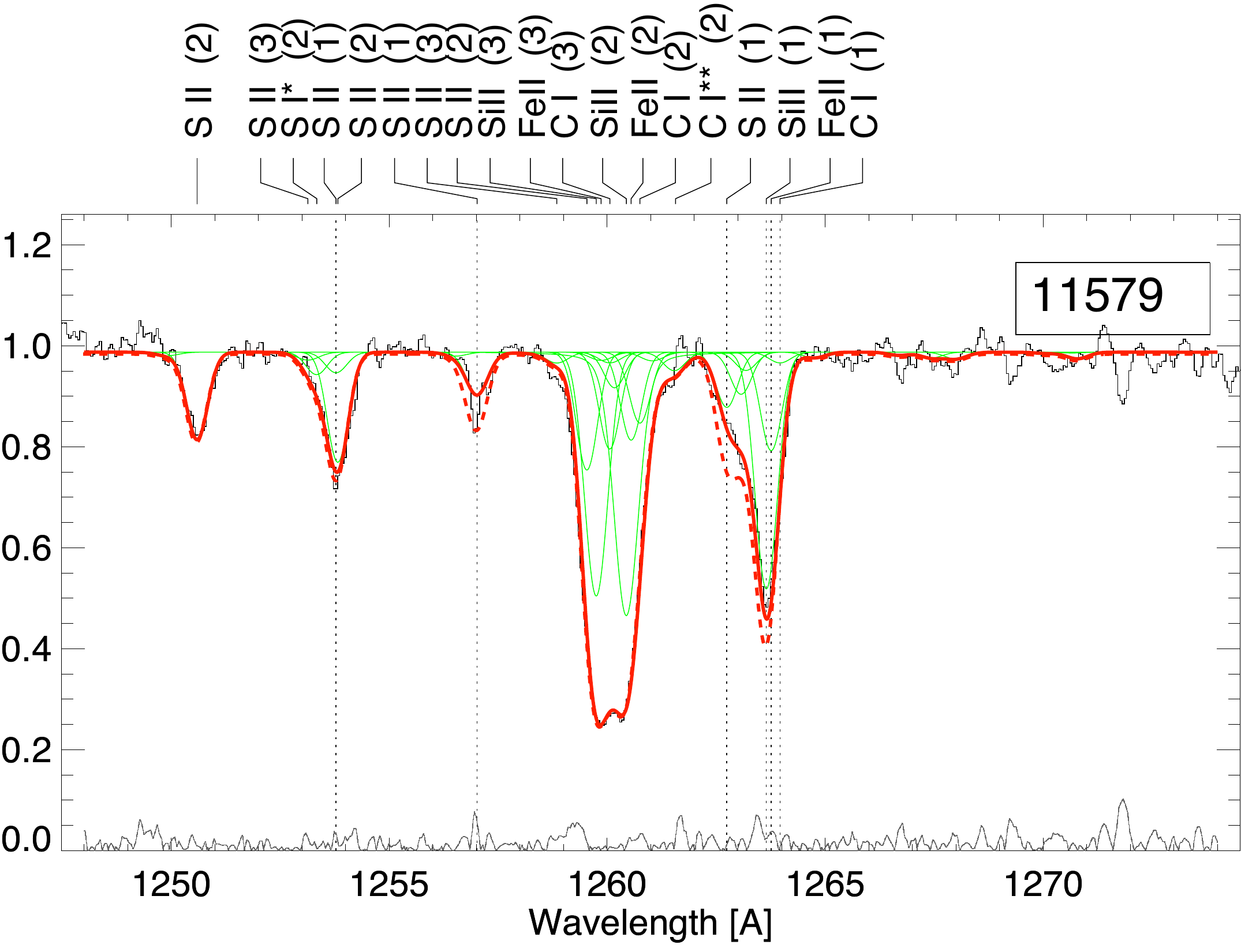}\\

\caption{See Fig.\,\ref{fig:fits1} for the plot description. The unlabelled line on the far right is Si\3\ $\lambda1206.5$  arising in I\,Zw\,18. }
\label{fig:fits2}
\end{figure*}

\begin{figure*}

\hspace{9.3cm}
\includegraphics[angle=90,width=9.3cm,height=7cm,clip=true]{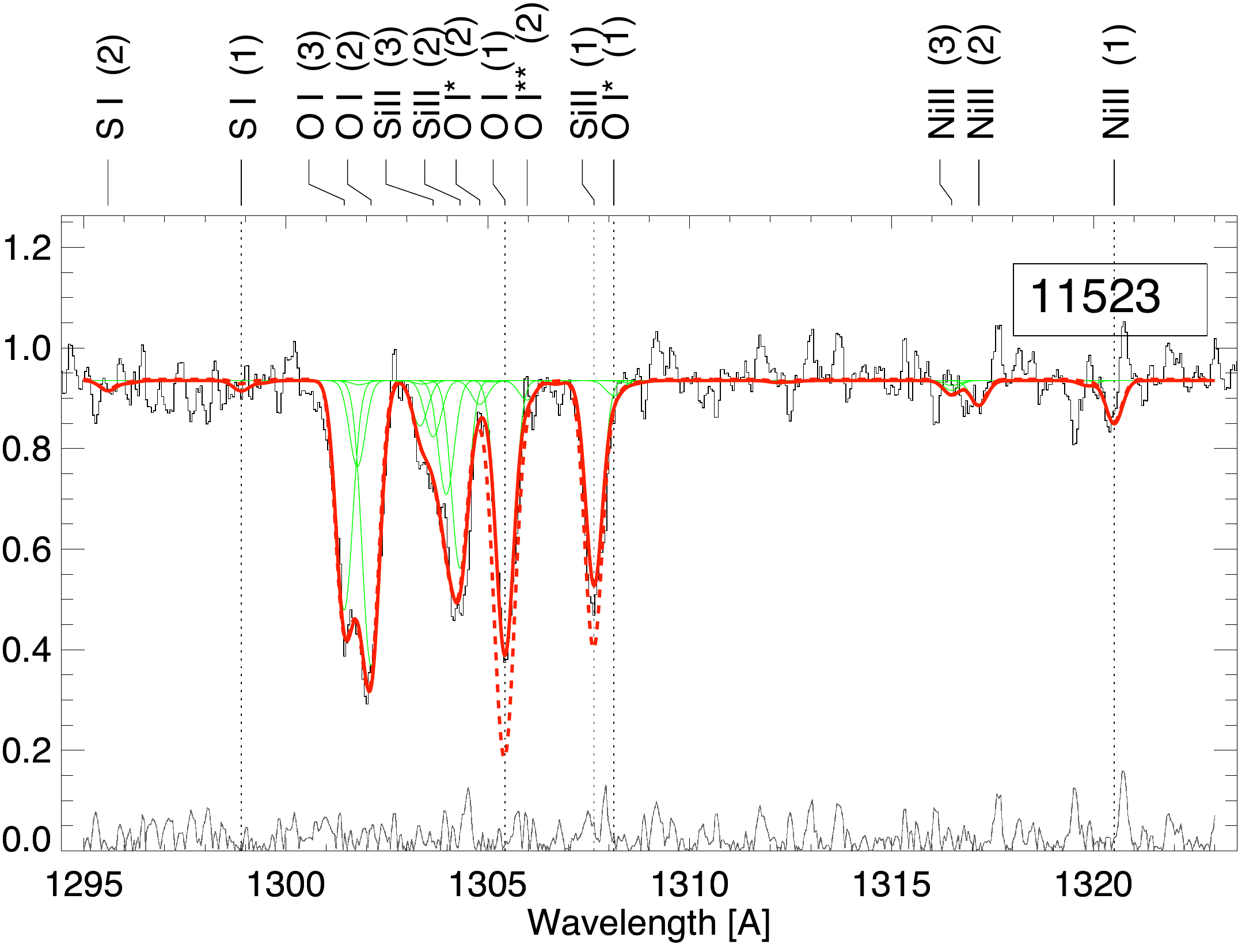}\\

\includegraphics[angle=90,width=9.3cm,height=7cm,clip=true]{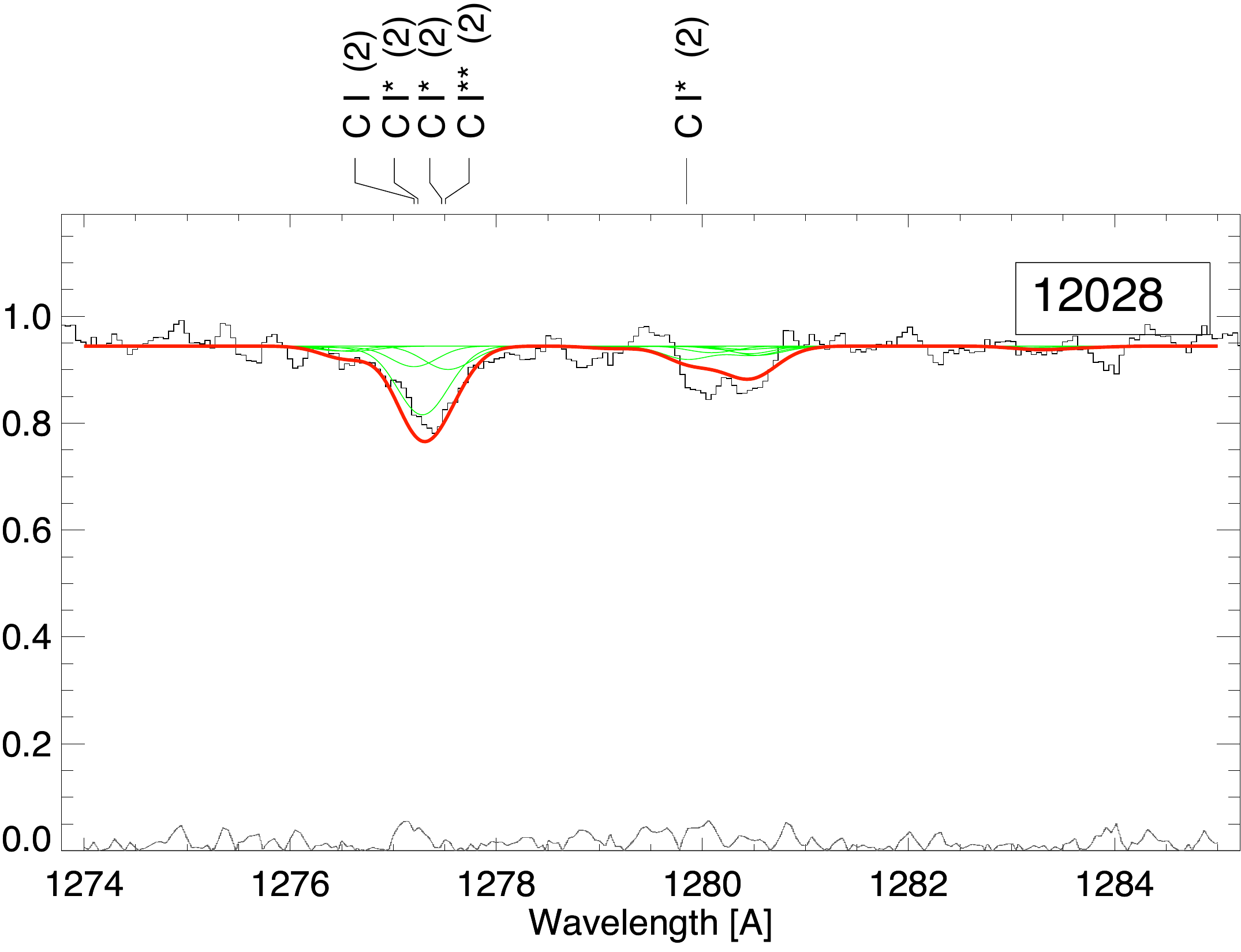}
\hspace{9.3cm}\\

\hspace{9.3cm}
\includegraphics[angle=90,width=9.3cm,height=7cm,clip=true]{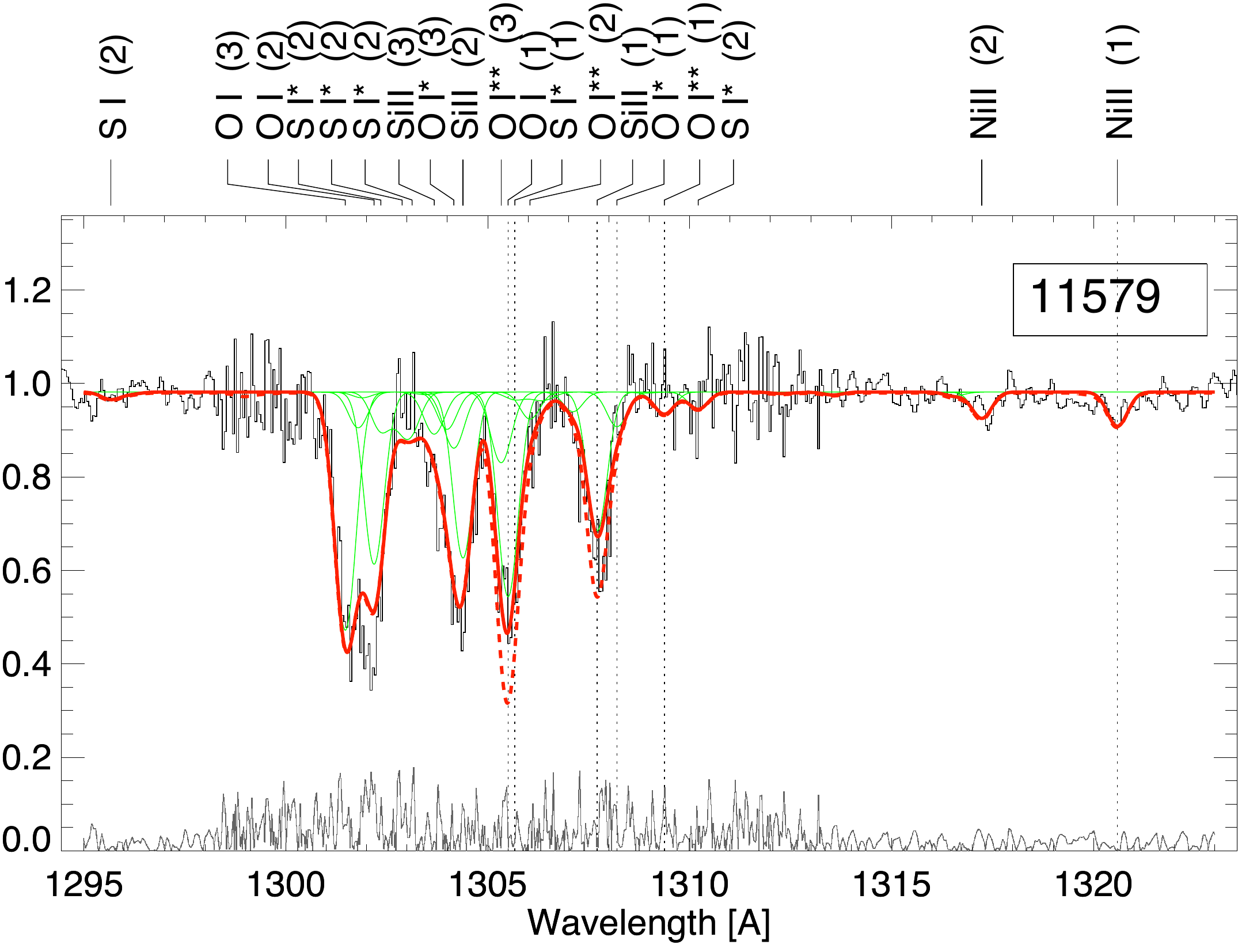}\\

\caption{See Fig.\,\ref{fig:fits1} for the plot description. }
\label{fig:fits3}
\end{figure*}

\begin{figure*}

\hspace{9.3cm}
\includegraphics[angle=90,width=9.3cm,height=7cm,clip=true]{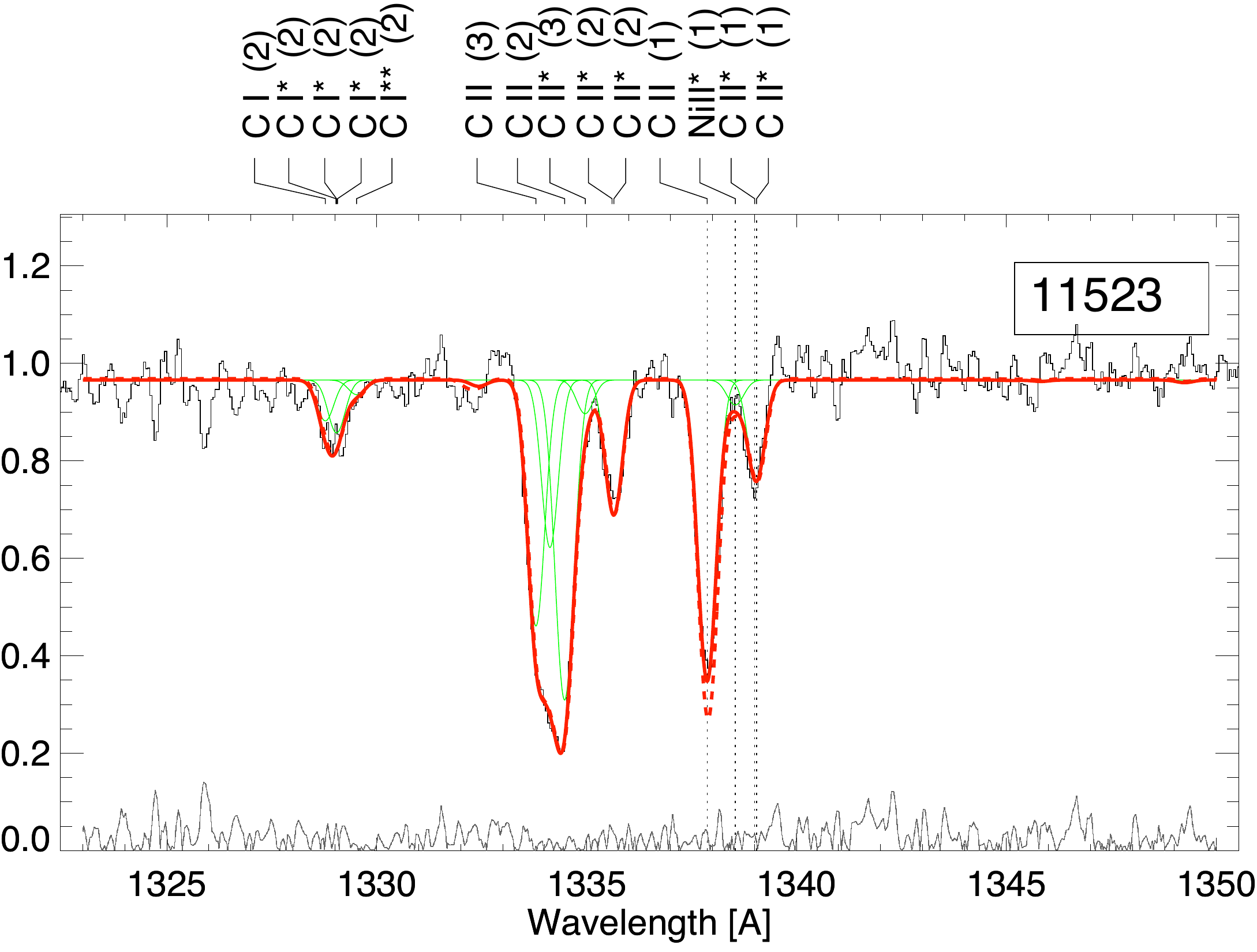}\\

\includegraphics[angle=90,width=9.3cm,height=7cm,clip=true]{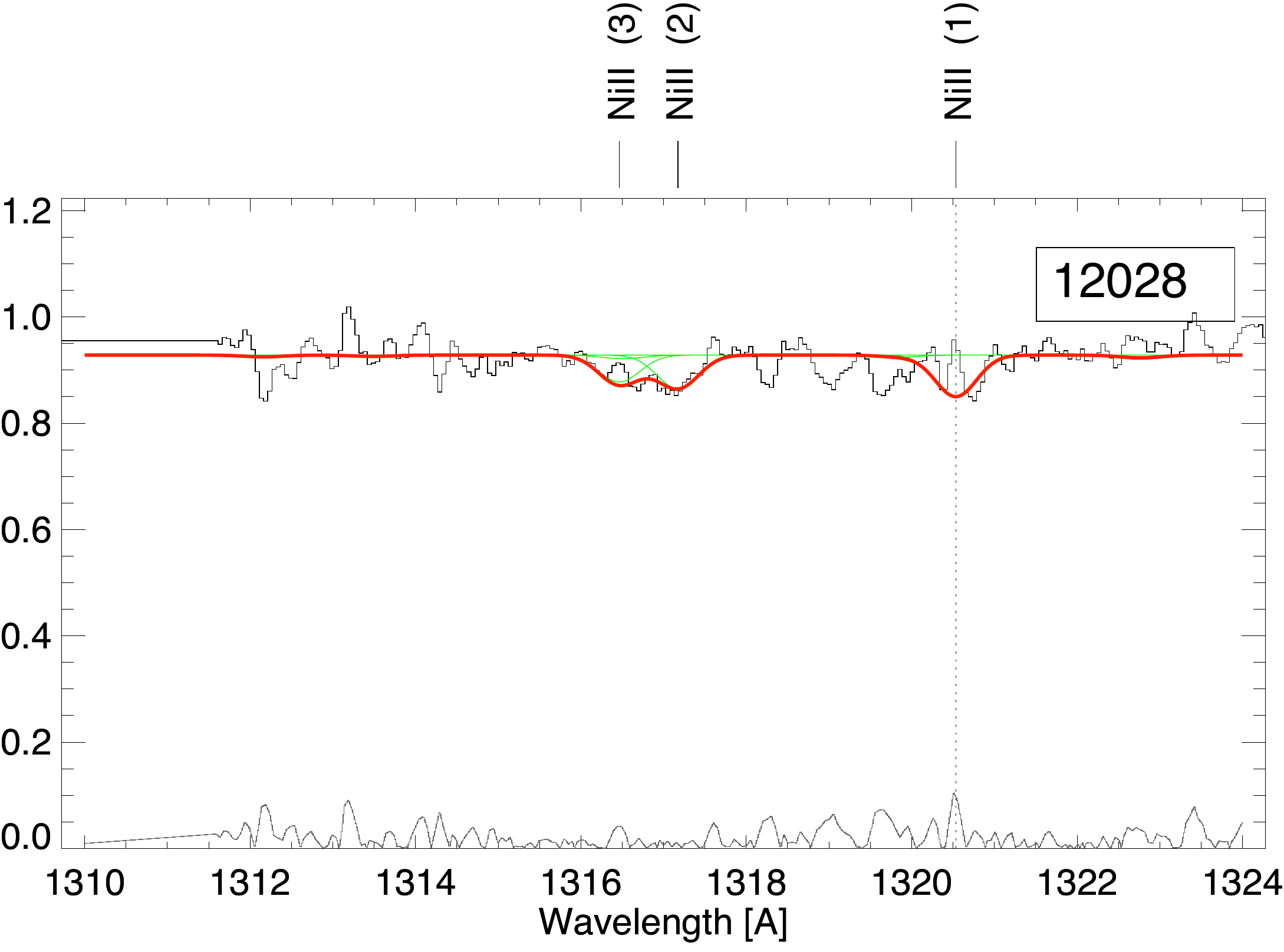}
\includegraphics[angle=90,width=9.3cm,height=7cm,clip=true]{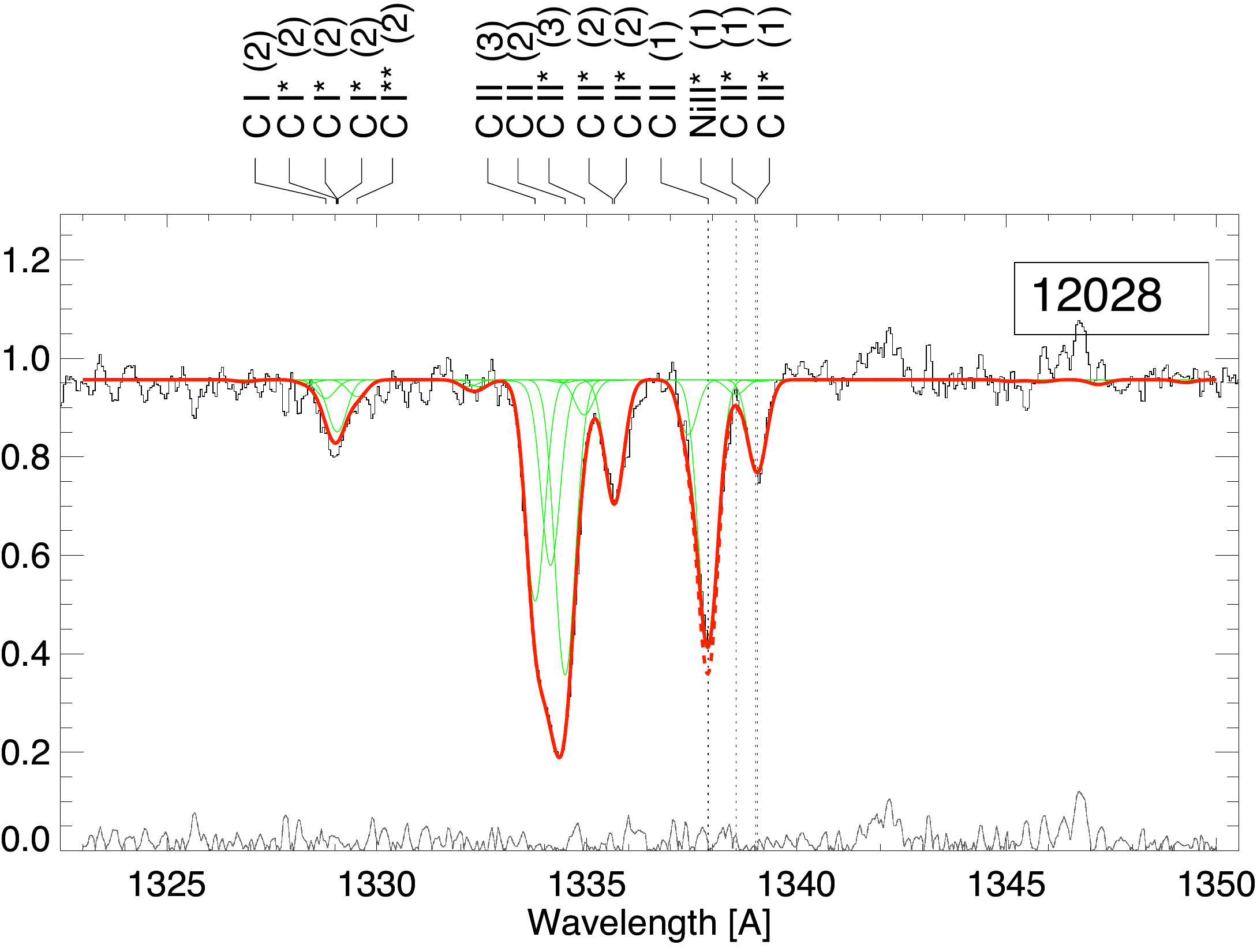}\\

\hspace{9.3cm}
\includegraphics[angle=90,width=9.3cm,height=7cm,clip=true]{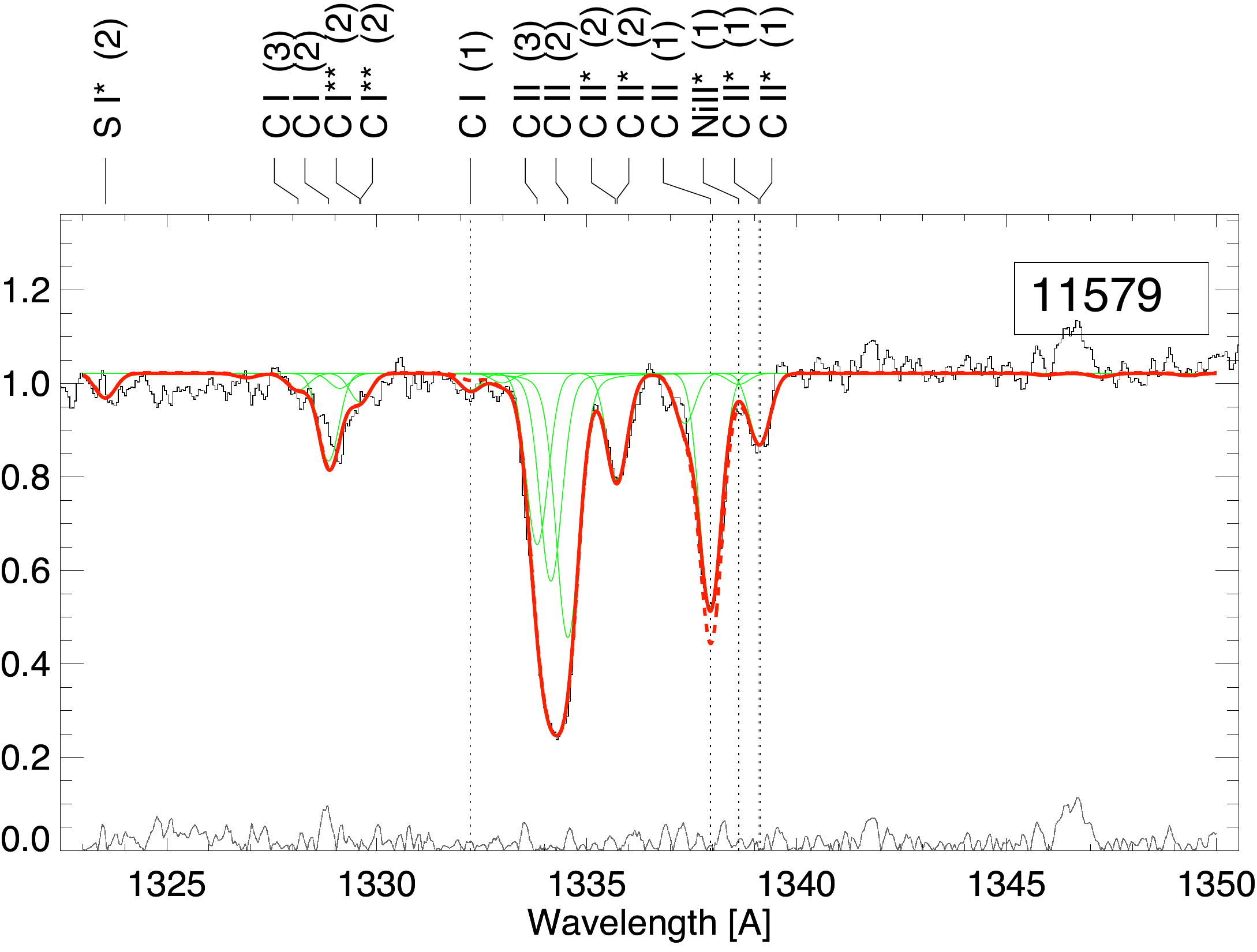}\\

\caption{See Fig.\,\ref{fig:fits1} for the plot description. }
\label{fig:fits4}
\end{figure*}

\begin{figure*}

\includegraphics[angle=90,width=9.3cm,height=7cm,clip=true]{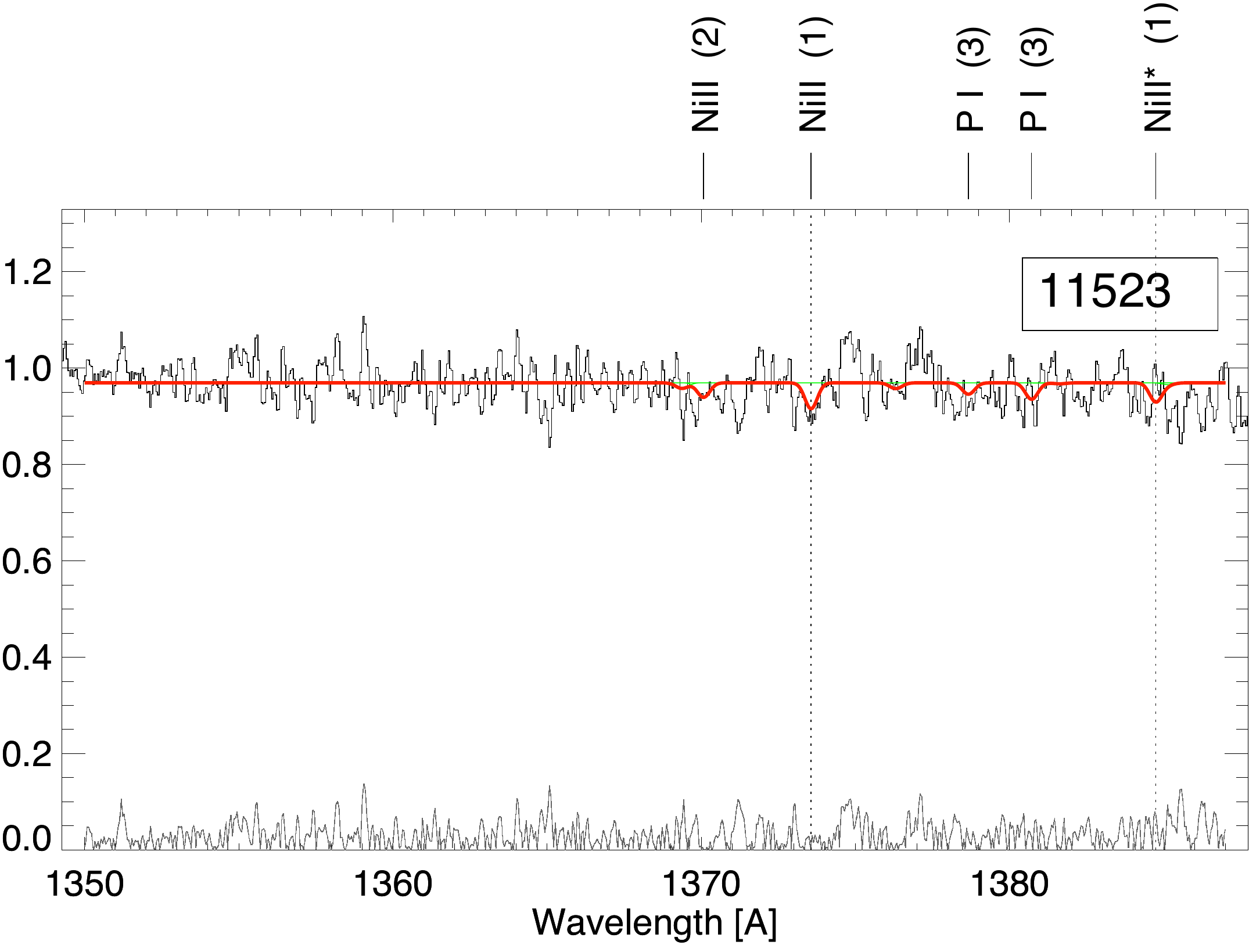}
\includegraphics[angle=90,width=9.3cm,height=6.8cm,clip=true]{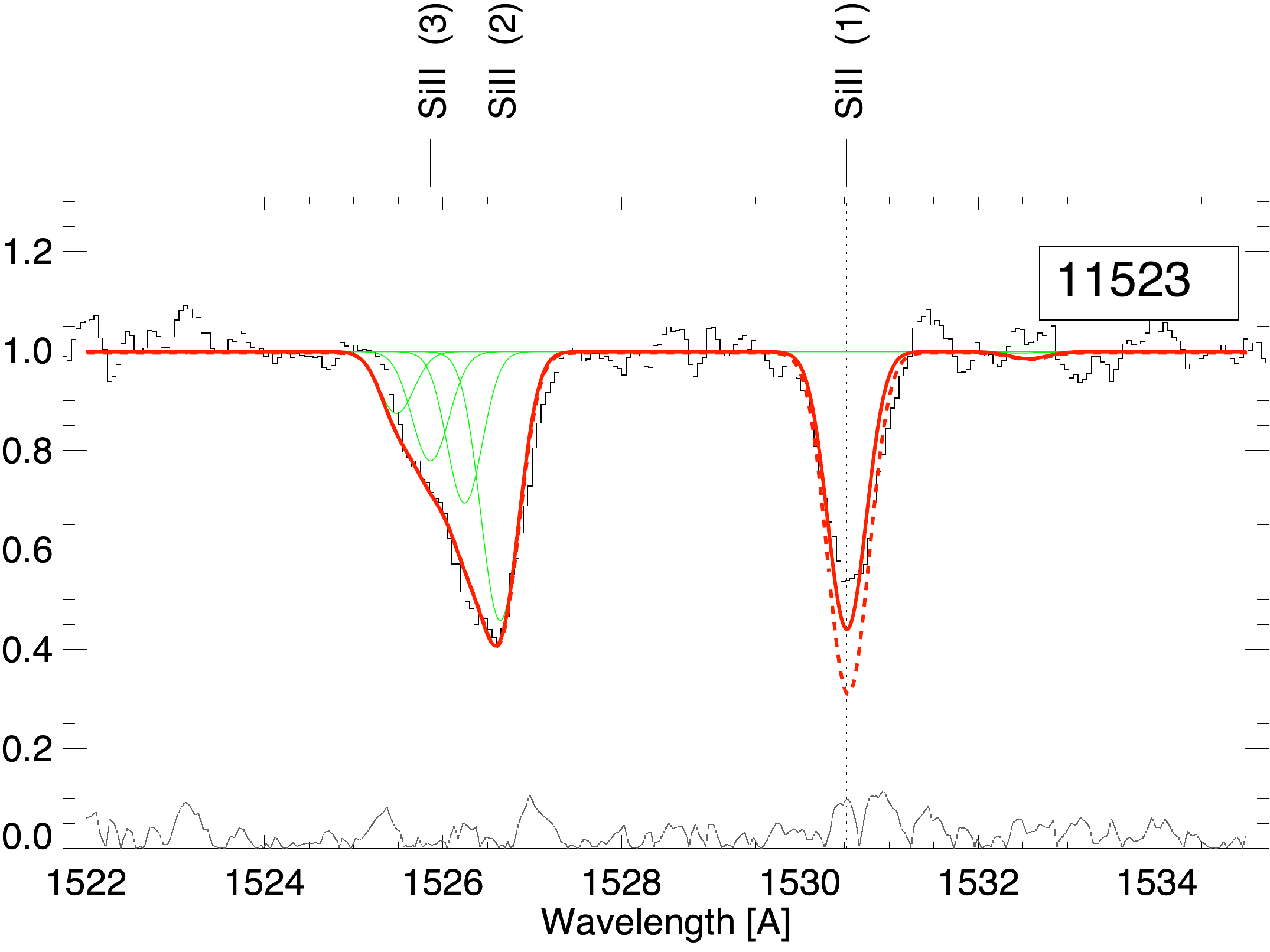}\\

\includegraphics[angle=90,width=9.3cm,height=7cm,clip=true]{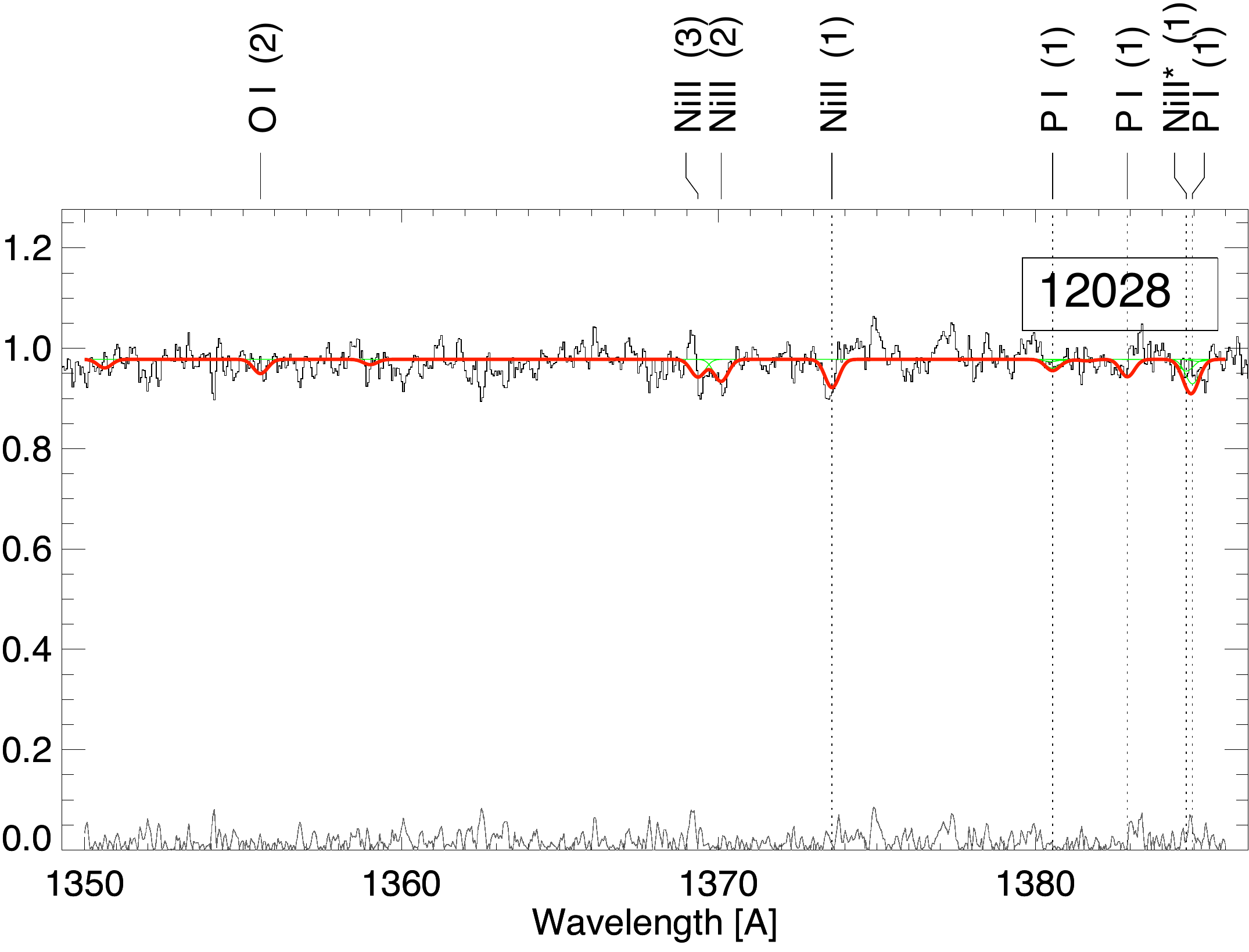}
\includegraphics[angle=90,width=9.3cm,height=7cm,clip=true]{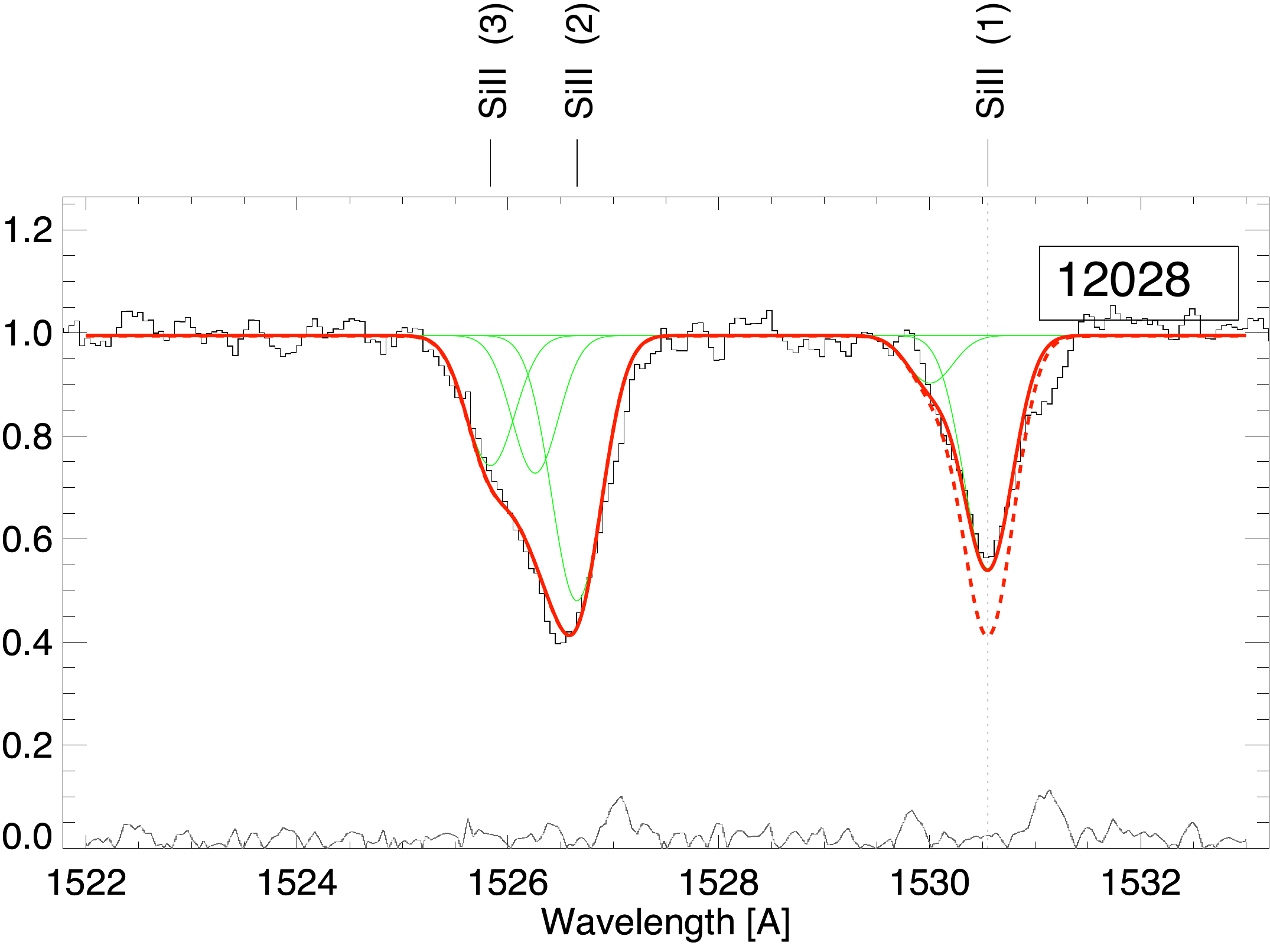}\\

\includegraphics[angle=90,width=9.3cm,height=7cm,clip=true]{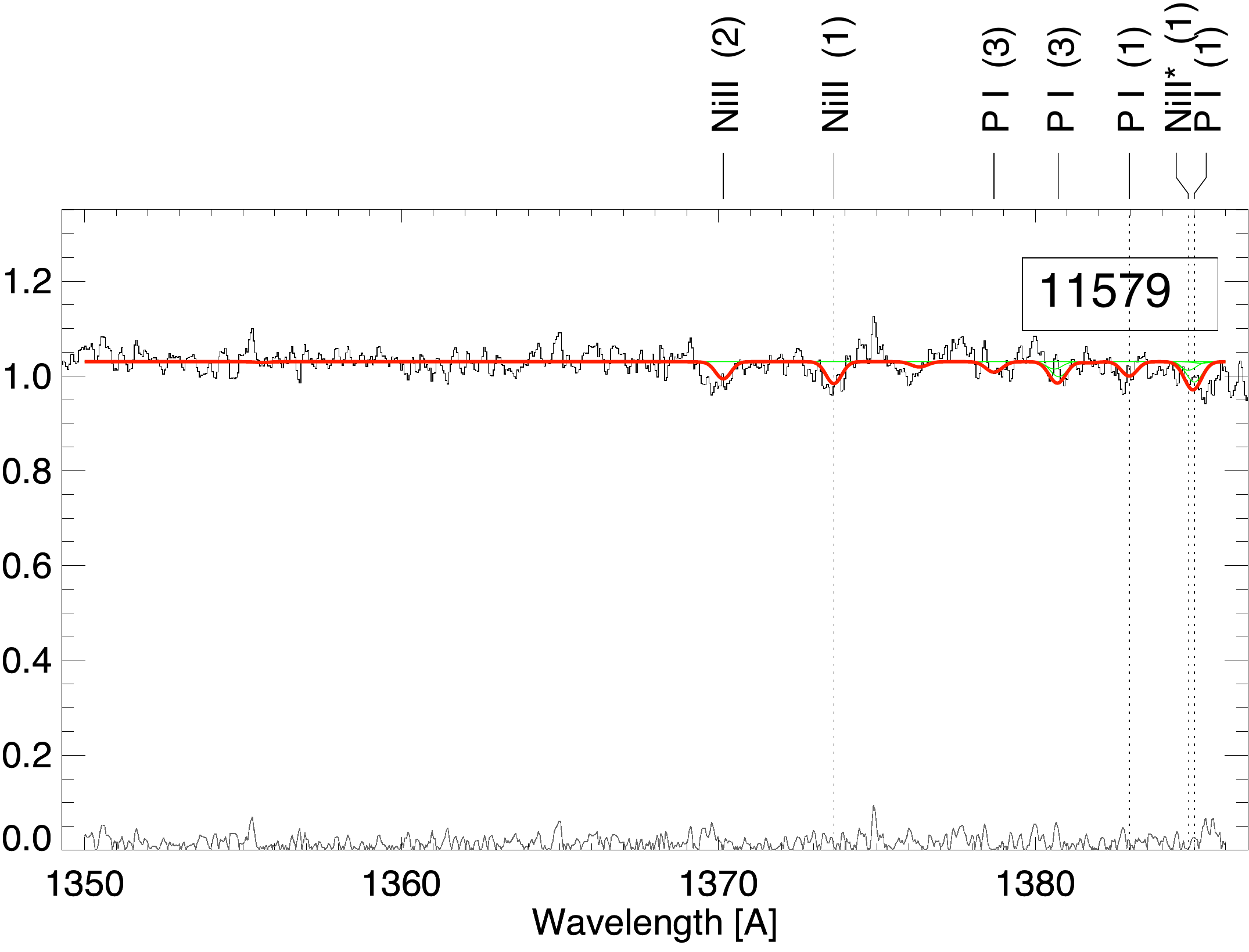}
\hspace{9.3cm}\\

\caption{See Fig.\,\ref{fig:fits1} for the plot description. }
\label{fig:fits5}
\end{figure*}

\begin{figure*}

\includegraphics[angle=90,width=9.3cm,height=7cm,clip=true]{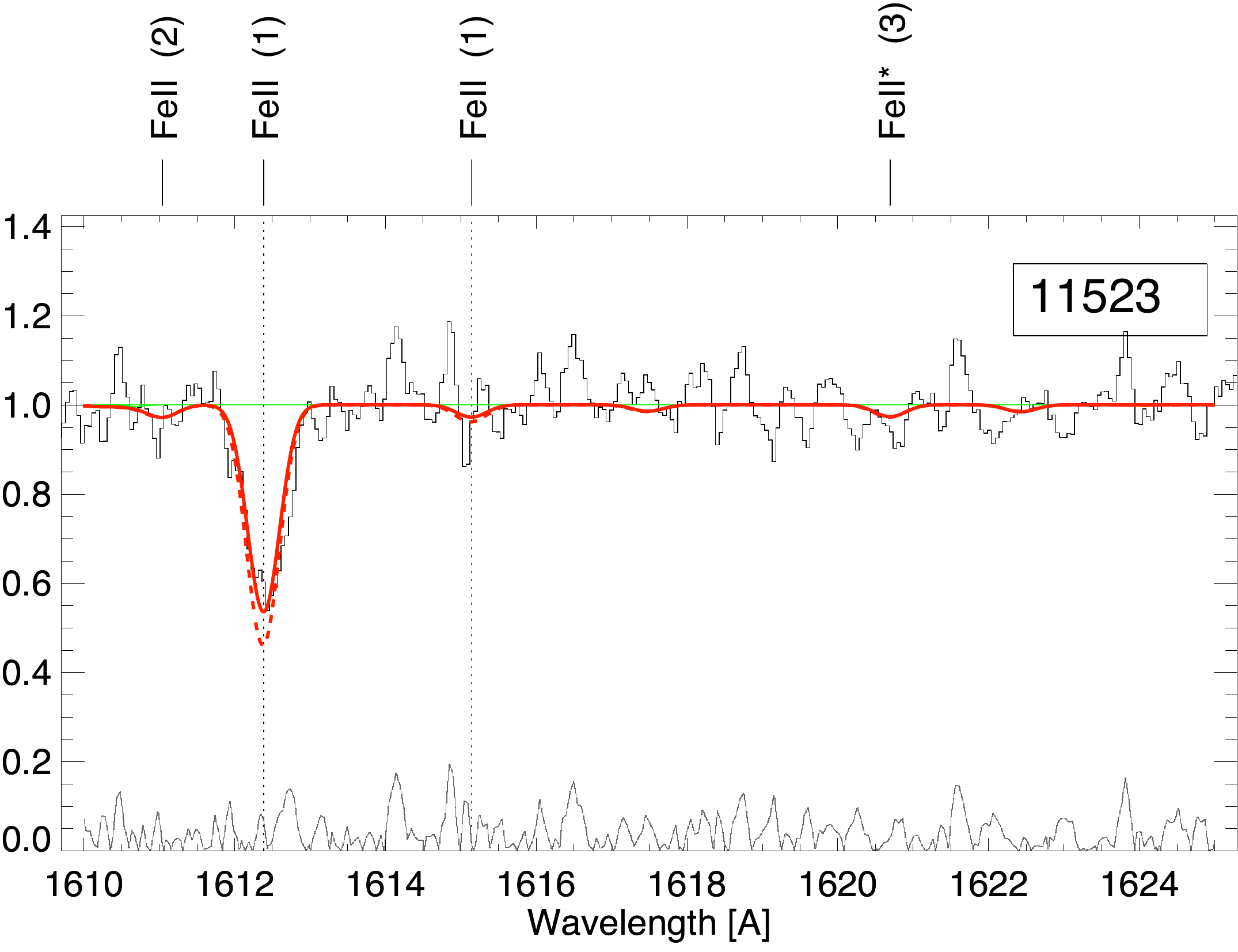}
\includegraphics[angle=90,width=9.3cm,height=7cm,clip=true]{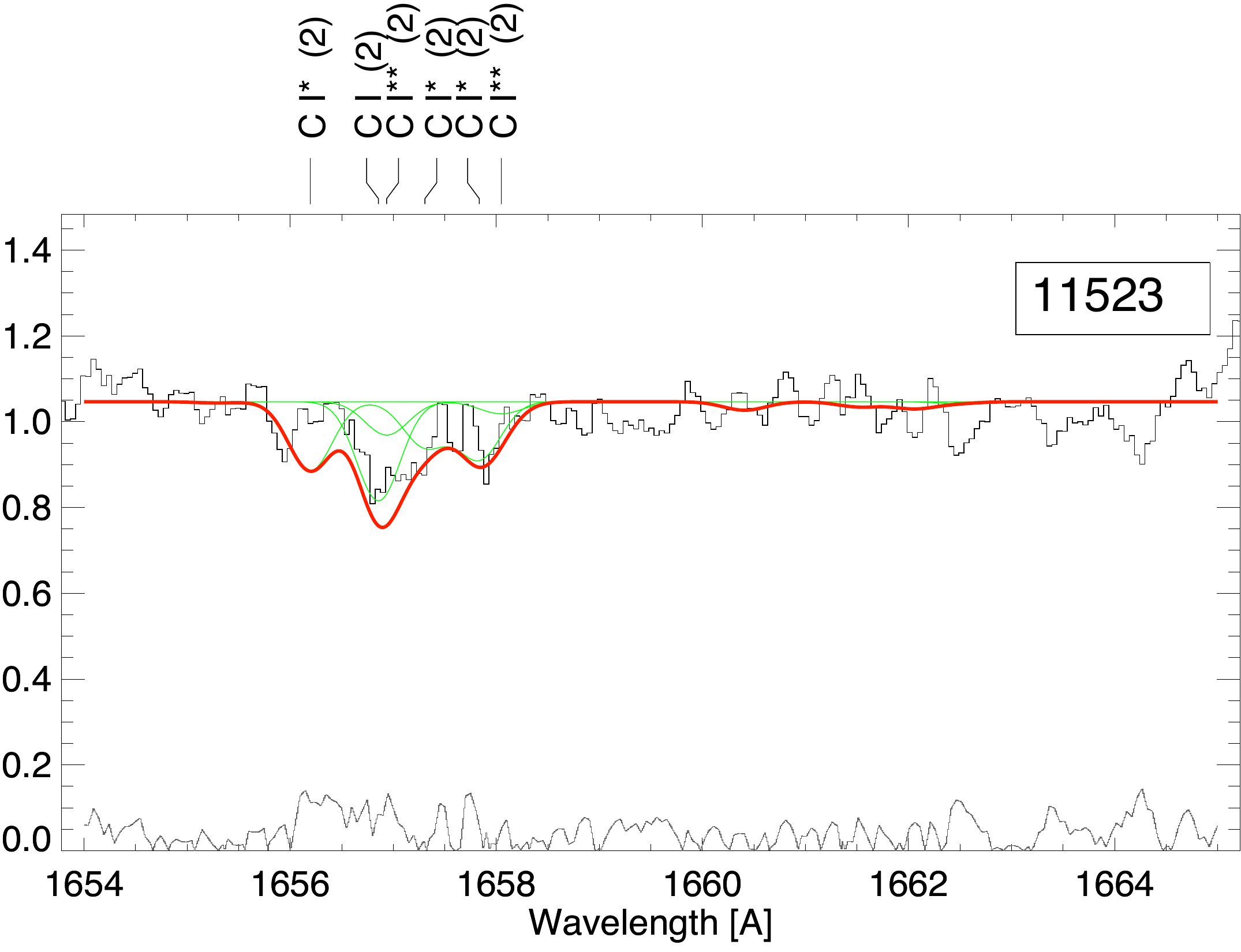}\\

\includegraphics[angle=90,width=9.3cm,height=7cm,clip=true]{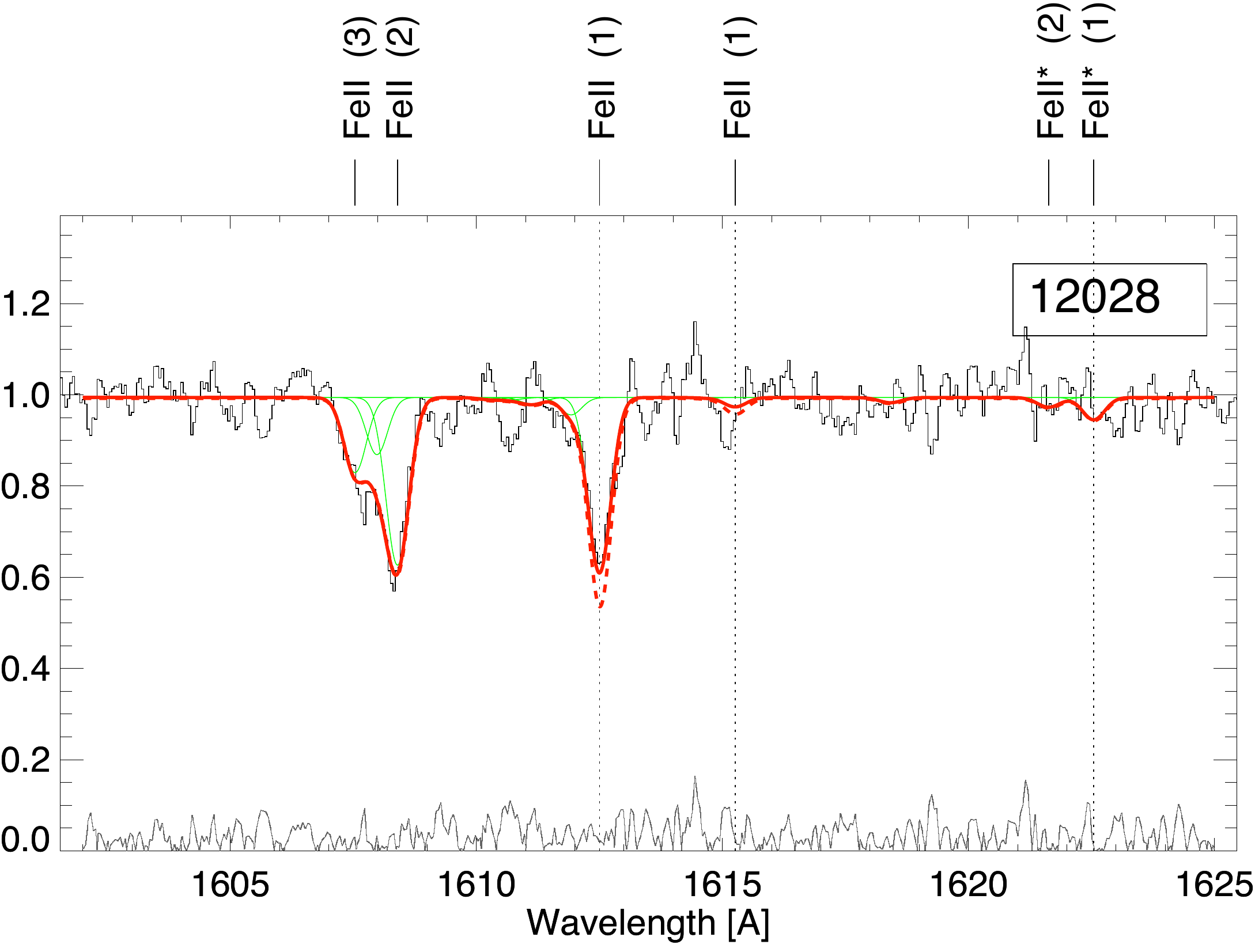}
\includegraphics[angle=90,width=9.3cm,height=7cm,clip=true]{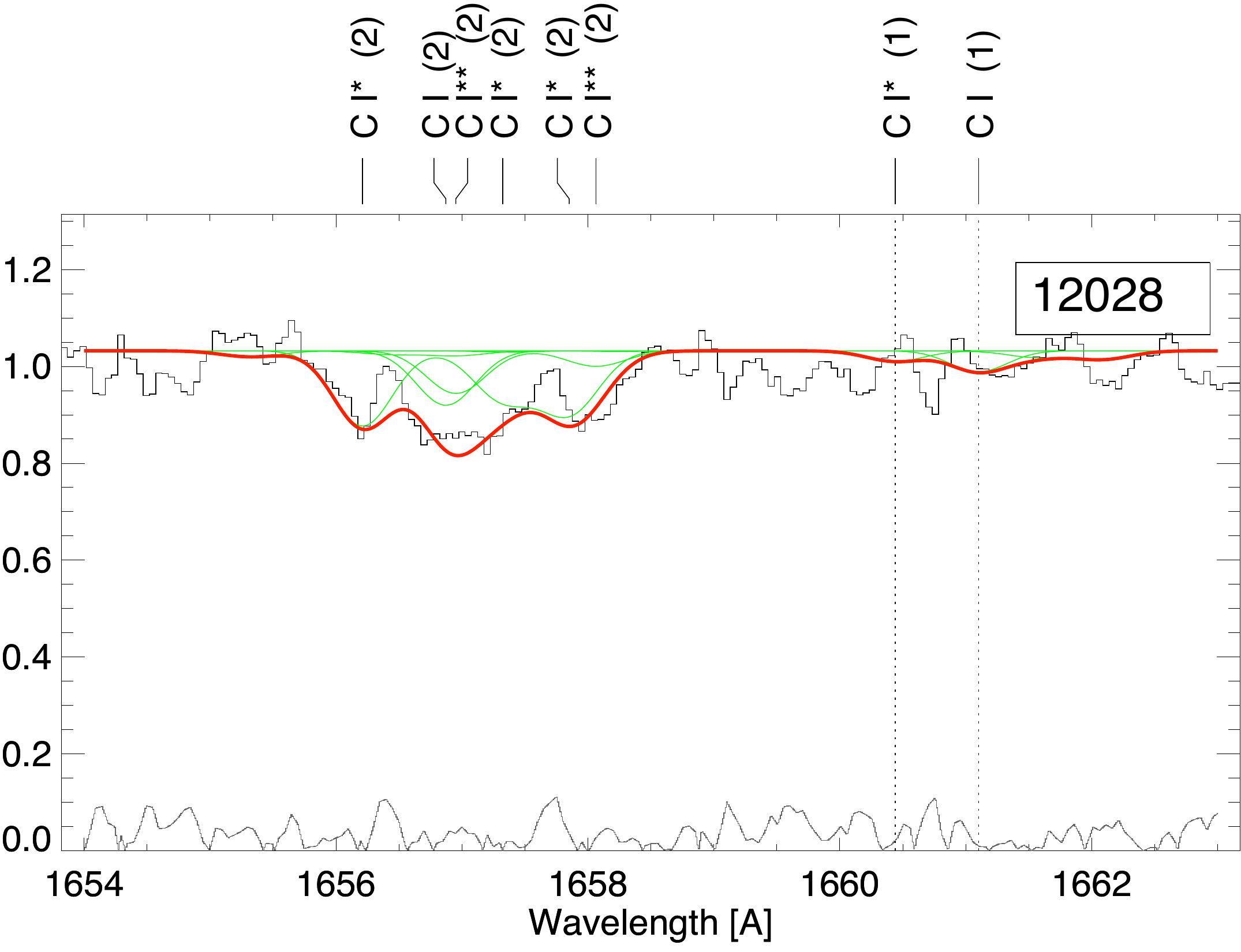}\\

\hspace{9.3cm}
\hspace{9.3cm}\\

\caption{See Fig.\,\ref{fig:fits1} for the plot description. }
\label{fig:fits6}
\end{figure*}

\begin{figure*}

\includegraphics[angle=90,width=9.3cm,height=7cm,clip=true]{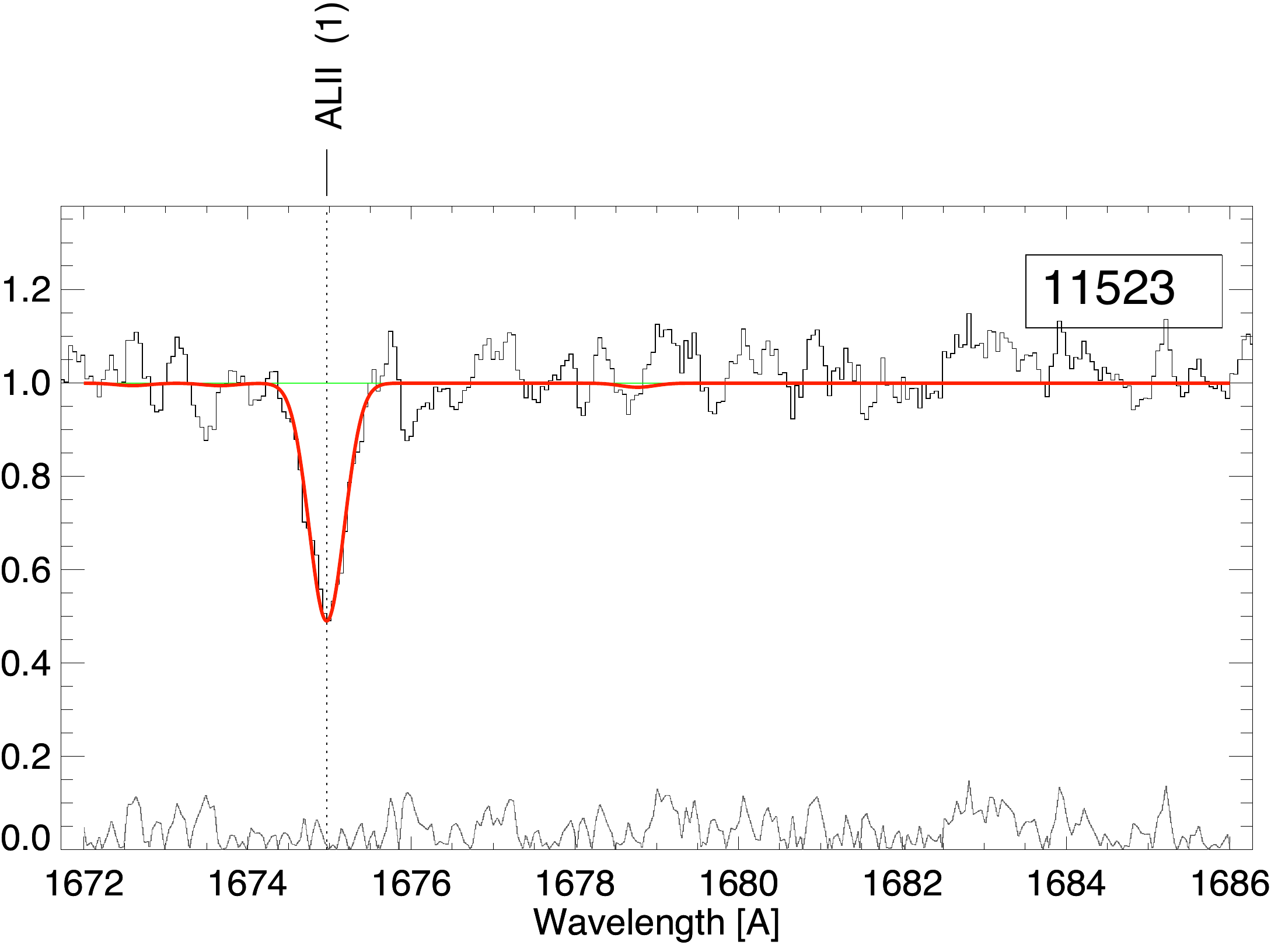}
\hspace{9.3cm}\\

\includegraphics[angle=90,width=9.3cm,height=7cm,clip=true]{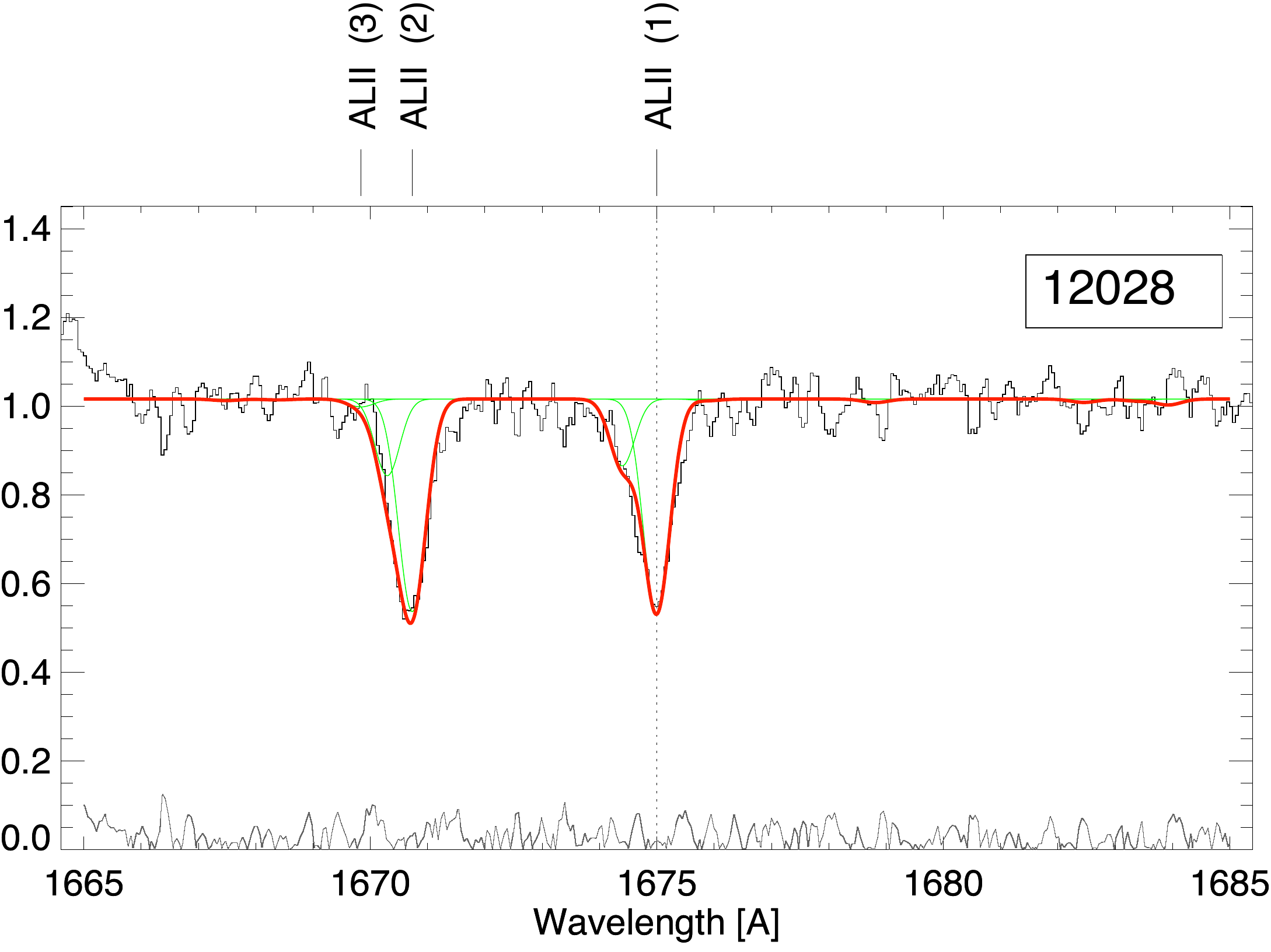}
\includegraphics[angle=90,width=9.3cm,height=7cm,clip=true]{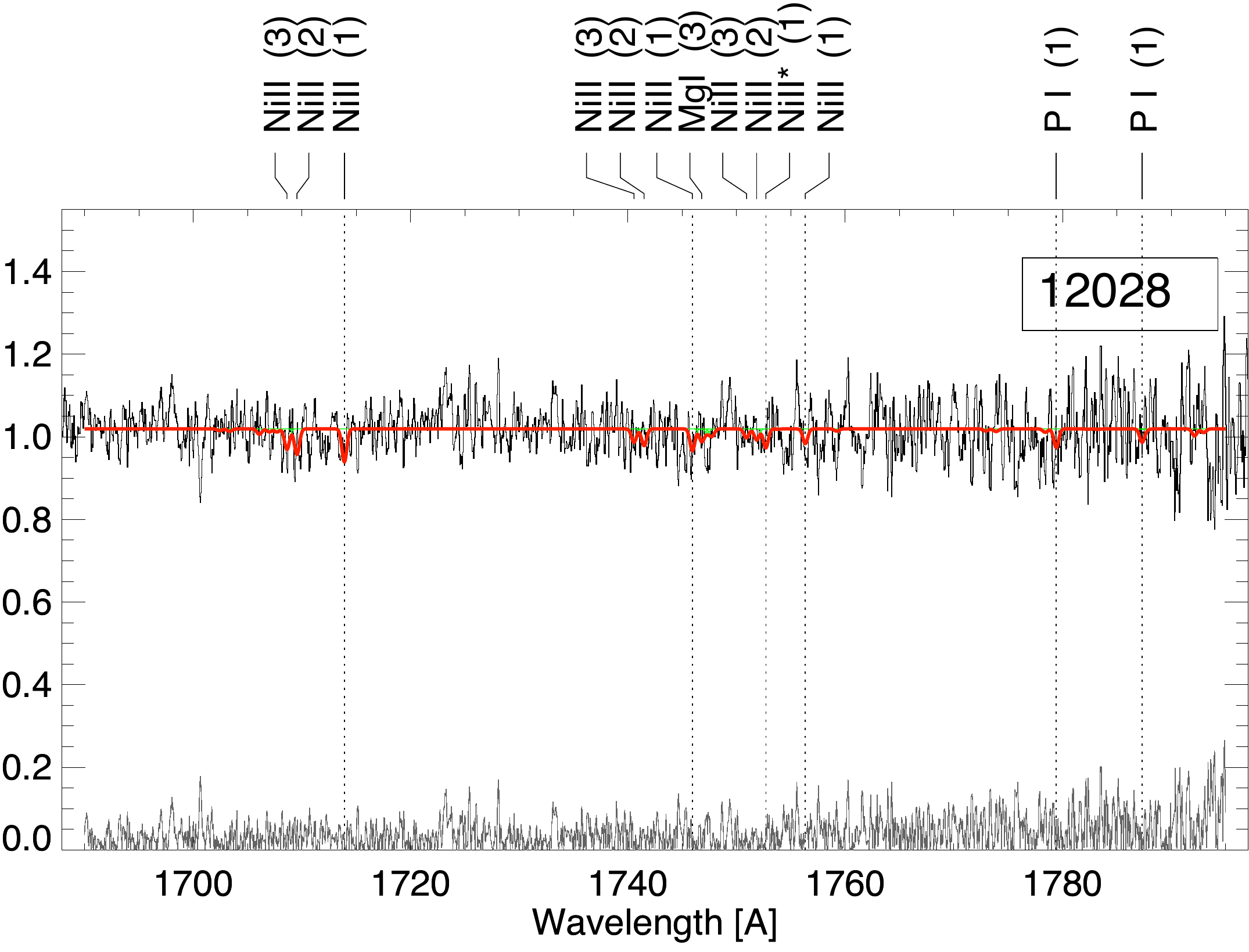}\\

\hspace{9.3cm}
\hspace{9.3cm}\\

\caption{See Fig.\,\ref{fig:fits1} for the plot description. }
\label{fig:fits7}
\end{figure*}

\end{document}